\documentclass[nature,twocolumn,longbibliography,superscriptaddress]{revtex4-2}

\usepackage{graphicx}
\usepackage[usenames,dvipsnames]{xcolor}
\usepackage{amssymb,amsfonts,amsmath}
\usepackage[colorlinks=true,citecolor=Cerulean,linkcolor=RubineRed,urlcolor=Cerulean]{hyperref}
\usepackage{epstopdf}
\usepackage{bm}
\usepackage{bbold}
\usepackage{upgreek}
\usepackage[inline]{enumitem}
\usepackage{booktabs}
\usepackage{multirow}
\usepackage[normalem]{ulem}
\usepackage{IEEEtrantools}
\usepackage{makecell}

\usepackage{tabularx}
\usepackage{float}

\usepackage{afterpage}

\usepackage{verbatim}

\usepackage{babel}
\usepackage{lipsum}
\usepackage{sidecap}
\usepackage{hyperref}
\hypersetup{
    colorlinks=true,
    linkcolor=magenta,
    filecolor=magenta,      
    urlcolor=cyan,
    pdfpagemode=FullScreen,
    }

\usepackage{xr}
\makeatletter
\newcommand*{\addFileDependency}[1]{
  \typeout{(#1)}
  \@addtofilelist{#1}
  \IfFileExists{#1}{}{\typeout{No file #1.}}
}
\makeatother
\newcommand*{\myexternaldocument}[1]{%
    \externaldocument{#1}%
    \addFileDependency{#1.tex}%
    \addFileDependency{#1.aux}%
}

\renewcommand{\vec}[1]{\mathbf{#1}}

\newcommand{\EXP}[1]{~\times~10^{#1}}
\newcommand{\Ang}{${\rm \AA ngstr\ddot{o}m}$}

\usepackage{txfonts}
\usepackage{latexsym}

\graphicspath{ {./figz/} }

\setcitestyle{super} 

 
\newcommand*{\citen}[1]{%
  \begingroup
  \romannumeral-`\x 
    \setcitestyle{numbers,open={},close={}}
    \cite{#1}%
  \endgroup   
}

\newlist{custominlist}{enumerate*}{1}
\setlist[custominlist]{label=(\roman*)}

\myexternaldocument{supplementLINK}

\begin{document}

\title{Quantum Electrometer for Time-Resolved Material Science at the Atomic Lattice Scale}

\author{Gregor Pieplow}
\altaffiliation{These authors contributed equally to this work}
\affiliation{Department of Physics, Humboldt-Universit\"{a}t zu Berlin, 12489 Berlin, Germany}

\author{Cem G\"{u}ney Torun}
\altaffiliation{These authors contributed equally to this work}
\affiliation{Department of Physics, Humboldt-Universit\"{a}t zu Berlin, 12489 Berlin, Germany}

\author{Charlotta Gurr}
\affiliation{Department of Physics, Humboldt-Universit\"{a}t zu Berlin, 12489 Berlin, Germany}

\author{Joseph H. D. Munns}
\affiliation{Department of Physics, Humboldt-Universit\"{a}t zu Berlin, 12489 Berlin, Germany}

\author{Franziska Marie Herrmann}
\affiliation{Department of Physics, Humboldt-Universit\"{a}t zu Berlin, 12489 Berlin, Germany}

\author{Andreas Thies}
\affiliation{Ferdinand-Braun-Institut (FBH), Gustav-Kirchhoff-Str. 4, 12489 Berlin, Germany}

\author{Tommaso Pregnolato}
\affiliation{Department of Physics, Humboldt-Universit\"{a}t zu Berlin, 12489 Berlin, Germany}
\affiliation{Ferdinand-Braun-Institut (FBH), Gustav-Kirchhoff-Str. 4, 12489 Berlin, Germany}

\author{Tim Schr\"{o}der}
\email[Corresponding author: ]{tim.schroeder@physik.hu-berlin.de}
\affiliation{Department of Physics, Humboldt-Universit\"{a}t zu Berlin, 12489 Berlin, Germany}
\affiliation{Ferdinand-Braun-Institut (FBH), Gustav-Kirchhoff-Str. 4, 12489 Berlin, Germany}

\begin{abstract}
The detection of individual charges plays a crucial role in fundamental material science and the advancement of classical and quantum high-performance technologies that operate with low noise. However, resolving charges at the lattice scale in a time-resolved manner has not been achieved so far. Here, we present the development of an electrometer with 60 ns acquisition steps, leveraging on the spectroscopy of an optically-active spin defect embedded in a solid-state material with a non-linear Stark response.
By applying our approach to diamond, a widely used platform for quantum technology applications, we can distinguish the distinct charge traps at the lattice scale, quantify their impact on transport dynamics and noise generation, analyze relevant material properties, and develop strategies for material optimization.  

\end{abstract}

\maketitle

\begin{figure*}[]
\centering
\includegraphics[width=\linewidth]{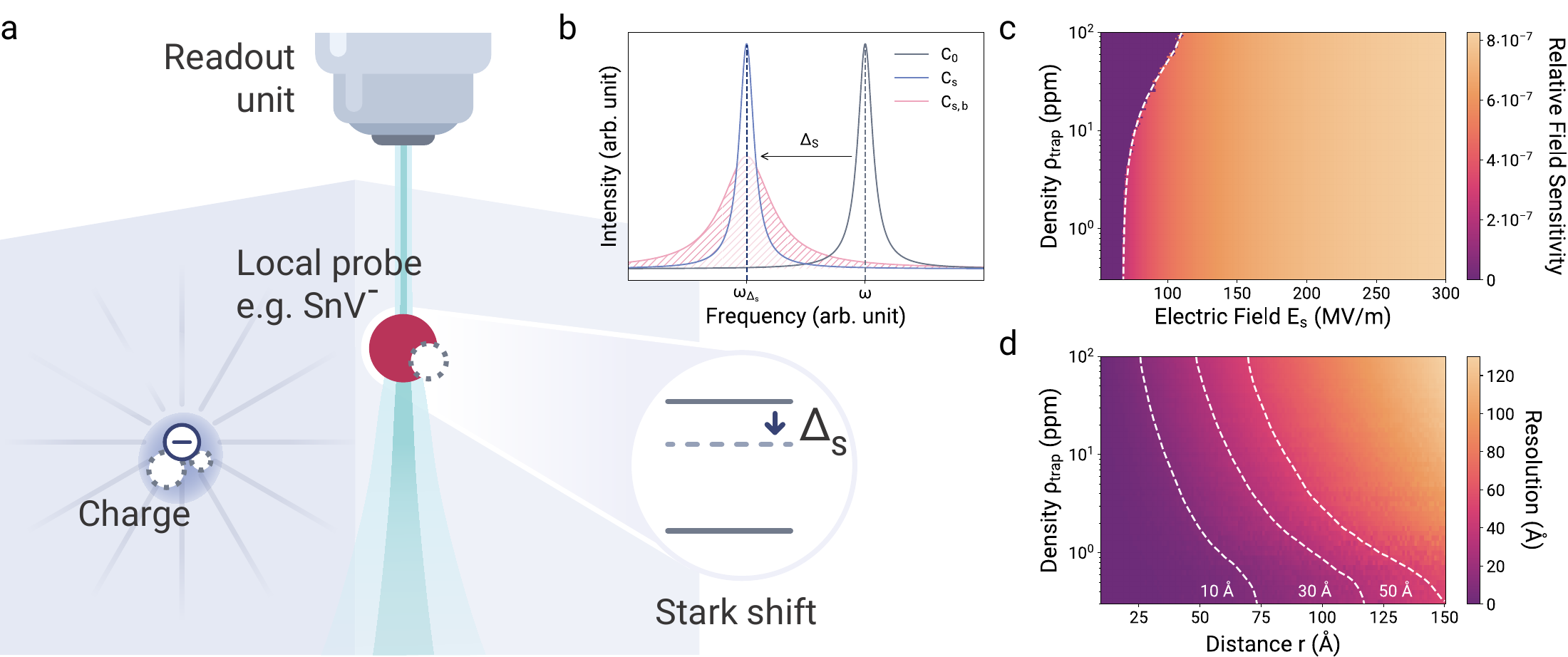}
\caption{
\protect\hypertarget{fig:principleL}{}
\textbf{Working principle of the quantum electrometer.} 
\textbf{a} 
The local probe is an optically-active atomic defect with non-linear Stark-sensitive energy-levels.
The read-out unit is a photoluminescence excitation spectroscopy microscope. 
\textbf{b} A nearby charge shifts the optical transition from C$_0$ to C$_{\rm s}$ by $\Delta_{\rm s}(\rm r)$ depending on its distance $r$. 
Additionally, an ensemble of remote fluctuating charges broadens the signal from C$_{\rm s}$ to C$_{\rm s,b}$ depending on the charge density $\rho_{\rm trap}$.
\textbf{c} Relative electric field sensitivity $|\Delta E| / E$ to electric field changes as function of electric field $E_{\rm s}$ and trap density $\rho_{\rm trap}$. Left of the white dashed line Stark-shifts are not large enough to be resolved by the Rayleigh-criterion. Larger field strengths are correlated with larger inhomogeneous broadening.  Eq.~\eqref{eq:nonlin_stark} assumes $\Delta\mu = 6.1 \times 10^{-4}$~GHz/({\rm MV/m})$^2$,  $\Delta\alpha= - 5.1 \times 10^{- 5}$~GHz/({\rm MV/m})$^2$,  $ \Delta\beta = - 5.5\times 10 ^{- 8}$ GHz/(MV/m)$^3$ and $\Delta\gamma=-2.2\times 10 ^{- 10}$ GHz/(MV/m)$^4$ \cite{Santis2021PRL}.
\textbf{d} The sensor's resolution in determining the distance of an elementary charge based on differentiating two distinct charge traps as a function of the charge trap density and distances. 
The resolution was determined for a trap with variable distance $r$ and bias-field equivalent to a trap distance $0.8$~nm. The dashed white lines indicate the inscribed resolution thresholds.}
\label{fig:principle}
\end{figure*}

\begin{figure*}[]
\centering
\includegraphics[width=\linewidth]{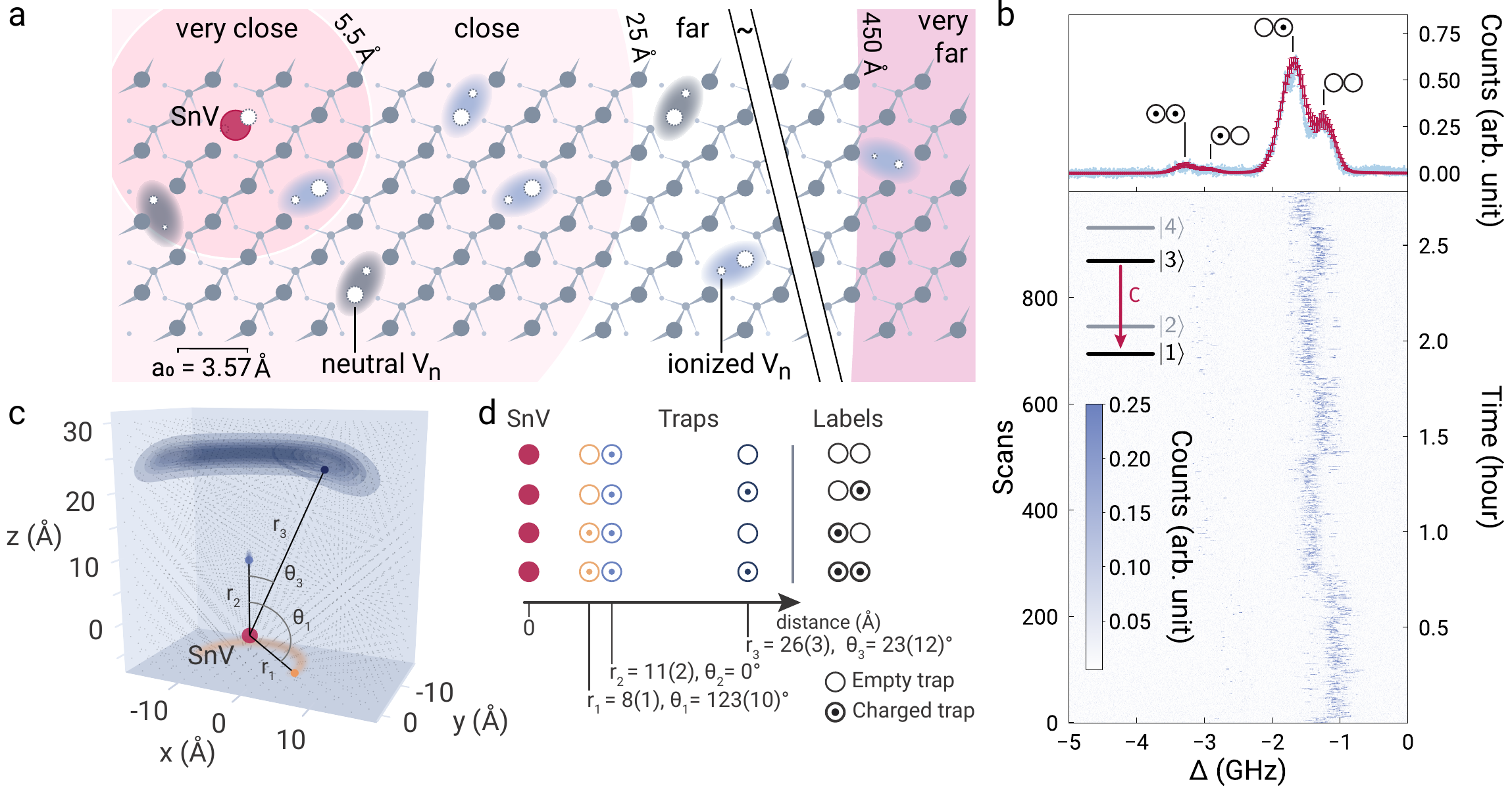}
\caption{
\textbf{Lattice scale localization of charge traps.} 
\protect\hypertarget{fig:sensorL}{}
\textbf{a} Illustration of a diamond lattice including SnV and defects, identified as multi-vacancy complexes (V$_{\rm n}$). Charges localized in these traps induce a Stark-shift of the energy levels of the atomic sensor probe. From very close to far, respectively, the spectral impact  of an elementary charge can be categorized as follows: Spectral shifts larger than 30 GHz are detectable by photoluminescence spectroscopy, spectral shifts $\sim$GHz are detectable by photoluminescence excitation spectroscopy (PLE), inhomogeneous broadenings are detectable by PLE. Charges in the very far region have negligible effects.
\textbf{b} TOP: An integrated multimodal PLE spectrum recorded with the SnV sensor S1 (blue) and modeled with Monte-Carlo simulations (red, error bars represent statistical standard deviation) to identify a proximity charge trap configuration (states labeled above the peaks, $\odot$ and $\medcirc$ represent ionized and neutral traps respectively) and the surrounding charge density.  
BOTTOM: Time-resolved individual linescans. INSET: SnV level scheme and the probed transition.
\textbf{c} LEFT: Identified charge trap configuration corresponding to (B), their relative positions and their probability distribution with respect to the SnV probe. Distributions resemble a donut shape due to the direction-independent calibration of the sensor. RIGHT: Table indicating the charge states and position of the identified traps.}
\label{fig:sensor}
\end{figure*}

\begin{figure*}[]
    \centering
    \includegraphics[width=\linewidth]{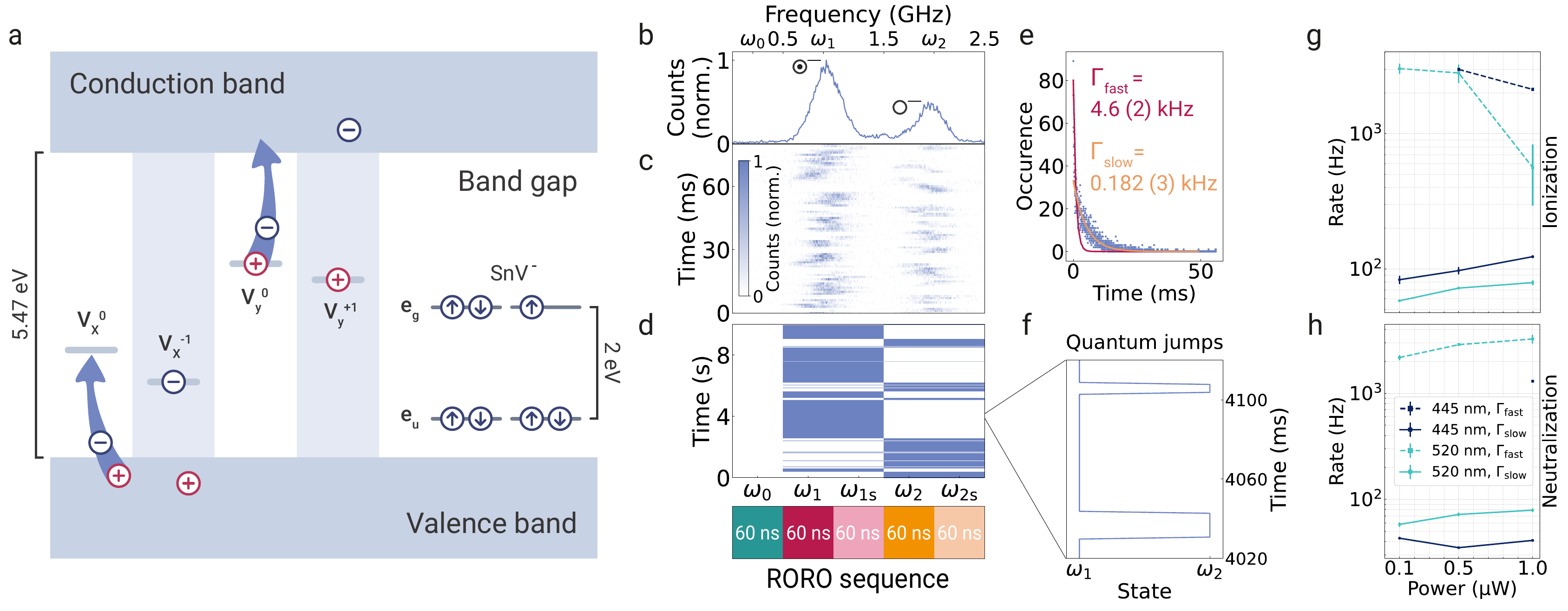}
    \caption{
    \textbf{Nanoscale charge dynamics in diamond.} 
   \protect\hypertarget{fig:dynamicsNewL}{}
    \textbf{a} Illustration of an SnV, ionized (V$_{\rm i}^{\pm1}$) and neutral (V$_{\rm i}^0$) defects, i.e. the charge traps,  in the diamond bandgap. Ionization occurs when either an electron is promoted from the valence band to the charge trap, or an electron is promoted from the trap to the conduction band by laser irradiation. Neutralization occurs when the trap either catches a hole from the valence band or an electron from the conduction band.
    \textbf{b} Integrated PLE spectrum and \textbf{c} time-resolved linescans acquired by the SnV sensor S2.
    \textbf{d} Using the specific resonance for each charge state obtained from the PLE spectrum, the laser frequency is periodically switched between the off-resonant $\omega_0$ and the resonances $\omega_1$, $\omega_2$. Each resonance is probed in two 60 ns steps with an additional frequency offset to compensate for spectral diffusion. The entire rapid optical readout (RORO) sequence takes 300 ns. Counts registered during each acquisition step are analyzed to determine how long a particular resonance remains bright.
    \textbf{e} An example histogram of the bright times of the resonance $\omega_1$ before a spectral jump to $\omega_2$ occurred. The data is fitted with a bi-exponential function that has two rates with constants $\varGamma_{\rm fast}$ and $\varGamma_{\rm slow}$.
    \textbf{f} A zoom-in of the spectral jumping plot from subfigure d reveals fast quantum jumps occurring on the order of ms.
    \textbf{g},\textbf{h} To investigate the influence of photon irradiation energies on the recharging dynamics, spectral jumping rates are acquired under distinct laser wavelengths and powers. Rates are extracted separately for the investigated nearby trap's spectral jumping directions corresponding to ionization (redshift) and neutralization (blueshift) processes. Dark blue data points are acquired under 445 nm and green data points under 520 nm laser illumination. For each wavelength the measurement is repeated at three different powers. Data points connected by solid lines correspond to the extracted $\varGamma_{\rm slow}$ exponential rate constant. Fast rates $\varGamma_{\rm fast}$ are connected by dashed lines and shown for histograms in which a fit to a bi-exponential distribution was possible. The error bars represent 95\% confidence intervals extracted from the fits.
    } 
    \label{fig:dynamics}
\end{figure*}

\begin{figure}[]
    \centering
    \includegraphics[width=0.8\linewidth]{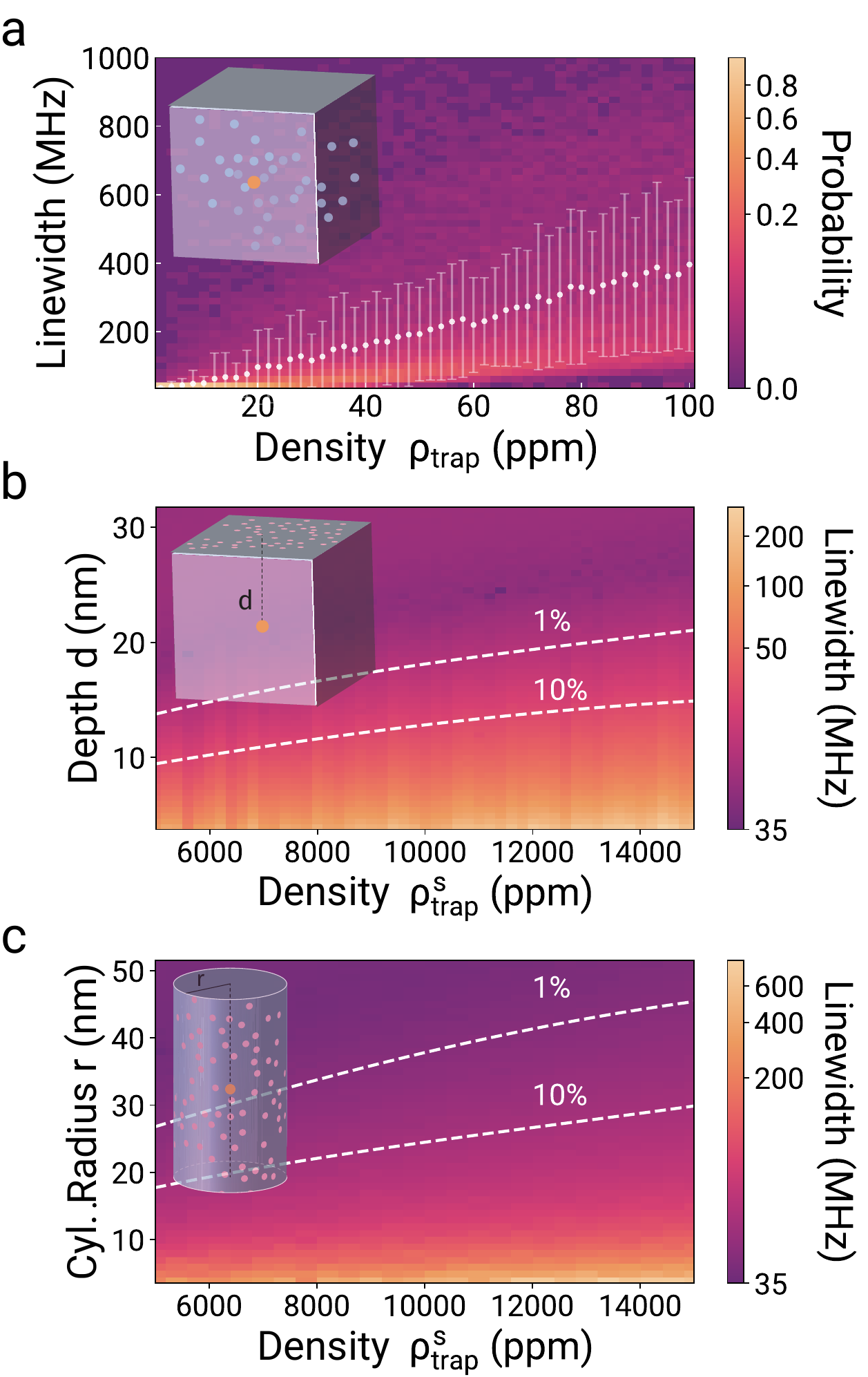}
    \caption{
    \textbf{Simulations of the SnV zero phonon line inhomogeneous broadening due to bulk and surface charges.} 
    \protect\hypertarget{fig:broadeningL}{}
    \textbf{a}  Inhomogeneous broadening of an SnV with a lifetime-limited linewidth of $35$~MHz
    as a function of $\rho_{\rm trap}$ in units of ppm in bulk diamond. The colors indicate the linewidth distribution. The mean and variance of the distribution are shown by the white dots and error bars.
    \textbf{b} Inhomogeneous broadening as a function of the distance of an SnV to a planar surface and the surface trap density $\rho^{\rm s}_{\rm trap}$ defined as fraction of surface lattice sites. The dashed lines show the threshold of 1\% and 10\% of broadening compared to the lifetime limited linewidth of 35~MHz 
    \textbf{c} Inhomogeneous broadening of an SnV centrally located in a cylinder with radius $r$ as a function of $\rho^{\rm s}_{\rm trap}$. The dashed lines signify $1\%$ and $10\%$ broadening.}
    \label{fig:broadening} 
\end{figure}

\begin{figure*}[]  
    \centering
    \includegraphics[width=\linewidth]{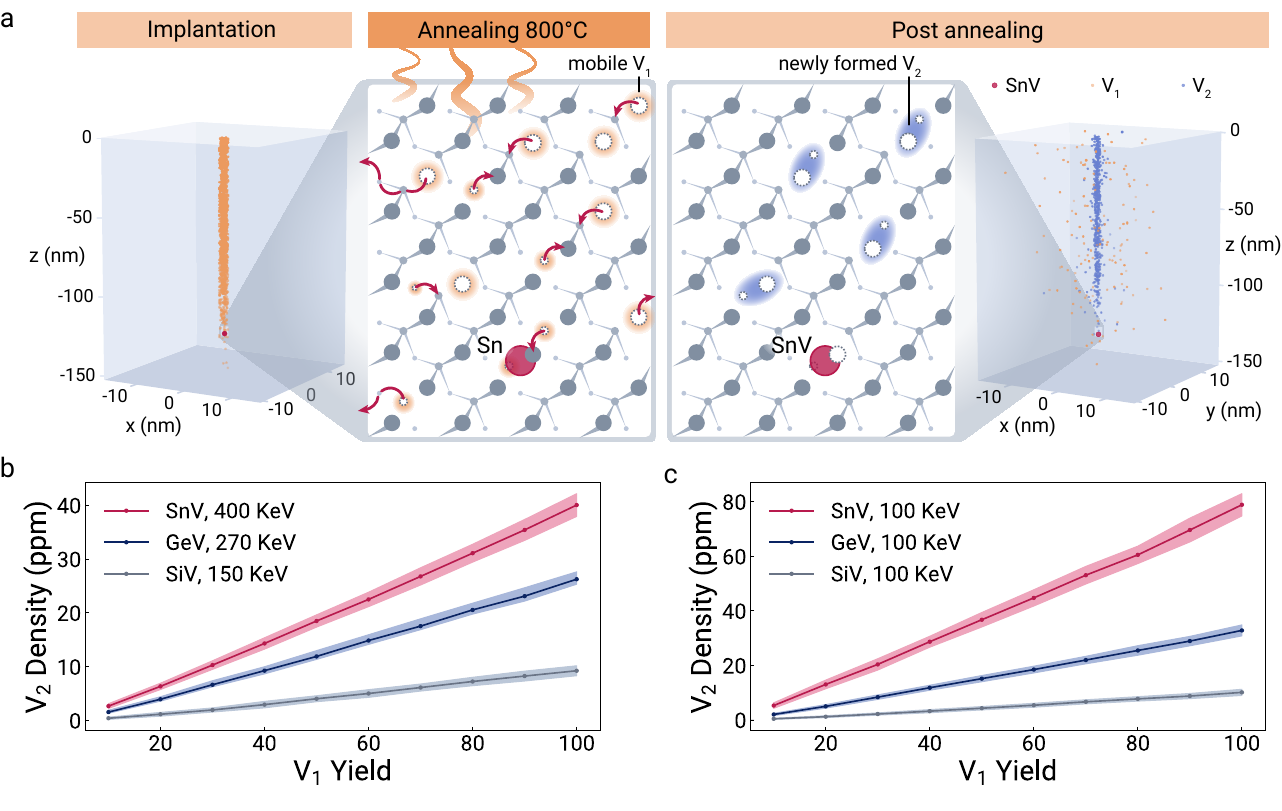}
    \caption{\label{fig:fig4}
    \textbf{Simulations of divacancy (V$_2$) creation during annealing.}
    \protect\hypertarget{fig:annealingL}{}
    \textbf{a} Implantation panel: Spatial distribution of created monovacancies (V$_1$) by 400~keV Sn implantation predicted by SRIM \cite{Ziegler2010NIM} simulations.
    Annealing panel: V$_2$ formation during annealing at 800~$^\circ$C. At elevated temperatures, V$_1$ start diffusing. Then, V$_1$ immobilize either by moving to the boundaries, recombining with interstitial carbons, or forming V$_2$. 
    Post annealing panel: Distribution of V$_1$ and V$_2$ which are distributed in the vicinity of the damage channel caused by the Sn implantation. Zoom in: V$_2$ in the vicinity of the SnV.      
   \textbf{b} Simulated densities of V$_2$ as a function of the V$_1$ implantation yield (\% of participating V$_1$ estimated by SRIM simulations) for the three different species Sn, Ge, and Si. A reduced V$_1$ yield is attributed to recombination with interstitial carbons. The implantation energies are selected so that an average implantation depth of 100~nm is reached. 
   \textbf{c} Densities of V$_2$ for 100~keV implantation energy. The error bands represent the statistical standard deviation.
   }
   \label{fig:annealing}
\end{figure*}

\section*{Introduction}

Free charge carriers such as electrons are essential components of our modern-day world. They facilitate devices such as smartphones and computers. Uncontrolled or undesired charges, on the other hand, can cause damage and reduce the performance of such devices. Prominent examples are gate-oxide breakdown in flash memory \cite{kaczer_impact_2002} and charge-noise on the nanoscopic level \cite{stacey_evidence_2019, Rehman2022}. 
The detection and quantification of desired and undesired charge carriers with electrometers \cite{Defo2023PRB,Mittiga2018PRL} holds significant technological importance on the nanoscale. 

Despite significant progress \cite{Vamivakas2011,dolde2011,dolde2014,Chen2016,Iwasaki2017ACS,Mittiga2018PRL,broadway2018,li_2020,barson_nanoscale_2021,Qiu2022npj, delord_correlated_2024-1, Ji2024}, electrometers have to date been unable to provide time-resolved access to elementary charges with subnanometer resolution.
Precicely resolving charges and performing temporal analysis at atomic lattice scales, however, is becoming increasingly important. For example, the investigation of 2D ferroelectric systems would greatly benefit from the use of a highly sensitive electrometer, which could provide critical insights into the unresolved foundational aspects of their physical characteristics \cite{song_reply_2023}. Furthermore, as silicon transistors reduce to sizes of a few nanometers they become more susceptible to charge-induced noise \cite{uren1985, Stampfer2020phd}. 

Particularly, quantum technology applications face challenges: in ion based quantum computers localized electronic states are suspected to cause decoherence due to motional heating \cite{hite2013}; superconducting qubits suffer from defect-induced charge-noise \cite{faoro2006,reshef2021}; in atom-like spin qubits in wide-bandgap semiconductors, charge-noise leads to optical and spin decoherence \cite{Mittiga2018PRL,barson_nanoscale_2021,Morioka2020NatComm,White2021Opt,Vajner2022AQT},
significantly limiting the development of quantum networking and sensing \cite{OrphalKobin2023PRX,Pompili2021Sci}.

Understanding the underlying mechanisms of such platform-specific detrimental processes is a necessity to improve the performance and application range of nanoscale electronic and photonic devices including open questions around decoherence processes, electron dynamics, and material questions of lattice defect formation.

Here, we introduce a quantum electrometer that enables the detection of electric fields produced by single and multiple elementary charges with a relative sensitivity of $10^{-7}$ and resolves individual charge state dynamics down to tens of nanoseconds at the \Ang-scale. 

The electrometer consists of an electric field sensitive, optically-active, local probe and a read-out unit (Fig.~\ref{fig:principle}\hyperlink{fig:principleL}{a}). In this demonstration, the local probe is a negatively-charged tin-vacancy color center (SnV) in diamond  \cite{Iwasaki2017PRL}, a solid-state defect with fluorescent transitions and a
non-linear electric field response, which is typical for defects in the D$_{\rm 3d}$ point group \cite{Goss1996PRL, Iwasaki2015SciRep, Niaz2016JPCC, Iwasaki2017PRL, Wang2021ACS, ShkarinPRL2021}. 
The optical transition energies directly depend on the local electric field via the DC-Stark-effect 
$\Delta_{\rm Stark} = - \boldsymbol{\mu}_{\rm ind}(E_{\rm s}) E_{\rm s}$, 
where ${\mu}_{\rm ind}$ is the induced dipole moment of the atomic defect and $E_{\rm s}$ is the sum of all static electric fields produced by surrounding charges, which shift the optical energies \cite{Stark1914} (Fig.~\ref{fig:principle}\hyperlink{fig:principleL}{b}).

The read-out unit is a microscope which is used to perform photoluminescence excitation spectroscopy on the local atomic probe, and therefore does not require magnetic resonance methods \cite{Mittiga2018PRL}. 
Measuring the energy shift reveals the magnitude of the electric field at the sensor probe $E_{\rm s}$ via the DC-Stark-shift
\begin{align}
    \begin{aligned}
        \Delta_{\rm Stark} & = -\Delta\mu E_{\rm s} - \frac{1}{2} \Delta\alpha E_{\rm s}^2 - \frac{1}{3!} \Delta\beta E_{\rm s}^3 - \frac{1}{4!} \Delta\gamma E_{\rm s}^4 ~,
    \end{aligned}\label{eq:nonlin_stark}
\end{align}
with $\Delta \mu$ the change in the dipole moment and $\Delta\alpha$, $\Delta\beta$, and $\Delta\gamma$ the differences between the higher-order polarizabilities \cite{Santis2021PRL}.
Importantly, in contrast to non-inversion symmetric configurations of color centers, such as the nitrogen-vacancy center in diamond \cite{Tamarat2006PRL} and the silicon-vacancy center in silicon carbide \cite{Ruhl2020ACS}, 
the negligible linear- and strong non-linear response due to inversion symmetry
makes the sensor applicable to typical semiconductor dopant and defect densities.
If $\Delta\alpha$ dominates and the observed $\Delta_{\rm Stark}$ arises from a localized elementary charge $e$ at a distance $r$ from the sensor, then:
\begin{equation}
    \Delta_{\rm Stark}(r) \sim \Delta \alpha / r^{4}~.
    \label{eq:charge_distance}
\end{equation}
Decreasing distances of charges to the sensor causes increasingly higher spectral shifts. 
This characteristic makes sensors that have an inversion center remarkably sensitive to charges in close proximity and insensitive to electric field background noise. 

The $\sim10^{-7}$ relative electric field sensitivity (Fig.~\ref{fig:principle}\hyperlink{fig:principleL}{c}) allows for a spectral sensor read-out that provides exceptionally high spatial resolution, reaching down to few \Ang, even for charge densities up to hundred ppm (Fig.~\ref{fig:principle}\hyperlink{fig:principleL}{d}).

In this work, the SnV local probe is stationarily located inside a bulk crystal housed in a cryostat at 4 K, however, it could also be integrated into the tip of a scanning probe microscope \cite{Huxter2023NatPhys}
for position dependent measurements well established in magnetometry \cite{Maletinsky2012NatNT,Budker2007NatPhys} 
or into a nanodiamond for integration with other materials \cite{Foy2020ACS} or even biological samples \cite{Kucsko2013Nat}. 
Alternatively to the SnV, also other D$_{\rm 3d}$ symmetric defects such as the silicon- \cite{Goss1996PRL} or germanium-vacancy \cite{Iwasaki2015SciRep, li_atomic_2024} and other inversion symmetric defects in other materials, for example, in silicon \cite{Hourahine2000} could be applied as local probe. 
To demonstrate the non-linear sensor principle, we utilize a single SnV that was created upon ion implantation and annealing. 

\section*{Results}
\subsection*{Determining charge trap positions at the atomic lattice scale}

The electrometer and its environment are depicted in Fig.~\ref{fig:sensor}\hyperlink{fig:sensorL}{a}.
To demonstrate its sensing capability, we analyze the time-varying  quasi-static electric field that is caused by charging and neutralization of crystal defects in the surrounding lattice under laser irradiation \cite{gaubas_spectroscopy_2016, girolami_transport_2019}.
Using the recorded field magnitude at different configurations of the charge distribution,  we can resolve the surrounding crystal defects at the lattice scale.

If all traps are neutral, the total field at the position of the probe is zero and the optical transition of the SnV is unperturbed. A charged trap will induce an electric field $\Vec{E}_{\rm s}$, which Stark-shifts the optical transition energy according to Eq.~\eqref{eq:nonlin_stark}. 
If a single elementary charge is located in proximity to the probe, the C transition is shifted more than its own linewidth producing a spectral jump.  (Fig.~\ref{fig:principle}\hyperlink{fig:principleL}{b}). The magnitude of the spectral shift can be determined by comparing with the unperturbed case, adding both resonances in one spectrum leads to a unique optical fingerprint with two peaks.

To collect the spectra, we measure the fluorescence from the C transition of the sensor S1 under photoluminescence excitation (PLE) with a narrowband laser (Fig.~\ref{fig:sensor}\hyperlink{fig:sensorL}{b}).
The experiment is conducted under 0.5 nW (a power much below the expected saturation power \cite{Trusheim2020PRL} for excluding power broadening effects) of 619 nm resonant light. Additionally, a 2 µW (CW power) 450 nm 4 ms pulse is applied between line scans for SnV charge state reinitialization.
From the spectral shift we extract the charge induced electric field. Knowing the local field and using the polarizability enables us to estimate the trap-probe distance. 

For $N$ charged traps in the probe’s vicinity, the electric fields add up to $\Vec{E}_{\rm s}^{'}$ and the individual charges cannot directly be separated.
To distinguish the $2^{N}$ charge states, the Stark-shifted PLE spectra are recorded repeatedly. Due to laser irradiation, the traps will be ionized and neutralized randomly. By sampling a large set of configurations complex trap distributions can be analyzed.

In addition to nearby charges that cause significant spectral line shifts, the numerous randomly distributed traps in the distant surrounding also contribute. 
These traps exhibit fluctuating charge states, resulting in a fluctuating electric field $\delta \Vec{E}_{\rm s}$ that causes inhomogeneous broadening. 
Consequently, the density of charge traps $\rho_{\rm trap}$ within the lattice can be determined using linewidth measurements.
We find that traps can be resolved with subnanometer resolution. For trap densities $\rho_{\rm trap}{\approx}0.3~{\rm ppm}$ detection volumes of $150^3$~\AA$^3$ are feasible. Fluctuating charge traps at larger distances primarily contribute to inhomogeneous broadening.

Finally, to fully calibrate the electrometer, we consider its non-linear response to external fields causing an interdependence of the different external field components. 
For example, the effective Stark-shift induced by two charges does not equal their sum. 
This phenomenon enables high resolution but makes the analysis of recorded fingerprints highly complex. 
We therefore build a theoretical database of simulated spectra for a large variety of discrete proximity charge positions and remote trap densities using Eq.~\eqref{eq:nonlin_stark}. 

We now analyze the complex experimental four peak fingerprint from Fig.~\ref{fig:sensor}\hyperlink{fig:sensorL}{b} 
quantitatively. We use experimentally determined polarizabilities \cite{Santis2021PRL}.
By comparing experimental and simulated fingerprints, we  find several possible trap configurations. From these possible configurations we identify the most plausible by specific physical considerations (Supplementary Fig.~\ref{fig:biasAnalysis}).

The most likely configuration of nearby traps consists of a permanent $\vec{E}_{\rm bias}$, generated for example by a permanently ionized trap, and two additional traps inducing spectral jumps. We assign labels to the spectral peaks in Fig.~\ref{fig:sensor}\hyperlink{fig:sensorL}{b} based on the charge state of the two additional proximity traps $\{\medcirc\medcirc, \medcirc\odot, \odot\medcirc, \odot\odot\}$, where $\medcirc$ signifies an uncharged trap, and $\odot$ a charged trap.
Subsequently, we determine the position of these charge traps up to an azimuthal angle using Monte-Carlo simulations. We extract the relative Stark-shifts corresponding to the proximity trap distances $r_{\rm 1} =8(1)$~\AA, $r_{\rm 2} = 11(2)$~\AA,   $r_{\rm 3} = 26(3)$~\AA~(Fig.~\ref{fig:sensor}\hyperlink{fig:sensorL}{c, d}) and a remote trap density of 74(22)~ppm. Furthermore we find for two charges, azimuthal angles with $\theta_1 = 123(10)^\circ$ and  $\theta_3 = 23(12)^\circ$. The polar angles cannot be further confined because of the cylindrical symmetry of the problem setup.

\subsection*{Charge dynamics}

For identifying the positions of charge traps, we have used accumulated spectral fingerprints that reflect the integrated spectrum for the entire set of charge states. However, understanding time-resolved charge transfer dynamics requires comparing individual read-out events of our electrometer, and therefore single PLE linescans. 
We interpret the charge state changes with a simplified charge transfer picture (Fig.~\ref{fig:dynamics}\hyperlink{fig:dynamicsNewL}{a}): 
Charge traps, later identified as multi-vacancy complexes V$_{\rm n}$, can be ionized under laser illumination through two distinct processes: negative charging, which occurs when the trap captures an electron promoted from the valence band leaving a positively-charged hole in the band; and positive charging when an electron is promoted from the trap to the conduction band. Generated holes and promoted electrons then diffuse and recombine with other charge traps leading to an overall charge-neutral environment. We denote the event ${\medcirc\rightarrow\odot}$ as ionization and the inverse ${\odot\rightarrow\medcirc}$ as neutralization event.
The charge transfer picture \cite{Defo2023PRB,Goerlitz2022NPJ} is consistent with the autocorrelation measurements we performed on the sensor emission (Supplementary Fig.~\ref{fig:bunching}).

To showcase the characterization of a local charge environment and dynamics of the charge transfer processes, we acquire a high timing-resolution set of data from a second sensor S2. The temporal analysis of the data acquired from S1 is provided in the supplementary materials. S2's charge dynamics are dominated by a single spectral jump as shown in Fig.~\ref{fig:dynamics}\hyperlink{fig:dynamicsNewL}{b}. We label the two states as the neutral $\medcirc$ and ionized $\odot$ states following our previous description. 

The individual lines in Fig.~\ref{fig:dynamics}\hyperlink{fig:dynamicsNewL}{c} are acquired with a GHz/µs scanning speed by using a chirped frequency modulation of an EOM generated sideband of a narrowband laser.

After identifying the resonance peaks in the integrated spectrum using the chirped frequency modulation, we perform a rapid optical read-out (RORO) of the charge trap (Fig.~\ref{fig:dynamics}\hyperlink{fig:dynamicsNewL}{d}). We generate a readout sequence by repeatedly modulating the sideband targeting five frequencies with 60 ns acquisition steps: off-resonant $\omega_0$ ,  the peaks of two resonances $\omega_{\rm 1,2}$ and some pre-selected frequency offset $\omega_{\rm 1s,2s}$ to probe the resonances even if some spectral diffusion has taken place. The entire RORO sequence lasts a total of 300 ns.
Using a single-shot read-out sequence, we extract a specific charge state of the sensor. We then visualize the distribution of duration the sensor stayed on one of the resonances in a histogram, as exemplified for the first resonance $\omega_1$ in  Fig.~\ref{fig:dynamics}\hyperlink{fig:dynamicsNewL}{e}. The histograms are fitted with a bi-exponential distribution featuring two charge transfer rates $\varGamma_{\rm fast}$ and $\varGamma_{\rm slow}$ reflecting slow and fast charge dynamics as detailed below.

To analyze the charge dynamics on the shortest time scale available in the specific measurement configuration, we zoom into the data of Fig.~\ref{fig:dynamics}\hyperlink{fig:dynamicsNewL}{d} and look closely at the quantum jumps (as shown in Fig.~\ref{fig:dynamics}\hyperlink{fig:dynamicsNewL}{f}). We observe that although a time period can look bright on the scale of seconds in Fig.~\ref{fig:dynamics}\hyperlink{fig:dynamicsNewL}{d}, faster jumps occur on the millisecond scale. We further investigate the source of the bi-stability of the rates by selectively analyzing windows from an extended dataset acquired under the same conditions as in Fig.~\ref{fig:dynamics}\hyperlink{fig:dynamicsNewL}{d}. We find that if a time interval is predominantly bright for one resonance, the jumping rate from that state is closer to the extracted slow rate. For example, a predominantly bright time window of a total of $\sim$31 s for $\omega_1$ yields a jumping rate $\varGamma_{\omega_1\rightarrow\omega_2}=$ 26 (5) Hz. On the other hand, when the time interval seems mainly dark the jumping rate decays quicker, e.g., during a predominantly dark time window of $\sim$13 s for $\omega_1$ yields $\varGamma_{\omega_1\rightarrow\omega_2}=$ 1.91 (13) kHz. While identifying the underlying physical process of the rate change requires further investigation, one possible explanation could be the change of a distant charge donor's trap state. The rate of the spectral jump may be modified by the availability and quantity of the required free charged particles.

To further experimentally investigate the origin of the charge dynamics, we repeat the measurement with a second laser applied during the off-resonant part of the readout sequence to characterize the charge dynamics induced by the properties of the illumination.
Ultimately, such a measurement allows us to spectroscopically investigate the local charge environment and, more specifically, how its dynamics are influenced by external laser irradiation or other sources of noise in the nanoscopic local environment.
We investigate six conditions: for both 455 nm and 520 nm wavelengths, three laser power levels are set. We plot the extracted rates for ionization and neutralization processes, respectively, in  Fig.~\ref{fig:dynamics}\hyperlink{fig:dynamicsNewL}{g} and \ref{fig:dynamics}\hyperlink{fig:dynamicsNewL}{h}.

An interesting and at the first sight unexpected observation for the neutralization process is observed. Green illumination induces faster jumps than the higher energy blue illumination. This may be caused by a two-photon process involving a defect level within the bandgap that is more likely to interact with the green laser than direct ionization from or to the valance and conduction bands. Acquiring data at additional power levels,  illumination wavelengths and conducting measurements on different emitters with varying trap environments would enable extending this proof-of-principle experiment to a more detailed study.

Given the exceptional timing resolution of our RORO method, we can determine charge processes exceeding kHz rates, which are orders of magnitude faster than the typical $\sim$1 Hz  modulation rates of tunable lasers. Using conventional PLE by applying a voltage signal is not quick enough to observe the charge dynamics in the sample. Moreover, RORO could reveal charge processes with several MHz rates by utilizing the 60 ns minimum read-out steps. Reaching this limit requires estimated photon detection rates of $\sim$17 MHz. By further reducing emitter lifetimes, for example through the Purcell-enhancement by cavity integration, the 60 ns read-out resolution could be reduced to several nanoseconds (see Supplementary Table \ref{tab:countRates}).

\subsection*{Evaluating spectral diffusion} \label{sec:specdiff}

The charge transfer induced spectral dynamics are highly detrimental for quantum technological applications. Spectral diffusion, a term indicating the probabilistic nature of the observed spectral dynamics, leads to optical decoherence \cite{Morioka2020NatComm,White2021Opt,Vajner2022AQT,OrphalKobin2023PRX}, which results in reduced entanglement fidelity in quantum network nodes \cite{Kambs2018NJP,Pompili2021Sci}.

Knowing the non-linear receptiveness of our quantum electrometer to charge-noise, we now make predictions of how a specific charge distribution influences the spectral properties of a color center. Based on our model, we provide an overview of the inhomogeneous broadening caused by a certain charge trap density $\rho_{\rm trap}$. The details of the calculation are provided in supplementary information. 

We first focus on the bulk case (Fig.~\ref{fig:broadening}\hyperlink{fig:broadeningL}{a}) and then analyze surface charge traps ($\rho^{\rm s}_{\rm trap}$) for two different surface geometries, planar (Fig.~\ref{fig:broadening}\hyperlink{fig:broadeningL}{b}) and cylindrical (Fig.~\ref{fig:broadening}\hyperlink{fig:broadeningL}{c}). 
We find that an implantation depth of $d{ >} 21$~nm and a cylinder with a radius of $r{ > }$45~nm will warrant that surface charges do not deteriorate the spectral properties of an SnV color center with linewidth broadening  of less than 1\%. Such broadening leads to 90\% interference visibility and more than  $87\%$ entanglement fidelity \cite{Kambs2018NJP}. Similar estimations can be performed for any defect given a known polarizability. 
Control measurements concerning spectral diffusion with respect to the illumination field are provided in the supplementary information.

The estimated minimum detrimental distances make SnVs and similar color centers well suited for the integration into nanostructures that enhance photon collection efficiencies \cite{Torun2021APL} and provide tailoring emission properties via the Purcell-effect \cite{Li2015NatComm,Bopp2022Arxiv} for quantum information applications. 

It must be noted that an increase in the absolute lifetime limited linewidth due to the Purcell effect would decrease the depth and radius criteria. Furthermore, these analyses do not take band-bending and Fermi level modification effects into account, which could have additional implications on the charge state stability and brightness of the color centers.

\begin{table*} []
    \centering
    \caption{\label{tab:literatureReview}
            \textbf{Summary of previously developed electrometers or related works.} 
            NV - Nitrogen-Vacancy Center, 
            RT - Room Temperature, 
            MW - Microwave, 
            QD - Quantum Dot, 
            DC - Direct Current,
            AC - Alternating Current,
            FID - Free Induction Decay, 
            PLE - Photoluminescence Excitation, 
            CDD - Continuous Dynamical Decoupling
            ODMR - Optically Detected Magnetic Resonance, 
            GeV - Germanium-Vacancy Center, 
            PL - Photoluminescence, 
            SnV - Tin-Vacancy Center, 
            MCS - Monte-Carlo Simulations, 
            RORO - Rapid Optical Readout.           
            `Ramsey, FID, CDD' and `XY4, Hahn-Echo, Qdyne' are MW spin control-based phase-sensitive methods suitable for DC and AC measurements, respectively. 
        }  
        \begin{tabular}{c c c c c c c c}
        \toprule 
            \bf{Reference} & \bf{Platform} & \thead{\bf{Single Charge}\\\bf{Localization}} & \bf{Precision} & \thead{\bf{Sensitivity in} \\ \bf{V~/~m~/$\sqrt{\textnormal{Hz}}$}} & \bf{Temporal Analysis} & \bf{Based on} & \thead{\bf{Operating} \\ \bf{Temperature}} \\ \midrule
            \citen{Mittiga2018PRL} & NV & \checkmark & $\sim$ nm & $\times$  & $\times$ & \thead{MW Polarization\\Dependent ODMR} & RT\\
            \citen{Vamivakas2011} & QD & $\checkmark$ & $\times$ & 5 (DC), 140 (AC) & $\times$ & Modulation spectroscopy & 4.2 K\\
            \citen{dolde2011} & NV & $\times$ & $\times$ & 891 (DC), 202 (AC)  & $\times$ & \thead{FID (DC), Hahn-Echo (AC)} & RT \\ 
            \citen{dolde2014} & NV & $\checkmark$ & 2 nm & $\times$  & $\times$ & Ramsey & RT \\ 
            \citen{Chen2016} & QD & $\checkmark$ & 70 nm volume  & $\times$ & $\times$ & PLE, Photon statistics & 5 K\\
            \citen{Iwasaki2017ACS} & NV & $\times$ & $\times$ & $\times$  & $\times$ &  ODMR & RT\\ 

\citen{broadway2018} & NV & $\times$ & $\times$ & $\times$  & $\times$ &  ODMR & RT\\ 
\citen{li_2020} & NV & $\times$ & $\times$ & $\times$  & $\times$ &  CDD & RT\\ 
\citen{barson_nanoscale_2021} & NV & $\times$ & $\times$ & $\times$  & $\times$ &  Qdyne & RT\\ 
\citen{Qiu2022npj} & NV & $\times$ & $\times$ & $26\times10^3$ (AC)  & $\times$ &  \thead{Ramsey Lock-In (DC)\\ XY4 Lock-In (AC)} & RT\\

            \citen{delord_correlated_2024-1} & NV & \checkmark & 1.6 nm & $\times$  & \checkmark, $\sim$1 Hz  &  PLE, correlated analysis & 9 K\\
            \citen{Ji2024} & NV & \checkmark & 1.7 nm & $\times$  & \checkmark, $\sim$1 Hz  &  PLE, correlated analysis & 11 K\\
            \citen{li_atomic_2024} & GeV & $\checkmark$ & $\times$ & $\times$  & \checkmark, $\sim$1 Hz &  PLE & 4 K\\
            \citen{Bluvstein2019PRL} & NV & $\times$ & $\times$ & $\times$  & \checkmark, 1 kHz &  PL & RT\\ \midrule
            Our work & SnV & \checkmark & 1 $\mathring{\textnormal{A}}$ & $\times$  & \checkmark, $\sim$ MHz &  PLE, MCS, RORO & 4 K \\ \bottomrule
        \end{tabular}
        
\end{table*}

\subsection*{Identifying material properties: divacancy formation}

Based on the electrometers ability to quantify charge trap densities, we extend our investigation of material properties and combine our sensor data with additional simulations to answer further open research questions.
Specifically, we determine the physical origin of charge traps in implanted diamond.
Considering a sample with less than 1~ppb of nitrogen and boron, and even lower lattice defect concentrations \cite{ElementSix2022}, the estimated charge trap density of 74(22) ppm must originate from the Sn-ion implantation damage and the subsequent annealing process. 
Implantation of ions produces Frenkel pairs: a pair of one monovacancy V$_1$ and one dislocated interstitial carbon atom.
During annealing V$_1$ becomes mobile and can form vacancy complexes, a process not well understood and an active area of research \cite{yamamoto_extending_2013, Slepetz2014PhysChem, onoda_diffusion_2017} 
(Fig.~\ref{fig:annealing}\hyperlink{fig:annealingL}{a}). 

Here, we estimate the V$_1$ to divacancy V$_2$ conversion yield using a kinetic Monte-Carlo simulation \cite{favaro_de_oliveira_tailoring_2017} 
in combination with a simple stochastic diffusion model described in the supplementary information. 

We consider the V$_2$ density as a proxy for higher-order vacancy complexes V$_{\rm n}$. 
Annealing up to 1100~$^\circ$C primarily converts V$_2$ into V$_3$ and V$_4$ \cite{Lomer1973,yamamoto_extending_2013}. 
We indeed observe wavelength-dependent spectral diffusion and jumps (Supplementary Fig.~\ref{fig:greenVSblue}) indicating different ionization energies of the multiple trap species.
We interpret the estimated density of V$_2$ as both an order-of-magnitude-approximation and an upper limit for the overall charge trap density. We estimate and compare $\rho_{\rm V_2} = 40.0(2.1)$~ppm to the experimentally estimated trap density of $\rho_{\rm Exp} = 74.1(22.5)$~ppm. Due to charge neutrality, the overall charge density $\rho_{\rm Sim}$ would correspond to twice the density of V$_2$ with $\rho_{\rm Sim}=\rho_{\rm V_2}\times2$=80.0(4.2)~ppm. We attribute the small mismatch to a reduction in the density of the V$_{\rm n}$ compared to the V$_2$ estimation.

Understanding the origin of the charge traps also provides a clear path on how to create optically noise free group-4 vacancy defects in diamond.
Single-peak fingerprints, indicating a low V$_{\rm n}$ density, are more frequently observed in high pressure high temperature (HPHT) annealed samples at 2000~$^\circ$C \cite{Goerlitz2022NPJ}, which is in agreement with electron spin resonance measurements \cite{Lomer1973}. 

Spectral jumps have been reported before for group-4 vacancy defects \cite{Nguyen2019PRB, Maity2018PRApp, Trusheim2020PRL}. Comparing the V$_{\rm n}$ density for the atomic species Si, Ge, and Sn and varying implantation energies (Fig.~\ref{fig:annealing}\hyperlink{fig:annealingL}{b,c}) indicates that Si implantation leads to the lowest V$_2$ density.
This observation is in accordance with the more frequent reports of spectrally stable SiV \cite{Sipahigil2014PRL} compared to SnV, now explained with our analysis that heavier ions cause increasing V$_{\rm n}$ densities.

\section*{Discussion}

In conclusion, we use the SnV in diamond, a representative of an inversion symmetric point defect in wide-bandgap materials, as local probe to detect localized charges with Angstrom resolution at the lattice scale. For this work, we constrained some degrees of freedom in the positions of the traps for the Monte Carlo simulation to make it more time-efficient. The main constraints were the confinement of the remote charges to a conical volume $z>z_0$, where $z_0$ is the assumed implantation depth of the SnV. The conical volume modeled the lattice damage remaining after annealing. Furthermore, we confined the three charge traps responsible for the multimodal spectrum to a plane. These limitations are not fundamental to this method and can be relaxed. In principle, a more extensive parameter space can be explored, but this would require significantly longer simulation times and increased memory demands. The addition of a more in-depth calibration of the probe would ameliorate this shortcoming. Particularly, implementing a probe-specific multi-directional calibration process, with the inclusion of extracting direction dependent polarizability coefficients, would allow us to construct suitable polarizability tensors. Using such a calibrated probe in combination with the Monte Carlo simulations would help enable azimuthal localization for full positioning.
Alternatively, correlated sensing \cite{delord_correlated_2024-1, Ji2024} can be used to reduce the uncertainty of charge traps' absolute position in our work.

We utilize a rapid spectroscopy technique based on frequency modulation with electro-optical modulators, to enable time-resolved access for recording and measuring charge dynamics of single nonfluorescent defects under laser irradiation with MHz readout rates, demonstrating a time-resolved quantum electrometer working at the atomic scale (see Table~\ref{tab:literatureReview}).
Such observations on the single-charge scale open the possibility to further understand the origin and type of charge defects and charge transport phenomena. Intriguingly, a sensor such as the one proposed in this study can be used to study topological quantum phenomena of ferroelectrics, for example the detection of ferroelectric vortices or polar skyrmions \cite{WangNatMat2023}.

An exciting secondary application is to use our sensor for sub-diffraction resolution position estimation. The electrometer's sensitivity to background charge-noise can be used to sense the position of an illumination laser with sub-diffraction precision. We estimate that a resolution below 1~nm can be achieved (Supplementary Fig.~\ref{fig:misalignmentDrift}).

From the analysis of the local charge environment, we are able to understand the nanoscopic origin of spectral diffusion of SnVs and formulate mitigation strategies.
We identify that the local defect density of $V_{\rm n}$ should be reduced, and quantify precisely the maximally allowed charge trap density for reaching optical coherence. 

Building on the insights of our work we believe that our electrometer opens up an exciting direction in material science, enabling the time-resolved study of single elementary charges with \Ang~spatial and 100 ns temporal resolution. 

The integration of the sensor into a scanning-probe tip or a nanodiamond will allow for the study of single and multiple lattice defects in silicon transistors, optically active quantum memories, and defect-induced charge-noise in on-chip ion and superconducting computers, potentially mitigating these detrimental effects and contributing to the further optimization of materials for the application in quantum technology.

\section*{Methods}

\subsection*{Monte Carlo simulations}
\label{ap:MCSoverview}

Here, we provide an overview of the general methodology of simulating single and multimodal spectra (Fig.~\ref{fig:sensor}\hyperlink{fig:sensorL}{b}).
The simulations begin with distributing charge traps uniformly within a specified volume or surface. 
For multimodal spectra, such as the one depicted in Fig.~\ref{fig:sensor}\hyperlink{fig:sensorL}{b}, the distribution of charge traps is divided into two categories: proximity traps and remote traps. Proximity traps are positioned at fixed locations, while remote traps are distributed with a fixed density within a prescribed volume. The SnV$^{-1}$ is always located in the origin of the coordinate space.  

Once a spatial trap configuration is created, a single iteration of the Monte Carlo simulation can be performed. It consists of assigning charges to the trap locations (charging of traps). A fixed number of charges are distributed assuming charge neutrality $-e + e\sum_i q_i = 0$, where $e$ is the elementary charge and $q_i$ a charge state $q_i \in \{-1, 0, +1\}$. The field strength at the location of the SnV$^{-1}$ then becomes:
\begin{equation}
 \vec{E} = \sum_{ i} \vec{E}(q_i, \vec{r}_{\rm i})~,  
 \label{eq:mc_spectrum}
\end{equation}
where $\vec{r}_i$ is the position of a trap and $\vec{E}(q_i,\vec{r}_i)$ is the electric field of a point charge in the medium, chosen such that it adequately reflects boundary conditions for the solution of Maxwell's equations. The non-linear Stark shift $\Delta_{\rm Stark}$ corresponding to the magnitude of the field $\vec{E}(q,\vec{r}_i)$ is calculated using \eqref{eq:nonlin_stark}, where $E_{\rm s}= |\vec{E}_{\rm s}|$ and the parameters $\Delta\mu = 6.1\times 10^{- 4}$~GHz/( MV/m)$^2$,  $\Delta\alpha= - 5.1 \times 10^{- 5}$~GHz/( MV/m)$^2$,  $ \Delta\beta = - 5.5\times 10 ^{- 8}$ GHz/(MV /m)$^3$ and $\Delta\gamma=-2.2\times 10 ^{- 10}$ GHz/(MV/m)$^4$ \cite{Santis2021PRL}. Unless explicitly stated otherwise, the procedure is repeated 1000 times and the spectrum corresponding to the distribution of Stark shifts is generated using       
\begin{equation}
S(\omega) = \frac{1}{N}\sum_{\rm n} L_\gamma\left(\omega - \Delta_{\rm Stark, n}\right)~, 
\label{eq:app_sim_spectrum}
\end{equation}
where $N$ is a normalization constant (${\rm max} S(\omega) = 1$), $n$ is the simulation step index and $L_\gamma(\omega)$ is a Lorentzian line profile with full width half maximum $\gamma = 35$ MHz corresponding to the lifetime limited linewidth of the SnV$^{-1}$ \cite{Trusheim2020PRL}. We assume that there is no additional power-broadening, nor a lifetime reduction due to Purcell enhancement. 

The process of determining the most probable trap configuration is divided into three main steps. First, we utilize the four peaks observed in the measured multimodal spectrum to identify the positions of proximity traps that generate Stark shifts consistent with the experimental observations. This initial step provides a rough estimation of the proximity trap positions.
Next, we employ an optimization procedure to fine-tune the predetermined positions of the proximity traps. By optimizing the relevant parameters, we generate a comprehensive database of simulated spectra.
Finally, utilizing the objective function ($\chi^2$ test) employed during the optimization procedure, we analyze the large dataset of simulated spectra to identify the most likely proximity trap configuration. This objective function serves as a measure of the agreement between the simulated spectra and the experimental observations. By comparing the calculated spectra with the measured data, we can identify the configuration that best matches the experimental results. In the following we provide a detailed description of each individual step. 

The four peaks in the measured spectrum shown in Fig.~\ref{fig:sensor}\hyperlink{fig:sensorL}{b} are used as a reference to estimate the location of a charged proximity trap relative to the SnV using Eq.~\eqref{eq:nonlin_stark}.

We deem the scenario of three traps contributing to the multimodal spectrum the most likely due to the temporal analysis of the PLE data (see Supplementary Information). The trap situated at $\vec{r}_2 = (0,0,r_2)$ is assumed to be permanently charged. The charge state of the other two traps are then given by $\{\medcirc\medcirc, \medcirc\odot, \odot\medcirc, \odot\odot\}$,
where the left circle represents a trap at position $\vec{r}_1$ and the right circle a trap at position $\vec{r}_3$. An empty circle represents a neutral trap, whereas a filled circle marks a trap with a charge. The peaks corresponding to each charge state are shown in Fig.~\ref{fig:sensor}\hyperlink{fig:sensorL}{b}. The negative charge located at a distance $\vec{r}_2$ magnifies the response of the SnV to remote charges and produces the observed inhomogeneous broadening.
The choice $\vec{r}_2 = (0,0,r_2)$ is of course not the most general, but we select it due to the anisotropy expected from the implantation procedure. Furthermore, including the position $\vec{r}_2$ with two more degrees of freedom would have made the free parameter space too large and increased the computational time.

We limit the placement of the three proximity traps to a plane, therefore further reducing the complexity of the problem. We approximate the initial positions $\vec{r}_1, \vec{r}_2, \vec{r}_3$ by solving the simultaneous set of equations:
\begin{align}
    \Delta_{\medcirc\medcirc} &= -\frac{\alpha^2}{2} E(-1, \vec{r}_2)^2 \\
    \Delta_{\medcirc\odot}    &= -\frac{\alpha^2}{2} [E(-1, \vec{r}_2) + E(-1, \vec{r}_1)]^2\\
    \Delta_{\odot\medcirc}    &= -\frac{\alpha^2}{2} [E(-1, \vec{r}_2) + E(-1, \vec{r}_3)]^2\\
    \Delta_{\odot\odot}       &= -\frac{\alpha^2}{2} E(-1, \vec{r}_2) + E(-1, \vec{r}_1)+ E(-1, \vec{r}_3)]^2
\end{align}
The positions are parameterized according to:
\begin{align}
    \vec{r}_1 &= r_1[\cos(\theta_1), 0, \sin(\theta_1)] \\
    \vec{r}_3 &= r_3[\cos(\theta_3), 0, \sin(\theta_3)]
\end{align}
The above equations can be solved such that $r_1(\theta_1),~r_3(\theta_1)$ and $\theta_3(\theta_1)$. The relative shifts $\Delta_{\medcirc\odot}, \Delta_{\odot\medcirc}, \Delta_{\odot\odot}$ are estimated from the central peak positions using a fitting procedure, where the integrated spectrum in Fig.~\ref{fig:sensor}\hyperlink{fig:sensorL}{b} is fitted simultaneously with four Voigt profiles. The choice of $\vec{r}_2 = (r_2 \sin(\theta_2),0, r_2 \cos(\theta_2)r_2)$ was based on a coarse scan of the Monte Carlo procedure by varying the angle $\theta_2$. We found that in conjunction with the remote traps $\theta_2 =0$ produced the most consistent results. 

The fine tuning of the proximity trap positions in the second step is again done using a Monte Carlo simulation in combination with an optimization procedure. For the optimization procedure, remote traps are distributed randomly in a conical volume $z > 0$~nm with an opening angle of $45^\circ$ at a fixed density $\rho_{\rm trap}$, to mimic the non-isotropic distribution of traps, that is expected to result from implantation damage. The volume is capped at $30$~nm. We exclude a spherical volume with a radius of $r_q < 2.5$ nm, for placing the proximity charges. A charged trap is assumed to contribute to the total field
\begin{equation}
 \vec{E} = \sum_{ i} \vec{E}(q_i, \vec{r}_{\rm i})~, 
 \label{eq:mc_spectrum}
\end{equation}
with 
\begin{equation}
    \vec{E}(q_i, \vec{r}) =  \frac{q_i}{4 \pi \epsilon_0 \epsilon_r} \frac{\vec{r}}{r^{3}}~,
    \label{eq:app_bulk_electric_field}
\end{equation}
where $e$ is the elementary charge and $q_i$ a charge state $q_i \in \{-1, 0, +1\}$, $\epsilon_0$ is the vacuum permittivity and $\epsilon_r = 5.5$ is the relative permittivity of diamond. 

For each choice of $\rho_{\rm trap}$, $r_2$ and $\theta_1$ we perform a Monte Carlo simulation of the spectral fingerprint as described in the overview of the methods section of the main text.

To adequately account for the charge state of the proximity traps, they are charged with a probability $p_i$ according to the relative peak heights in each individual step of the simulation. We use $p_{\medcirc\medcirc} = 0.31$, $p_{\medcirc\odot} = 0.63$, $p_{\odot\medcirc} = 0.017$ and $p_{\odot\odot} = 0.041$.     

The optimization of the trap positions for a given $\rho_{\rm trap}$, $r_2$ and $\theta_1$ is done by minimizing $\chi^2$
\begin{align}
    \sum_i \chi(\boldsymbol{\theta},i) = \sum_{n = 0, i}^N \frac{[O_{\rm n}(\boldsymbol{\theta}, i) - E_{\rm n, i}]^2}{E_{\rm n}} ~,
    \label{eq:objective}
\end{align}
with the simplicial homology global optimization (shgo) algorithm. We use an implementation of the shgo algorithm provided by the python library \emph{SciPy} \cite{2020SciPy-NMeth}. In Eq.~\eqref{eq:objective},  $\boldsymbol{\theta} = [a,b]$, where $a, b$ fine tune $r_2' = a r_2 $ and $r_3' = b r_3$. 

We split the spectrum into three parts associated to $ i \in \{{\medcirc\medcirc}, {\medcirc\odot}, {{\odot\medcirc} + {\odot\odot}\}}$. For each part, we use the respective single and double Voigt profile fits for comparison with the simulated spectra by using Eq.~\eqref{eq:objective}. In Eq.~\eqref{eq:objective} $E_{\rm n, i}$ are the expected counts in the $n$'th bin, which is found by binning the (normalized) single and double Voigt profiles fitted to the measured spectrum into 170 equally sized bins over an interval containing the profile with a width of $4$~GHz. $O_{\rm n}(\boldsymbol{\theta}, i)$ is number of expected counts of the respective $i$ for the simulated spectrum in the $n$'th bin.

Finally, the values $\chi^2$, $r_1'$, $r_3'$ are then tabulated for $\rho_{\rm trap} \in [35, 100]$ ppm, $\Delta_{\odot\odot} \in [0.5, 1.7]$ GHz and $\theta_1 \in [0, 0.6]$ rad. We perform 500 iterations of the optimization over distinct spatial configurations of the remote traps for each value of $\rho_{\rm trap}$, $r_2$ and $\theta_1$. We only use the 50 lowest values of $\chi^2$ (the others are considered outliers), and perform a weighted average for determining $\langle \chi^2 \rangle$.    

We find the 68\% confidence intervals for $\rho_{\rm trap},~\Delta_{\odot\odot}$ and $\theta_1$ by $\min\{\langle \chi^2 \rangle\} + 3.5$ \cite{Avni1976}. The results are shown in Fig.~\ref{fig:sensor}\hyperlink{fig:sensorL}{c, d}. The statistical error shown in Fig.~\ref{fig:sensor} is produced by all the simulated spectra within the $68$\% confidence interval.

\subsection*{Samples} \label{ap:sample}
The samples used in this work (E001 for S1 and E013 for S2) are a single-crystal electronic grade diamonds grown by chemical vapor deposition (supplied by Element Six Technologies Ltd. (UK) and with {100} faces). Both substrates are initially cleaned for about one hour in a boiling triacid solution (H$_2$SO$_4$:HNO$_3$:HClO$_4$, 1:1:1)\cite{Brown2019} and then etched in Cl$_2$/He and O$_2$/CF$_4$ plasmas, in order to remove any organic contaminants and structural defects from the surface.\cite{Atikian2014} The diamond hosting S1 (E001) is then implanted with $^{120}$Sn (spin-0) ions, using a fluence of $5\EXP{10}$ atoms${\rm~cm^{-2}}$ and an implantation energy of $400{\rm~keV}$, which corresponds to a penetration depth of 100 nm, as estimated by SRIM simulations~\cite{Ziegler2010NIM}. The formation of the SnV color centers is finally achieved by low pressure-high temperature (LPHT) annealing step at the temperature of 1050$^\circ$C for about 12 hours in vacuum (pressure $< 7.5\EXP{-8}$ mbar). The sample hosting S2 (E013) is instead initially implanted with sulphur ions, with a nominal fluence and implantation energy of $5\EXP{12}$ atoms${\rm~cm^{-2}}$ and $160{\rm~keV}$, respectively. A LPHT annealing step (T= 1050$^\circ$C and P$\approx 1\EXP{-7}$ mbar), for about 12 hours, is then used to heal the lattice damage caused by implantation process. After a second triacid cleaning step, Sn ions are implanted in the same substrate, with nominal values $5\EXP{10}$ atoms${\rm~cm^{-2}}$ and $400{\rm~keV}$ for fluence and energy, respectively. According to SRIM simulations, the expected implantation depth is about 100 nm for both S and Sn. After a second LPHT annealing step, the substrate is annealed at 2100$^\circ$C at P$\approx$ 6-8 GPa for a total of 2 hours and finally cleaned in a boiling triacid solution.

Nanopillars are fabricated on both samples by e-beam lithography and plasma etching. In the case of S1 (E001), 200 nm of SiN$_x$ were deposited on the surface of the diamond in an inductively coupled-plasma (ICP) enhanced chemical vapor deposition system. After coating the sample with 300nm of electro-sensitive resist (ZEP520A) and a few nm of a charge-dissipating layer (ESpacer$\rm{TM}$), pillars with nominal diameters ranging from 180 nm to 340 nm (in steps of 40 nm) are exposed by means of electron-beam lithography. After development, the pattern was transferred into the SiN$_x$ layer by a reactive ion etching (RIE) plasma (10 sccm CF$_4$, RF power = 100 W, P = 1 Pa) and then etched into the diamond during an ICP process in O$_2$ plasma (80 sccm, ICP power = 750 W, RF power = 200 W, P = 0.3 Pa). The sample hosting S2 (E013) is coated with a similar stack of layers, but with the introduction of 10 nm of Ti between SiN$_x$ and resist to replace the ESpacer$\rm{TM}$ layer. Moreover, a different range of nominal diameters (from 140 nm to 260 nm, in steps of 20 nm) is used in this case for the patterned pillars. After development, the hard mask layer was etched by a ICP-RIE process in a F-based plasma (SF$_6$:Ar, 30:15 sccm, ICP power = 500 W, RF power = 35 W, P = 1 Pa), followed by an ICP process in O$_2$ plasma (99 sccm, ICP power = 1000 W, RF power = 200 W, P = 1 Pa) for the complete transfer of the pattern into the diamond substrate. The final height of the nanopillars is approximately 500 nm for both processed samples. The remaining nitride layer is finally removed in a solution of buffered HF and the diamond surface is exposed.

\subsection*{Measurement setup}
The sample is cooled to 4~K in a closed-cycle helium cryostat (Montana s50).
    A home-built confocal scanning microscope is utilized to locate and optically address nanopillars with SnVs.
    PL spectra are measured via a spectrometer (Princeton Instruments HR500) with a CCD camera (Princeton Instruments Excelon ProEM:400BX3). 
    Photons collected from the cryogenic setup are coupled into a fiber, converted into digital signals via avalanche photo diodes (APD, Excelitas SPCM-AQ4C or SPCM-AQRH) and counted via a data acquisition divide (NIDAQ USB 6363).

    Non-resonant confocal microscopy measurements are  done with a green diode laser at 520~nm (DLnsec).
    The SnV's charge state is initialized by a blue diode laser (450 nm, Thorlabs LP450-SF15 or 445 nm Hübner Cobolt 06-MLD). 
   {Both green and blue lasers are employed for inducing charge dynamics in the investigated traps.
   For PLE scans and readouts two 619 nm orange lasers, a tunable dye laser ~nm (Sirah Matisse, DCM in EPL/EG solution) and an SHG laser source (TOPTICA SHG DLC PRO) are employed.
   Orange lasers are used in the experiment by scanning the frequency of the resonant excitation laser across the C transition of an SnV center and collecting the phonon sideband of the fluorescence.
   The orange lasers' frequency is measured and stabilized using a wavemeter (High Finesse W7)
    
  For the rapid PLE scans and optical readout sequences (RORO), a system based on sideband generation via an electro-optical modulator (EOM, Jenoptik AM635b) is utilized.  When driven by a sinusoidal signal $\Omega_{\rm i}$, the EOM creates two sidebands with frequency offsets $\omega_L\pm\Omega_{\rm i}$, where $\omega_L$ is the laser's frequency. By chirping the modulation, it is possible to modulate the frequency of the sidebands for scanning or probing resonances. We generate the voltage signal using an arbitrary waveform generator (AWG, Keysight M8195A).   
    We employ the blueshifted sideband to excite the SnV.

    We connect a pick-off path from the output of the EOM to the photodiode embedded into a modulation bias controller (MBC, OZ Optics Mini-MBC-1B0). MBC applies a bias voltage to the EOM and stabilizes it to its interferometric minimum. The MBC stabilization diminishes the central frequency component of the laser and any higher-order even sidebands. An optimized choice for the driving amplitude maximizes the first-order sideband. 

    The frequency-modulated sidebands are directed to the sample through the confocal microscope.
    Emitted photons from the SnVs are turned into digital signals via the APDs and timestamped (arrival time of a photon) using a timetagger (quTools qutag).
    A pulse streamer (Swabian Instruments 8/2) is used to trigger the AWG to start the sequence. It also generates a periodic calibration signal sent to one of the channels of the timetagger.
    A signal generator (Rigol DG952) streams a signal throughout the experiment to synchronize the clocks of the pulse streamer, timetagger, and AWG.
    Experiments are controlled with the software suite Qudi \cite{Binder2017SoftwareX}.

\section*{Data and Code Availability} 
The data and code that support the findings of this study have been deposited in Zenodo repository with DOI: 10.5281/zenodo.15704695      \cite{Zenodo2025}.

\bibliography{refz}

\begin{thebibliography}{10}
\expandafter\ifx\csname url\endcsname\relax
  \def\url#1{\texttt{#1}}\fi
\expandafter\ifx\csname urlprefix\endcsname\relax\def\urlprefix{URL }\fi
\providecommand{\bibinfo}[2]{#2}
\providecommand{\eprint}[2][]{\url{#2}}

\bibitem{Santis2021PRL}
\bibinfo{author}{de~Santis, L.}, \bibinfo{author}{Trusheim, M.~E.},
  \bibinfo{author}{Chen, K.~C.} \& \bibinfo{author}{Englund, D.~R.}
\newblock \bibinfo{title}{{Investigation of the Stark Effect on a
  Centrosymmetric Quantum Emitter in Diamond}}.
\newblock \emph{\bibinfo{journal}{Physical Review Letters}}
  \textbf{\bibinfo{volume}{127}}, \bibinfo{pages}{147402}
  (\bibinfo{year}{2021}).

\bibitem{Ziegler2010NIM}
\bibinfo{author}{Ziegler, J.~F.}, \bibinfo{author}{Ziegler, M.~D.} \&
  \bibinfo{author}{Biersack, J.~P.}
\newblock \bibinfo{title}{{SRIM -- The stopping and range of ions in matter}}.
\newblock \emph{\bibinfo{journal}{Nuclear Instruments and Methods in Physics
  Research Section B: Beam Interactions with Materials and Atoms}}
  \textbf{\bibinfo{volume}{268}}, \bibinfo{pages}{1818--1823}
  (\bibinfo{year}{2010}).

\bibitem{kaczer_impact_2002}
\bibinfo{author}{Kaczer, B.} \emph{et~al.}
\newblock \bibinfo{title}{Impact of {MOSFET} gate oxide breakdown on digital
  circuit operation and reliability}.
\newblock \emph{\bibinfo{journal}{IEEE Transactions on Electron Devices}}
  \textbf{\bibinfo{volume}{49}}, \bibinfo{pages}{500--506}
  (\bibinfo{year}{2002}).

\bibitem{stacey_evidence_2019}
\bibinfo{author}{Stacey, A.} \emph{et~al.}
\newblock \bibinfo{title}{Evidence for {Primal} sp2 {Defects} at the {Diamond}
  {Surface}: {Candidates} for {Electron} {Trapping} and {Noise} {Sources}}.
\newblock \emph{\bibinfo{journal}{Advanced Materials Interfaces}}
  \textbf{\bibinfo{volume}{6}}, \bibinfo{pages}{1801449}
  (\bibinfo{year}{2019}).

\bibitem{Rehman2022}
\bibinfo{author}{Rehman, A.} \emph{et~al.}
\newblock \bibinfo{title}{Nature of the 1/f noise in graphene-direct evidence
  for the mobility fluctuation mechanism}.
\newblock \emph{\bibinfo{journal}{Nanoscale}} \textbf{\bibinfo{volume}{14}},
  \bibinfo{pages}{7242--7249} (\bibinfo{year}{2022}).

\bibitem{Defo2023PRB}
\bibinfo{author}{{Kuate Defo}, R.}, \bibinfo{author}{Rodriguez, A.~W.},
  \bibinfo{author}{Kaxiras, E.} \& \bibinfo{author}{Richardson, S.~L.}
\newblock \bibinfo{title}{Theoretical investigation of charge transfer between
  two defects in a wide band gap semiconductor}.
\newblock \emph{\bibinfo{journal}{Physical Review B}}
  \textbf{\bibinfo{volume}{107}}, \bibinfo{pages}{125305}
  (\bibinfo{year}{2023}).

\bibitem{Mittiga2018PRL}
\bibinfo{author}{Mittiga, T.} \emph{et~al.}
\newblock \bibinfo{title}{Imaging the local charge environment of
  nitrogen-vacancy centers in diamond}.
\newblock \emph{\bibinfo{journal}{Physical Review Letters}}
  \textbf{\bibinfo{volume}{121}}, \bibinfo{pages}{246402}
  (\bibinfo{year}{2018}).

\bibitem{Vamivakas2011}
\bibinfo{author}{Vamivakas, A.~N.} \emph{et~al.}
\newblock \bibinfo{title}{Nanoscale optical electrometer}.
\newblock \emph{\bibinfo{journal}{Physical Review Letters}}
  \textbf{\bibinfo{volume}{107}}, \bibinfo{pages}{166802}
  (\bibinfo{year}{2011}).

\bibitem{dolde2011}
\bibinfo{author}{Dolde, F.} \emph{et~al.}
\newblock \bibinfo{title}{Electric-field sensing using single diamond spins}.
\newblock \emph{\bibinfo{journal}{Nature Physics}}
  \textbf{\bibinfo{volume}{7}}, \bibinfo{pages}{459--463}
  (\bibinfo{year}{2011}).

\bibitem{dolde2014}
\bibinfo{author}{Dolde, F.} \emph{et~al.}
\newblock \bibinfo{title}{Nanoscale {Detection} of a {Single} {Fundamental}
  {Charge} in {Ambient} {Conditions} {Using} the {NV} - {Center} in {Diamond}}.
\newblock \emph{\bibinfo{journal}{Physical Review Letters}}
  \textbf{\bibinfo{volume}{112}}, \bibinfo{pages}{097603}
  (\bibinfo{year}{2014}).

\bibitem{Chen2016}
\bibinfo{author}{Chen, D.}, \bibinfo{author}{Lander, G.~R.},
  \bibinfo{author}{Krowpman, K.~S.}, \bibinfo{author}{Solomon, G.~S.} \&
  \bibinfo{author}{Flagg, E.~B.}
\newblock \bibinfo{title}{Characterization of the local charge environment of a
  single quantum dot via resonance fluorescence}.
\newblock \emph{\bibinfo{journal}{Physical Review B}}
  \textbf{\bibinfo{volume}{93}}, \bibinfo{pages}{115307}
  (\bibinfo{year}{2016}).

\bibitem{Iwasaki2017ACS}
\bibinfo{author}{Iwasaki, T.} \emph{et~al.}
\newblock \bibinfo{title}{Direct nanoscale sensing of the internal electric
  field in operating semiconductor devices using single electron spins}.
\newblock \emph{\bibinfo{journal}{ACS Nano}} \textbf{\bibinfo{volume}{11}},
  \bibinfo{pages}{1238--1245} (\bibinfo{year}{2017}).

\bibitem{broadway2018}
\bibinfo{author}{Broadway, D.~A.} \emph{et~al.}
\newblock \bibinfo{title}{Spatial mapping of band bending in semiconductor
  devices using in situ quantum sensors}.
\newblock \emph{\bibinfo{journal}{Nature Electronics}}
  \textbf{\bibinfo{volume}{1}}, \bibinfo{pages}{502--507}
  (\bibinfo{year}{2018}).

\bibitem{li_2020}
\bibinfo{author}{Li, R.} \emph{et~al.}
\newblock \bibinfo{title}{Nanoscale {Electrometry} {Based} on a
  {Magnetic}-{Field}-{Resistant} {Spin} {Sensor}}.
\newblock \emph{\bibinfo{journal}{Physical Review Letters}}
  \textbf{\bibinfo{volume}{124}}, \bibinfo{pages}{247701}
  (\bibinfo{year}{2020}).

\bibitem{barson_nanoscale_2021}
\bibinfo{author}{Barson, M. S.~J.} \emph{et~al.}
\newblock \bibinfo{title}{Nanoscale {Vector} {Electric} {Field} {Imaging}
  {Using} a {Single} {Electron} {Spin}}.
\newblock \emph{\bibinfo{journal}{Nano Letters}} \textbf{\bibinfo{volume}{21}},
  \bibinfo{pages}{2962--2967} (\bibinfo{year}{2021}).

\bibitem{Qiu2022npj}
\bibinfo{author}{Qiu, Z.}, \bibinfo{author}{Hamo, A.}, \bibinfo{author}{Vool,
  U.}, \bibinfo{author}{Zhou, T.~X.} \& \bibinfo{author}{Yacoby, A.}
\newblock \bibinfo{title}{Nanoscale electric field imaging with an ambient
  scanning quantum sensor microscope}.
\newblock \emph{\bibinfo{journal}{npj Quantum Information}}
  \textbf{\bibinfo{volume}{8}}, \bibinfo{pages}{1--7} (\bibinfo{year}{2022}).

\bibitem{delord_correlated_2024-1}
\bibinfo{author}{Delord, T.}, \bibinfo{author}{Monge, R.} \&
  \bibinfo{author}{Meriles, C.~A.}
\newblock \bibinfo{title}{Correlated {Spectroscopy} of {Electric} {Noise} with
  {Color} {Center} {Clusters}}.
\newblock \emph{\bibinfo{journal}{Nano Letters}} \textbf{\bibinfo{volume}{24}},
  \bibinfo{pages}{6474--6479} (\bibinfo{year}{2024}).

\bibitem{Ji2024}
\bibinfo{author}{Ji, W.} \emph{et~al.}
\newblock \bibinfo{title}{Correlated sensing with a solid-state quantum
  multisensor system for atomic-scale structural analysis}.
\newblock \emph{\bibinfo{journal}{Nature Photonics}}
  \textbf{\bibinfo{volume}{18}}, \bibinfo{pages}{230--235}
  (\bibinfo{year}{2024}).

\bibitem{song_reply_2023}
\bibinfo{author}{Song, Q.} \emph{et~al.}
\newblock \bibinfo{title}{Reply to: {Dilemma} in optical identification of
  single-layer multiferroics}.
\newblock \emph{\bibinfo{journal}{Nature}} \textbf{\bibinfo{volume}{619}},
  \bibinfo{pages}{E44--E46} (\bibinfo{year}{2023}).

\bibitem{uren1985}
\bibinfo{author}{Uren, M.~J.}, \bibinfo{author}{Day, D.~J.} \&
  \bibinfo{author}{Kirton, M.~J.}
\newblock \bibinfo{title}{1/ \textit{f} and random telegraph noise in silicon
  metal‐oxide‐semiconductor field‐effect transistors}.
\newblock \emph{\bibinfo{journal}{Applied Physics Letters}}
  \textbf{\bibinfo{volume}{47}}, \bibinfo{pages}{1195--1197}
  (\bibinfo{year}{1985}).

\bibitem{Stampfer2020phd}
\bibinfo{author}{Stampfer, B.}
\newblock \bibinfo{title}{Advanced electrical characterization of charge
  trapping in mos transistors: Advanced electrical characterization of charge
  trapping in mos transistors}.
\newblock \emph{\bibinfo{journal}{{PhD Thesis, TU Wien}}}
  (\bibinfo{year}{2020}).

\bibitem{hite2013}
\bibinfo{author}{Hite, D.} \emph{et~al.}
\newblock \bibinfo{title}{Surface science for improved ion traps}.
\newblock \emph{\bibinfo{journal}{MRS Bulletin}} \textbf{\bibinfo{volume}{38}},
  \bibinfo{pages}{826--833} (\bibinfo{year}{2013}).

\bibitem{faoro2006}
\bibinfo{author}{Faoro, L.} \& \bibinfo{author}{Ioffe, L.~B.}
\newblock \bibinfo{title}{Quantum {Two} {Level} {Systems} and {Kondo}-{Like}
  {Traps} as {Possible} {Sources} of {Decoherence} in {Superconducting}
  {Qubits}}.
\newblock \emph{\bibinfo{journal}{Physical Review Letters}}
  \textbf{\bibinfo{volume}{96}}, \bibinfo{pages}{047001}
  (\bibinfo{year}{2006}).

\bibitem{reshef2021}
\bibinfo{author}{Reshef, A.} \& \bibinfo{author}{Caspary~Toroker, M.}
\newblock \bibinfo{title}{Method for assessing atomic sources of flicker noise
  in superconducting qubits}.
\newblock \emph{\bibinfo{journal}{npj Computational Materials}}
  \textbf{\bibinfo{volume}{7}}, \bibinfo{pages}{1--6} (\bibinfo{year}{2021}).

\bibitem{Morioka2020NatComm}
\bibinfo{author}{Morioka, N.} \emph{et~al.}
\newblock \bibinfo{title}{Spin-controlled generation of indistinguishable and
  distinguishable photons from silicon vacancy centres in silicon carbide}.
\newblock \emph{\bibinfo{journal}{Nature Communications}}
  \textbf{\bibinfo{volume}{11}}, \bibinfo{pages}{2516} (\bibinfo{year}{2020}).

\bibitem{White2021Opt}
\bibinfo{author}{White, S.} \emph{et~al.}
\newblock \bibinfo{title}{Phonon dephasing and spectral diffusion of quantum
  emitters in hexagonal boron nitride}.
\newblock \emph{\bibinfo{journal}{Optica}} \textbf{\bibinfo{volume}{8}},
  \bibinfo{pages}{1153} (\bibinfo{year}{2021}).

\bibitem{Vajner2022AQT}
\bibinfo{author}{Vajner, D.~A.}, \bibinfo{author}{Rickert, L.},
  \bibinfo{author}{Gao, T.}, \bibinfo{author}{Kaymazlar, K.} \&
  \bibinfo{author}{Heindel, T.}
\newblock \bibinfo{title}{Quantum communication using semiconductor quantum
  dots}.
\newblock \emph{\bibinfo{journal}{Advanced Quantum Technologies}}
  \textbf{\bibinfo{volume}{5}}, \bibinfo{pages}{2100116}
  (\bibinfo{year}{2022}).

\bibitem{OrphalKobin2023PRX}
\bibinfo{author}{Orphal-Kobin, L.} \emph{et~al.}
\newblock \bibinfo{title}{Optically coherent nitrogen-vacancy defect centers in
  diamond nanostructures}.
\newblock \emph{\bibinfo{journal}{Physical Review X}}
  \textbf{\bibinfo{volume}{13}}, \bibinfo{pages}{011042}
  (\bibinfo{year}{2023}).

\bibitem{Pompili2021Sci}
\bibinfo{author}{Pompili, M.} \emph{et~al.}
\newblock \bibinfo{title}{Realization of a multinode quantum network of remote
  solid-state qubits}.
\newblock \emph{\bibinfo{journal}{Science}} \textbf{\bibinfo{volume}{372}},
  \bibinfo{pages}{259--264} (\bibinfo{year}{2021}).

\bibitem{Iwasaki2017PRL}
\bibinfo{author}{Iwasaki, T.} \emph{et~al.}
\newblock \bibinfo{title}{Tin-vacancy quantum emitters in diamond}.
\newblock \emph{\bibinfo{journal}{Physical Review Letters}}
  \textbf{\bibinfo{volume}{119}}, \bibinfo{pages}{253601}
  (\bibinfo{year}{2017}).

\bibitem{Goss1996PRL}
\bibinfo{author}{Goss, J.~P.}, \bibinfo{author}{Jones, R.},
  \bibinfo{author}{Breuer, S.~J.}, \bibinfo{author}{Briddon, P.~R.} \&
  \bibinfo{author}{{\"O}berg, S.}
\newblock \bibinfo{title}{The twelve-line 1.682 ev luminescence center in
  diamond and the vacancy-silicon complex}.
\newblock \emph{\bibinfo{journal}{Physical Review Letters}}
  \textbf{\bibinfo{volume}{77}}, \bibinfo{pages}{3041--3044}
  (\bibinfo{year}{1996}).

\bibitem{Iwasaki2015SciRep}
\bibinfo{author}{Iwasaki, T.} \emph{et~al.}
\newblock \bibinfo{title}{Germanium-vacancy single color centers in diamond}.
\newblock \emph{\bibinfo{journal}{Scientific Reports}}
  \textbf{\bibinfo{volume}{5}}, \bibinfo{pages}{12882} (\bibinfo{year}{2015}).

\bibitem{Niaz2016JPCC}
\bibinfo{author}{Niaz, S.} \& \bibinfo{author}{Zdetsis, A.~D.}
\newblock \bibinfo{title}{Comprehensive ab initio study of electronic, optical,
  and cohesive properties of silicon quantum dots of various morphologies and
  sizes up to infinity}.
\newblock \emph{\bibinfo{journal}{The Journal of Physical Chemistry C}}
  \textbf{\bibinfo{volume}{120}}, \bibinfo{pages}{11288--11298}
  (\bibinfo{year}{2016}).

\bibitem{Wang2021ACS}
\bibinfo{author}{Wang, P.}, \bibinfo{author}{Taniguchi, T.},
  \bibinfo{author}{Miyamoto, Y.}, \bibinfo{author}{Hatano, M.} \&
  \bibinfo{author}{Iwasaki, T.}
\newblock \bibinfo{title}{Low-temperature spectroscopic investigation of
  lead-vacancy centers in diamond fabricated by high-pressure and
  high-temperature treatment}.
\newblock \emph{\bibinfo{journal}{ACS Photonics}} \textbf{\bibinfo{volume}{8}},
  \bibinfo{pages}{2947--2954} (\bibinfo{year}{2021}).

\bibitem{ShkarinPRL2021}
\bibinfo{author}{Shkarin, A.} \emph{et~al.}
\newblock \bibinfo{title}{Nanoscopic {Charge} {Fluctuations} in a {Gallium}
  {Phosphide} {Waveguide} {Measured} by {Single} {Molecules}}.
\newblock \emph{\bibinfo{journal}{Physical Review Letters}}
  \textbf{\bibinfo{volume}{126}}, \bibinfo{pages}{133602}
  (\bibinfo{year}{2021}).

\bibitem{Stark1914}
\bibinfo{author}{Stark, J.}
\newblock \bibinfo{title}{Beobachtungen \"{u}ber den effekt des elektrischen
  feldes auf spektrallinien. i. quereffekt}.
\newblock \emph{\bibinfo{journal}{Annalen der Physik}}
  \textbf{\bibinfo{volume}{348}}, \bibinfo{pages}{965--982}
  (\bibinfo{year}{1914}).

\bibitem{Tamarat2006PRL}
\bibinfo{author}{Tamarat, P.} \emph{et~al.}
\newblock \bibinfo{title}{Stark shift control of single optical centers in
  diamond}.
\newblock \emph{\bibinfo{journal}{Physical Review Letters}}
  \textbf{\bibinfo{volume}{97}}, \bibinfo{pages}{083002}
  (\bibinfo{year}{2006}).

\bibitem{Ruhl2020ACS}
\bibinfo{author}{R{\"u}hl, M.}, \bibinfo{author}{Bergmann, L.},
  \bibinfo{author}{Krieger, M.} \& \bibinfo{author}{Weber, H.~B.}
\newblock \bibinfo{title}{Stark tuning of the silicon vacancy in silicon
  carbide}.
\newblock \emph{\bibinfo{journal}{Nano Letters}} \textbf{\bibinfo{volume}{20}},
  \bibinfo{pages}{658--663} (\bibinfo{year}{2020}).

\bibitem{Huxter2023NatPhys}
\bibinfo{author}{Huxter, W.~S.}, \bibinfo{author}{Sarott, M.~F.},
  \bibinfo{author}{Trassin, M.} \& \bibinfo{author}{Degen, C.~L.}
\newblock \bibinfo{title}{Imaging ferroelectric domains with a single-spin
  scanning quantum sensor}.
\newblock \emph{\bibinfo{journal}{Nature Physics}}
  \textbf{\bibinfo{volume}{19}}, \bibinfo{pages}{644--648}
  (\bibinfo{year}{2023}).

\bibitem{Maletinsky2012NatNT}
\bibinfo{author}{Maletinsky, P.} \emph{et~al.}
\newblock \bibinfo{title}{A robust scanning diamond sensor for nanoscale
  imaging with single nitrogen-vacancy centres}.
\newblock \emph{\bibinfo{journal}{Nature Nanotechnology}}
  \textbf{\bibinfo{volume}{7}}, \bibinfo{pages}{320--324}
  (\bibinfo{year}{2012}).

\bibitem{Budker2007NatPhys}
\bibinfo{author}{Budker, D.} \& \bibinfo{author}{Romalis, M.}
\newblock \bibinfo{title}{Optical magnetometry}.
\newblock \emph{\bibinfo{journal}{Nature Physics}}
  \textbf{\bibinfo{volume}{3}}, \bibinfo{pages}{227--234}
  (\bibinfo{year}{2007}).

\bibitem{Foy2020ACS}
\bibinfo{author}{Foy, C.} \emph{et~al.}
\newblock \bibinfo{title}{Wide-field magnetic field and temperature imaging
  using nanoscale quantum sensors}.
\newblock \emph{\bibinfo{journal}{ACS Applied Materials {\&} Interfaces}}
  \textbf{\bibinfo{volume}{12}}, \bibinfo{pages}{26525--26533}
  (\bibinfo{year}{2020}).

\bibitem{Kucsko2013Nat}
\bibinfo{author}{Kucsko, G.} \emph{et~al.}
\newblock \bibinfo{title}{Nanometre-scale thermometry in a living cell}.
\newblock \emph{\bibinfo{journal}{Nature}} \textbf{\bibinfo{volume}{500}},
  \bibinfo{pages}{54--58} (\bibinfo{year}{2013}).

\bibitem{li_atomic_2024}
\bibinfo{author}{Li, Z.} \emph{et~al.}
\newblock \bibinfo{title}{Atomic optical antennas in solids}.
\newblock \emph{\bibinfo{journal}{Nature Photonics}}
  \textbf{\bibinfo{volume}{18}}, \bibinfo{pages}{1113--1120}
  (\bibinfo{year}{2024}).

\bibitem{Hourahine2000}
\bibinfo{author}{Hourahine, B.} \emph{et~al.}
\newblock \bibinfo{title}{Identification of the hexavacancy in silicon with the
  ${B}^4_{80}$ optical center}.
\newblock \emph{\bibinfo{journal}{Physical Review B}}
  \textbf{\bibinfo{volume}{61}}, \bibinfo{pages}{12594--12597}
  (\bibinfo{year}{2000}).

\bibitem{gaubas_spectroscopy_2016}
\bibinfo{author}{Gaubas, E.}, \bibinfo{author}{Ceponis, T.},
  \bibinfo{author}{Meskauskaite, D.}, \bibinfo{author}{Grigonis, R.} \&
  \bibinfo{author}{Sirutkaitis, V.}
\newblock \bibinfo{title}{Spectroscopy of defects in hpht and cvd diamond by
  esr and pulsed photo-ionization measurements}.
\newblock \emph{\bibinfo{journal}{Journal of Instrumentation}}
  \textbf{\bibinfo{volume}{11}}, \bibinfo{pages}{C01017--C01017}
  (\bibinfo{year}{2016}).

\bibitem{girolami_transport_2019}
\bibinfo{author}{Girolami, M.} \emph{et~al.}
\newblock \bibinfo{title}{Transport properties of photogenerated charge
  carriers in black diamond films}.
\newblock \emph{\bibinfo{journal}{Ceramics International}}
  \textbf{\bibinfo{volume}{45}}, \bibinfo{pages}{9544--9547}
  (\bibinfo{year}{2019}).

\bibitem{Trusheim2020PRL}
\bibinfo{author}{Trusheim, M.~E.} \emph{et~al.}
\newblock \bibinfo{title}{Transform-limited photons from a coherent tin-vacancy
  spin in diamond}.
\newblock \emph{\bibinfo{journal}{Physical Review Letters}}
  \textbf{\bibinfo{volume}{124}}, \bibinfo{pages}{023602}
  (\bibinfo{year}{2020}).

\bibitem{Goerlitz2022NPJ}
\bibinfo{author}{G{\"o}rlitz, J.} \emph{et~al.}
\newblock \bibinfo{title}{Coherence of a charge stabilised tin-vacancy spin in
  diamond}.
\newblock \emph{\bibinfo{journal}{npj Quantum Information}}
  \textbf{\bibinfo{volume}{8}}, \bibinfo{pages}{1--9} (\bibinfo{year}{2022}).

\bibitem{Kambs2018NJP}
\bibinfo{author}{Kambs, B.} \& \bibinfo{author}{Becher, C.}
\newblock \bibinfo{title}{Limitations on the indistinguishability of photons
  from remote solid state sources}.
\newblock \emph{\bibinfo{journal}{New Journal of Physics}}
  \textbf{\bibinfo{volume}{20}}, \bibinfo{pages}{115003}
  (\bibinfo{year}{2018}).

\bibitem{Torun2021APL}
\bibinfo{author}{Torun, C.~G.} \emph{et~al.}
\newblock \bibinfo{title}{Optimized diamond inverted nanocones for enhanced
  color center to fiber coupling}.
\newblock \emph{\bibinfo{journal}{Applied Physics Letters}}
  \textbf{\bibinfo{volume}{118}}, \bibinfo{pages}{234002}
  (\bibinfo{year}{2021}).

\bibitem{Li2015NatComm}
\bibinfo{author}{Li, L.} \emph{et~al.}
\newblock \bibinfo{title}{Coherent spin control of a nanocavity-enhanced qubit
  in diamond}.
\newblock \emph{\bibinfo{journal}{Nature Communications}}
  \textbf{\bibinfo{volume}{6}}, \bibinfo{pages}{6173} (\bibinfo{year}{2015}).

\bibitem{Bopp2022Arxiv}
\bibinfo{author}{Bopp, J.~M.} \emph{et~al.}
\newblock \bibinfo{title}{{`Sawfish' Photonic Crystal Cavity for Near--Unity
  Emitter--to--Fiber Interfacing in Quantum Network Applications}}.
\newblock \emph{\bibinfo{journal}{Advanced Optical Materials}}
  \textbf{\bibinfo{volume}{12}}, \bibinfo{pages}{2301286}
  (\bibinfo{year}{2024}).

\bibitem{Bluvstein2019PRL}
\bibinfo{author}{Bluvstein, D.}, \bibinfo{author}{Zhang, Z.} \&
  \bibinfo{author}{Jayich, A. C.~B.}
\newblock \bibinfo{title}{Identifying and mitigating charge instabilities in
  shallow diamond nitrogen-vacancy centers}.
\newblock \emph{\bibinfo{journal}{Physical Review Letters}}
  \textbf{\bibinfo{volume}{122}}, \bibinfo{pages}{076101}
  (\bibinfo{year}{2019}).

\bibitem{ElementSix2022}
\bibinfo{author}{{Element Six}}.
\newblock \bibinfo{title}{{CVD Diamond Handbook}} (\bibinfo{year}{2022}).
\newblock \bibinfo{note}{{https://e6cvd.com/media/wysiwyg/pdf/
  Element\_Six\_CVD\_Diamond\_handbook\_2022.pdf}}.

\bibitem{yamamoto_extending_2013}
\bibinfo{author}{Yamamoto, T.} \emph{et~al.}
\newblock \bibinfo{title}{Extending spin coherence times of diamond qubits by
  high-temperature annealing}.
\newblock \emph{\bibinfo{journal}{Physical Review B}}
  \textbf{\bibinfo{volume}{88}}, \bibinfo{pages}{075206}
  (\bibinfo{year}{2013}).

\bibitem{Slepetz2014PhysChem}
\bibinfo{author}{Slepetz, B.} \& \bibinfo{author}{Kertesz, M.}
\newblock \bibinfo{title}{Divacancies in diamond: a stepwise formation
  mechanism}.
\newblock \emph{\bibinfo{journal}{Physical Chemistry Chemical Physics}}
  \textbf{\bibinfo{volume}{16}}, \bibinfo{pages}{1515--1521}
  (\bibinfo{year}{2014}).

\bibitem{onoda_diffusion_2017}
\bibinfo{author}{Onoda, S.} \emph{et~al.}
\newblock \bibinfo{title}{Diffusion of {Vacancies} {Created} by {High}-{Energy}
  {Heavy} {Ion} {Strike} {Into} {Diamond}}.
\newblock \emph{\bibinfo{journal}{physica status solidi (a)}}
  \textbf{\bibinfo{volume}{214}}, \bibinfo{pages}{1700160}
  (\bibinfo{year}{2017}).

\bibitem{favaro_de_oliveira_tailoring_2017}
\bibinfo{author}{Fávaro~de Oliveira, F.} \emph{et~al.}
\newblock \bibinfo{title}{Tailoring spin defects in diamond by lattice
  charging}.
\newblock \emph{\bibinfo{journal}{Nature Communications}}
  \textbf{\bibinfo{volume}{8}}, \bibinfo{pages}{15409} (\bibinfo{year}{2017}).

\bibitem{Lomer1973}
\bibinfo{author}{Lomer, J.~N.} \& \bibinfo{author}{Wild, A. M.~A.}
\newblock \bibinfo{title}{Electron spin resonance in electron irradiated
  diamond annealed to high temperatures}.
\newblock \emph{\bibinfo{journal}{Radiation Effects}}
  \textbf{\bibinfo{volume}{17}}, \bibinfo{pages}{37--44}
  (\bibinfo{year}{1973}).

\bibitem{Nguyen2019PRB}
\bibinfo{author}{Nguyen, C.~T.} \emph{et~al.}
\newblock \bibinfo{title}{An integrated nanophotonic quantum register based on
  silicon-vacancy spins in diamond}.
\newblock \emph{\bibinfo{journal}{Physical Review B}}
  \textbf{\bibinfo{volume}{100}}, \bibinfo{pages}{165428}
  (\bibinfo{year}{2019}).

\bibitem{Maity2018PRApp}
\bibinfo{author}{Maity, S.} \emph{et~al.}
\newblock \bibinfo{title}{Spectral alignment of single-photon emitters in
  diamond using strain gradient}.
\newblock \emph{\bibinfo{journal}{Physical Review Applied}}
  \textbf{\bibinfo{volume}{10}}, \bibinfo{pages}{024050}
  (\bibinfo{year}{2018}).

\bibitem{Sipahigil2014PRL}
\bibinfo{author}{Sipahigil, A.} \emph{et~al.}
\newblock \bibinfo{title}{Indistinguishable photons from separated
  silicon-vacancy centers in diamond}.
\newblock \emph{\bibinfo{journal}{Physical Review Letters}}
  \textbf{\bibinfo{volume}{113}}, \bibinfo{pages}{113602}
  (\bibinfo{year}{2014}).

\bibitem{WangNatMat2023}
\bibinfo{author}{Wang, C.}, \bibinfo{author}{You, L.}, \bibinfo{author}{Cobden,
  D.} \& \bibinfo{author}{Wang, J.}
\newblock \bibinfo{title}{Towards two-dimensional van der waals
  ferroelectrics}.
\newblock \emph{\bibinfo{journal}{Nature Materials}}
  \textbf{\bibinfo{volume}{22}}, \bibinfo{pages}{542--552}
  (\bibinfo{year}{2023}).

\bibitem{2020SciPy-NMeth}
\bibinfo{author}{Virtanen, P.} \emph{et~al.}
\newblock \bibinfo{title}{{{SciPy} 1.0: Fundamental Algorithms for Scientific
  Computing in Python}}.
\newblock \emph{\bibinfo{journal}{Nature Methods}}
  \textbf{\bibinfo{volume}{17}}, \bibinfo{pages}{261--272}
  (\bibinfo{year}{2020}).

\bibitem{Avni1976}
\bibinfo{author}{Avni, Y.}
\newblock \bibinfo{title}{Energy spectra of x-ray clusters of galaxies}.
\newblock \emph{\bibinfo{journal}{The Astrophysical Journal}}
  \textbf{\bibinfo{volume}{210}}, \bibinfo{pages}{642} (\bibinfo{year}{1976}).

\bibitem{Brown2019}
\bibinfo{author}{Brown, K.~J.}, \bibinfo{author}{Chartier, E.},
  \bibinfo{author}{Sweet, E.~M.}, \bibinfo{author}{Hopper, D.~A.} \&
  \bibinfo{author}{Bassett, L.~C.}
\newblock \bibinfo{title}{{Cleaning diamond surfaces using boiling acid
  treatment in a standard laboratory chemical hood}}.
\newblock \emph{\bibinfo{journal}{J. Chem. Heal. Saf.}}
  \textbf{\bibinfo{volume}{26}}, \bibinfo{pages}{40--44}
  (\bibinfo{year}{2019}).

\bibitem{Atikian2014}
\bibinfo{author}{Atikian, H.~A.} \emph{et~al.}
\newblock \bibinfo{title}{{Superconducting nanowire single photon detector on
  diamond}}.
\newblock \emph{\bibinfo{journal}{Applied Physics Letters}}
  \textbf{\bibinfo{volume}{104}}, \bibinfo{pages}{122602}
  (\bibinfo{year}{2014}).

\bibitem{Binder2017SoftwareX}
\bibinfo{author}{Binder, J.~M.} \emph{et~al.}
\newblock \bibinfo{title}{{Qudi: A modular python suite for experiment control
  and data processing}}.
\newblock \emph{\bibinfo{journal}{SoftwareX}} \textbf{\bibinfo{volume}{6}},
  \bibinfo{pages}{85--90} (\bibinfo{year}{2017}).

\bibitem{Zenodo2025}
\bibinfo{author}{Pieplow, G.} \emph{et~al.}
\newblock \bibinfo{title}{{Quantum Electrometer for Time-Resolved Material
  Science at the Atomic Lattice Scale}}.
\newblock \emph{\bibinfo{journal}{Zenodo}}  (\bibinfo{year}{2025}).
\newblock \bibinfo{note}{{https://zenodo.org/records/15704695}}.

\end{thebibliography}


\begin{thebibliography}{10}
\expandafter\ifx\csname url\endcsname\relax
  \def\url#1{\texttt{#1}}\fi
\expandafter\ifx\csname urlprefix\endcsname\relax\def\urlprefix{URL }\fi
\providecommand{\bibinfo}[2]{#2}
\providecommand{\eprint}[2][]{\url{#2}}

\bibitem{rayleigh_xxxi_1879}
\bibinfo{author}{Rayleigh, J. W.~S.}
\newblock \bibinfo{title}{{XXXI}. {Investigations} in optics, with special
  reference to the spectroscope}.
\newblock \emph{\bibinfo{journal}{The London, Edinburgh, and Dublin
  Philosophical Magazine and Journal of Science}}
  \textbf{\bibinfo{volume}{8:49}}, \bibinfo{pages}{261--274}
  (\bibinfo{year}{1879}).

\bibitem{ShkarinPRL2021}
\bibinfo{author}{Shkarin, A.} \emph{et~al.}
\newblock \bibinfo{title}{Nanoscopic {Charge} {Fluctuations} in a {Gallium}
  {Phosphide} {Waveguide} {Measured} by {Single} {Molecules}}.
\newblock \emph{\bibinfo{journal}{Physical Review Letters}}
  \textbf{\bibinfo{volume}{126}}, \bibinfo{pages}{133602}
  (\bibinfo{year}{2021}).

\bibitem{Olivero1977}
\bibinfo{author}{Olivero, J.} \& \bibinfo{author}{Longbothum, R.}
\newblock \bibinfo{title}{Empirical fits to the voigt line width: A brief
  review}.
\newblock \emph{\bibinfo{journal}{Journal of Quantitative Spectroscopy and
  Radiative Transfer}} \textbf{\bibinfo{volume}{17}}, \bibinfo{pages}{233--236}
  (\bibinfo{year}{1977}).

\bibitem{Slepetz2014PhysChem}
\bibinfo{author}{Slepetz, B.} \& \bibinfo{author}{Kertesz, M.}
\newblock \bibinfo{title}{Divacancies in diamond: a stepwise formation
  mechanism}.
\newblock \emph{\bibinfo{journal}{Physical Chemistry Chemical Physics}}
  \textbf{\bibinfo{volume}{16}}, \bibinfo{pages}{1515--1521}
  (\bibinfo{year}{2014}).

\bibitem{Fu2019}
\bibinfo{author}{Fu, X.}, \bibinfo{author}{Xu, Z.}, \bibinfo{author}{He, Z.},
  \bibinfo{author}{Hartmaier, A.} \& \bibinfo{author}{Fang, F.}
\newblock \bibinfo{title}{Molecular dynamics simulation of silicon ion
  implantation into diamond and subsequent annealing}.
\newblock \emph{\bibinfo{journal}{Nuclear Instruments and Methods in Physics
  Research Section B: Beam Interactions with Materials and Atoms}}
  \textbf{\bibinfo{volume}{450}}, \bibinfo{pages}{51--55}
  (\bibinfo{year}{2019}).

\bibitem{koike1992}
\bibinfo{author}{Koike, J.}, \bibinfo{author}{Parkin, D.~M.} \&
  \bibinfo{author}{Mitchell, T.~E.}
\newblock \bibinfo{title}{Displacement threshold energy for type {IIa}
  diamond}.
\newblock \emph{\bibinfo{journal}{Applied Physics Letters}}
  \textbf{\bibinfo{volume}{60}}, \bibinfo{pages}{1450--1452}
  (\bibinfo{year}{1992}).

\bibitem{laidlaw2020}
\bibinfo{author}{Laidlaw, F. H.~J.}, \bibinfo{author}{Beanland, R.},
  \bibinfo{author}{Fisher, D.} \& \bibinfo{author}{Diggle, P.~L.}
\newblock \bibinfo{title}{Point defects and interstitial climb of 90° partial
  dislocations in brown type {IIa} natural diamond}.
\newblock \emph{\bibinfo{journal}{Acta Materialia}}
  \textbf{\bibinfo{volume}{201}}, \bibinfo{pages}{494--503}
  (\bibinfo{year}{2020}).

\bibitem{kiflawi2007}
\bibinfo{author}{Kiflawi, I.}, \bibinfo{author}{Collins, A.~T.},
  \bibinfo{author}{Iakoubovskii, K.} \& \bibinfo{author}{Fisher, D.}
\newblock \bibinfo{title}{Electron irradiation and the formation of
  vacancy–interstitial pairs in diamond}.
\newblock \emph{\bibinfo{journal}{Journal of Physics: Condensed Matter}}
  \textbf{\bibinfo{volume}{19}}, \bibinfo{pages}{046216}
  (\bibinfo{year}{2007}).

\bibitem{newton2002}
\bibinfo{author}{Newton, M.~E.}, \bibinfo{author}{Campbell, B.~A.},
  \bibinfo{author}{Twitchen, D.~J.}, \bibinfo{author}{Baker, J.~M.} \&
  \bibinfo{author}{Anthony, T.~R.}
\newblock \bibinfo{title}{Recombination-enhanced diffusion of self-interstitial
  atoms and vacancy–interstitial recombination in diamond}.
\newblock \emph{\bibinfo{journal}{Diamond and Related Materials}}
  \textbf{\bibinfo{volume}{11}}, \bibinfo{pages}{618--622}
  (\bibinfo{year}{2002}).

\bibitem{favaro_de_oliveira_tailoring_2017}
\bibinfo{author}{Fávaro~de Oliveira, F.} \emph{et~al.}
\newblock \bibinfo{title}{Tailoring spin defects in diamond by lattice
  charging}.
\newblock \emph{\bibinfo{journal}{Nature Communications}}
  \textbf{\bibinfo{volume}{8}}, \bibinfo{pages}{15409} (\bibinfo{year}{2017}).

\bibitem{Ziegler2010NIM}
\bibinfo{author}{Ziegler, J.~F.}, \bibinfo{author}{Ziegler, M.~D.} \&
  \bibinfo{author}{Biersack, J.~P.}
\newblock \bibinfo{title}{{SRIM -- The stopping and range of ions in matter}}.
\newblock \emph{\bibinfo{journal}{Nuclear Instruments and Methods in Physics
  Research Section B: Beam Interactions with Materials and Atoms}}
  \textbf{\bibinfo{volume}{268}}, \bibinfo{pages}{1818--1823}
  (\bibinfo{year}{2010}).

\bibitem{Kambs2018NJP}
\bibinfo{author}{Kambs, B.} \& \bibinfo{author}{Becher, C.}
\newblock \bibinfo{title}{Limitations on the indistinguishability of photons
  from remote solid state sources}.
\newblock \emph{\bibinfo{journal}{New Journal of Physics}}
  \textbf{\bibinfo{volume}{20}}, \bibinfo{pages}{115003}
  (\bibinfo{year}{2018}).

\bibitem{Cui_2006}
\bibinfo{author}{Cui, S.~T.}
\newblock \bibinfo{title}{Electrostatic potential in cylindrical dielectric
  media using the image charge method}.
\newblock \emph{\bibinfo{journal}{Mol. Phys.}} \textbf{\bibinfo{volume}{104}},
  \bibinfo{pages}{2993--3001} (\bibinfo{year}{2006}).

\bibitem{OrphalKobin2023PRX}
\bibinfo{author}{Orphal-Kobin, L.} \emph{et~al.}
\newblock \bibinfo{title}{Optically coherent nitrogen-vacancy defect centers in
  diamond nanostructures}.
\newblock \emph{\bibinfo{journal}{Physical Review X}}
  \textbf{\bibinfo{volume}{13}}, \bibinfo{pages}{011042}
  (\bibinfo{year}{2023}).

\bibitem{Kuruma2021}
\bibinfo{author}{Kuruma, K.} \emph{et~al.}
\newblock \bibinfo{title}{Coupling of a single tin-vacancy center to a photonic
  crystal cavity in diamond}.
\newblock \emph{\bibinfo{journal}{Applied Physics Letters}}
  \textbf{\bibinfo{volume}{118}}, \bibinfo{pages}{230601}
  (\bibinfo{year}{2021}).

\bibitem{Rugar2021PRX}
\bibinfo{author}{Rugar, A.~E.} \emph{et~al.}
\newblock \bibinfo{title}{Quantum photonic interface for tin-vacancy centers in
  diamond}.
\newblock \emph{\bibinfo{journal}{Physical Review X}}
  \textbf{\bibinfo{volume}{11}}, \bibinfo{pages}{031021}
  (\bibinfo{year}{2021}).

\bibitem{Burek2017}
\bibinfo{author}{Burek, M.~J.} \emph{et~al.}
\newblock \bibinfo{title}{Fiber-coupled diamond quantum nanophotonic
  interface}.
\newblock \emph{\bibinfo{journal}{Physical Review Applied}}
  \textbf{\bibinfo{volume}{8}}, \bibinfo{pages}{024026} (\bibinfo{year}{2017}).

\bibitem{Wang2019}
\bibinfo{author}{Wang, H.} \emph{et~al.}
\newblock \bibinfo{title}{Fast and high efficiency superconducting nanowire
  single-photon detector at 630 nm wavelength}.
\newblock \emph{\bibinfo{journal}{Applied Optics}}
  \textbf{\bibinfo{volume}{58}}, \bibinfo{pages}{1868--1872}
  (\bibinfo{year}{2019}).

\bibitem{Iwasaki2017PRL}
\bibinfo{author}{Iwasaki, T.} \emph{et~al.}
\newblock \bibinfo{title}{Tin-vacancy quantum emitters in diamond}.
\newblock \emph{\bibinfo{journal}{Physical Review Letters}}
  \textbf{\bibinfo{volume}{119}}, \bibinfo{pages}{253601}
  (\bibinfo{year}{2017}).

\bibitem{Wan2018}
\bibinfo{author}{Wan, N.~H.} \emph{et~al.}
\newblock \bibinfo{title}{Efficient extraction of light from a nitrogen-vacancy
  center in a diamond parabolic reflector}.
\newblock \emph{\bibinfo{journal}{Nano Letters}} \textbf{\bibinfo{volume}{18}},
  \bibinfo{pages}{2787--2793} (\bibinfo{year}{2018}).

\bibitem{Goerlitz2022NPJ}
\bibinfo{author}{G{\"o}rlitz, J.} \emph{et~al.}
\newblock \bibinfo{title}{Coherence of a charge stabilised tin-vacancy spin in
  diamond}.
\newblock \emph{\bibinfo{journal}{npj Quantum Information}}
  \textbf{\bibinfo{volume}{8}}, \bibinfo{pages}{1--9} (\bibinfo{year}{2022}).

\bibitem{Kitson1998PRA}
\bibinfo{author}{Kitson, S.~C.}, \bibinfo{author}{Jonsson, P.},
  \bibinfo{author}{Rarity, J.~G.} \& \bibinfo{author}{Tapster, P.~R.}
\newblock \bibinfo{title}{Intensity fluctuation spectroscopy of small numbers
  of dye molecules in a microcavity}.
\newblock \emph{\bibinfo{journal}{Physical Review A}}
  \textbf{\bibinfo{volume}{58}}, \bibinfo{pages}{620--627}
  (\bibinfo{year}{1998}).

\bibitem{Neu2011NJP}
\bibinfo{author}{Neu, E.} \emph{et~al.}
\newblock \bibinfo{title}{Single photon emission from silicon-vacancy colour
  centres in chemical vapour deposition nano-diamonds on iridium}.
\newblock \emph{\bibinfo{journal}{New Journal of Physics}}
  \textbf{\bibinfo{volume}{13}}, \bibinfo{pages}{025012}
  (\bibinfo{year}{2011}).

\bibitem{Gardill2021NanoLet}
\bibinfo{author}{Gardill, A.} \emph{et~al.}
\newblock \bibinfo{title}{Probing charge dynamics in diamond with an individual
  color center}.
\newblock \emph{\bibinfo{journal}{Nano Letters}} \textbf{\bibinfo{volume}{21}},
  \bibinfo{pages}{6960--6966} (\bibinfo{year}{2021}).

\bibitem{Bluvstein2019PRL}
\bibinfo{author}{Bluvstein, D.}, \bibinfo{author}{Zhang, Z.} \&
  \bibinfo{author}{Jayich, A. C.~B.}
\newblock \bibinfo{title}{Identifying and mitigating charge instabilities in
  shallow diamond nitrogen-vacancy centers}.
\newblock \emph{\bibinfo{journal}{Physical Review Letters}}
  \textbf{\bibinfo{volume}{122}}, \bibinfo{pages}{076101}
  (\bibinfo{year}{2019}).

\bibitem{Debroux2021PRX}
\bibinfo{author}{Debroux, R.} \emph{et~al.}
\newblock \bibinfo{title}{Quantum control of the tin-vacancy spin qubit in
  diamond}.
\newblock \emph{\bibinfo{journal}{Physical Review X}}
  \textbf{\bibinfo{volume}{11}}, \bibinfo{pages}{041041}
  (\bibinfo{year}{2021}).

\bibitem{Martinez2022PRL}
\bibinfo{author}{{Arjona Mart{\'i}nez}, J.} \emph{et~al.}
\newblock \bibinfo{title}{Photonic indistinguishability of the tin-vacancy
  center in nanostructured diamond}.
\newblock \emph{\bibinfo{journal}{Physical Review Letters}}
  \textbf{\bibinfo{volume}{129}}, \bibinfo{pages}{173603}
  (\bibinfo{year}{2022}).

\bibitem{Mortensen2010Nat}
\bibinfo{author}{Mortensen, K.~I.}, \bibinfo{author}{Churchman, L.~S.},
  \bibinfo{author}{Spudich, J.~A.} \& \bibinfo{author}{Flyvbjerg, H.}
\newblock \bibinfo{title}{Optimized localization analysis for single-molecule
  tracking and super-resolution microscopy}.
\newblock \emph{\bibinfo{journal}{Nature Methods}}
  \textbf{\bibinfo{volume}{7}}, \bibinfo{pages}{377--381}
  (\bibinfo{year}{2010}).

\bibitem{Narita.2023}
\bibinfo{author}{Narita, Y.} \emph{et~al.}
\newblock \bibinfo{title}{Multiple tin-vacancy centers in diamond with nearly
  identical photon frequency and linewidth}.
\newblock \emph{\bibinfo{journal}{Physical Review Applied}}
  \textbf{\bibinfo{volume}{19}}, \bibinfo{pages}{024061}
  (\bibinfo{year}{2023}).

\bibitem{li_atomic_2024}
\bibinfo{author}{Li, Z.} \emph{et~al.}
\newblock \bibinfo{title}{Atomic optical antennas in solids}.
\newblock \emph{\bibinfo{journal}{Nature Photonics}}
  \textbf{\bibinfo{volume}{18}}, \bibinfo{pages}{1113--1120}
  (\bibinfo{year}{2024}).

\bibitem{Cohen1998}
\bibinfo{author}{Cohen-Tannoudji, C.}, \bibinfo{author}{Dupont-Roc, J.} \&
  \bibinfo{author}{Grynberg, G.}
\newblock \emph{\bibinfo{title}{Atom---Photon Interactions}}
  (\bibinfo{publisher}{Wiley}, \bibinfo{year}{1998}).

\bibitem{Trusheim2020PRL}
\bibinfo{author}{Trusheim, M.~E.} \emph{et~al.}
\newblock \bibinfo{title}{Transform-limited photons from a coherent tin-vacancy
  spin in diamond}.
\newblock \emph{\bibinfo{journal}{Physical Review Letters}}
  \textbf{\bibinfo{volume}{124}}, \bibinfo{pages}{023602}
  (\bibinfo{year}{2020}).

\bibitem{anderson_absence_1958}
\bibinfo{author}{Anderson, P.~W.}
\newblock \bibinfo{title}{Absence of {Diffusion} in {Certain} {Random}
  {Lattices}}.
\newblock \emph{\bibinfo{journal}{Physical Review}}
  \textbf{\bibinfo{volume}{109}}, \bibinfo{pages}{1492--1505}
  (\bibinfo{year}{1958}).

\bibitem{lehtinen_molecular_2016}
\bibinfo{author}{Lehtinen, O.} \emph{et~al.}
\newblock \bibinfo{title}{Molecular dynamics simulations of shallow nitrogen
  and silicon implantation into diamond}.
\newblock \emph{\bibinfo{journal}{Physical Review B}}
  \textbf{\bibinfo{volume}{93}}, \bibinfo{pages}{035202}
  (\bibinfo{year}{2016}).

\bibitem{Santis2021PRL}
\bibinfo{author}{de~Santis, L.}, \bibinfo{author}{Trusheim, M.~E.},
  \bibinfo{author}{Chen, K.~C.} \& \bibinfo{author}{Englund, D.~R.}
\newblock \bibinfo{title}{{Investigation of the Stark Effect on a
  Centrosymmetric Quantum Emitter in Diamond}}.
\newblock \emph{\bibinfo{journal}{Physical Review Letters}}
  \textbf{\bibinfo{volume}{127}}, \bibinfo{pages}{147402}
  (\bibinfo{year}{2021}).

\bibitem{Ziegler2010NIM}
\bibinfo{author}{Ziegler, J.~F.}, \bibinfo{author}{Ziegler, M.~D.} \&
  \bibinfo{author}{Biersack, J.~P.}
\newblock \bibinfo{title}{{SRIM -- The stopping and range of ions in matter}}.
\newblock \emph{\bibinfo{journal}{Nuclear Instruments and Methods in Physics
  Research Section B: Beam Interactions with Materials and Atoms}}
  \textbf{\bibinfo{volume}{268}}, \bibinfo{pages}{1818--1823}
  (\bibinfo{year}{2010}).

\bibitem{kaczer_impact_2002}
\bibinfo{author}{Kaczer, B.} \emph{et~al.}
\newblock \bibinfo{title}{Impact of {MOSFET} gate oxide breakdown on digital
  circuit operation and reliability}.
\newblock \emph{\bibinfo{journal}{IEEE Transactions on Electron Devices}}
  \textbf{\bibinfo{volume}{49}}, \bibinfo{pages}{500--506}
  (\bibinfo{year}{2002}).

\bibitem{stacey_evidence_2019}
\bibinfo{author}{Stacey, A.} \emph{et~al.}
\newblock \bibinfo{title}{Evidence for {Primal} sp2 {Defects} at the {Diamond}
  {Surface}: {Candidates} for {Electron} {Trapping} and {Noise} {Sources}}.
\newblock \emph{\bibinfo{journal}{Advanced Materials Interfaces}}
  \textbf{\bibinfo{volume}{6}}, \bibinfo{pages}{1801449}
  (\bibinfo{year}{2019}).

\bibitem{Rehman2022}
\bibinfo{author}{Rehman, A.} \emph{et~al.}
\newblock \bibinfo{title}{Nature of the 1/f noise in graphene-direct evidence
  for the mobility fluctuation mechanism}.
\newblock \emph{\bibinfo{journal}{Nanoscale}} \textbf{\bibinfo{volume}{14}},
  \bibinfo{pages}{7242--7249} (\bibinfo{year}{2022}).

\bibitem{Defo2023PRB}
\bibinfo{author}{{Kuate Defo}, R.}, \bibinfo{author}{Rodriguez, A.~W.},
  \bibinfo{author}{Kaxiras, E.} \& \bibinfo{author}{Richardson, S.~L.}
\newblock \bibinfo{title}{Theoretical investigation of charge transfer between
  two defects in a wide band gap semiconductor}.
\newblock \emph{\bibinfo{journal}{Physical Review B}}
  \textbf{\bibinfo{volume}{107}}, \bibinfo{pages}{125305}
  (\bibinfo{year}{2023}).

\bibitem{Mittiga2018PRL}
\bibinfo{author}{Mittiga, T.} \emph{et~al.}
\newblock \bibinfo{title}{Imaging the local charge environment of
  nitrogen-vacancy centers in diamond}.
\newblock \emph{\bibinfo{journal}{Physical Review Letters}}
  \textbf{\bibinfo{volume}{121}}, \bibinfo{pages}{246402}
  (\bibinfo{year}{2018}).

\bibitem{Vamivakas2011}
\bibinfo{author}{Vamivakas, A.~N.} \emph{et~al.}
\newblock \bibinfo{title}{Nanoscale optical electrometer}.
\newblock \emph{\bibinfo{journal}{Physical Review Letters}}
  \textbf{\bibinfo{volume}{107}}, \bibinfo{pages}{166802}
  (\bibinfo{year}{2011}).

\bibitem{dolde2011}
\bibinfo{author}{Dolde, F.} \emph{et~al.}
\newblock \bibinfo{title}{Electric-field sensing using single diamond spins}.
\newblock \emph{\bibinfo{journal}{Nature Physics}}
  \textbf{\bibinfo{volume}{7}}, \bibinfo{pages}{459--463}
  (\bibinfo{year}{2011}).

\bibitem{dolde2014}
\bibinfo{author}{Dolde, F.} \emph{et~al.}
\newblock \bibinfo{title}{Nanoscale {Detection} of a {Single} {Fundamental}
  {Charge} in {Ambient} {Conditions} {Using} the {NV} - {Center} in {Diamond}}.
\newblock \emph{\bibinfo{journal}{Physical Review Letters}}
  \textbf{\bibinfo{volume}{112}}, \bibinfo{pages}{097603}
  (\bibinfo{year}{2014}).

\bibitem{Chen2016}
\bibinfo{author}{Chen, D.}, \bibinfo{author}{Lander, G.~R.},
  \bibinfo{author}{Krowpman, K.~S.}, \bibinfo{author}{Solomon, G.~S.} \&
  \bibinfo{author}{Flagg, E.~B.}
\newblock \bibinfo{title}{Characterization of the local charge environment of a
  single quantum dot via resonance fluorescence}.
\newblock \emph{\bibinfo{journal}{Physical Review B}}
  \textbf{\bibinfo{volume}{93}}, \bibinfo{pages}{115307}
  (\bibinfo{year}{2016}).

\bibitem{Iwasaki2017ACS}
\bibinfo{author}{Iwasaki, T.} \emph{et~al.}
\newblock \bibinfo{title}{Direct nanoscale sensing of the internal electric
  field in operating semiconductor devices using single electron spins}.
\newblock \emph{\bibinfo{journal}{ACS Nano}} \textbf{\bibinfo{volume}{11}},
  \bibinfo{pages}{1238--1245} (\bibinfo{year}{2017}).

\bibitem{broadway2018}
\bibinfo{author}{Broadway, D.~A.} \emph{et~al.}
\newblock \bibinfo{title}{Spatial mapping of band bending in semiconductor
  devices using in situ quantum sensors}.
\newblock \emph{\bibinfo{journal}{Nature Electronics}}
  \textbf{\bibinfo{volume}{1}}, \bibinfo{pages}{502--507}
  (\bibinfo{year}{2018}).

\bibitem{li_2020}
\bibinfo{author}{Li, R.} \emph{et~al.}
\newblock \bibinfo{title}{Nanoscale {Electrometry} {Based} on a
  {Magnetic}-{Field}-{Resistant} {Spin} {Sensor}}.
\newblock \emph{\bibinfo{journal}{Physical Review Letters}}
  \textbf{\bibinfo{volume}{124}}, \bibinfo{pages}{247701}
  (\bibinfo{year}{2020}).

\bibitem{barson_nanoscale_2021}
\bibinfo{author}{Barson, M. S.~J.} \emph{et~al.}
\newblock \bibinfo{title}{Nanoscale {Vector} {Electric} {Field} {Imaging}
  {Using} a {Single} {Electron} {Spin}}.
\newblock \emph{\bibinfo{journal}{Nano Letters}} \textbf{\bibinfo{volume}{21}},
  \bibinfo{pages}{2962--2967} (\bibinfo{year}{2021}).

\bibitem{Qiu2022npj}
\bibinfo{author}{Qiu, Z.}, \bibinfo{author}{Hamo, A.}, \bibinfo{author}{Vool,
  U.}, \bibinfo{author}{Zhou, T.~X.} \& \bibinfo{author}{Yacoby, A.}
\newblock \bibinfo{title}{Nanoscale electric field imaging with an ambient
  scanning quantum sensor microscope}.
\newblock \emph{\bibinfo{journal}{npj Quantum Information}}
  \textbf{\bibinfo{volume}{8}}, \bibinfo{pages}{1--7} (\bibinfo{year}{2022}).

\bibitem{delord_correlated_2024-1}
\bibinfo{author}{Delord, T.}, \bibinfo{author}{Monge, R.} \&
  \bibinfo{author}{Meriles, C.~A.}
\newblock \bibinfo{title}{Correlated {Spectroscopy} of {Electric} {Noise} with
  {Color} {Center} {Clusters}}.
\newblock \emph{\bibinfo{journal}{Nano Letters}} \textbf{\bibinfo{volume}{24}},
  \bibinfo{pages}{6474--6479} (\bibinfo{year}{2024}).

\bibitem{Ji2024}
\bibinfo{author}{Ji, W.} \emph{et~al.}
\newblock \bibinfo{title}{Correlated sensing with a solid-state quantum
  multisensor system for atomic-scale structural analysis}.
\newblock \emph{\bibinfo{journal}{Nature Photonics}}
  \textbf{\bibinfo{volume}{18}}, \bibinfo{pages}{230--235}
  (\bibinfo{year}{2024}).

\bibitem{song_reply_2023}
\bibinfo{author}{Song, Q.} \emph{et~al.}
\newblock \bibinfo{title}{Reply to: {Dilemma} in optical identification of
  single-layer multiferroics}.
\newblock \emph{\bibinfo{journal}{Nature}} \textbf{\bibinfo{volume}{619}},
  \bibinfo{pages}{E44--E46} (\bibinfo{year}{2023}).

\bibitem{uren1985}
\bibinfo{author}{Uren, M.~J.}, \bibinfo{author}{Day, D.~J.} \&
  \bibinfo{author}{Kirton, M.~J.}
\newblock \bibinfo{title}{1/ \textit{f} and random telegraph noise in silicon
  metal‐oxide‐semiconductor field‐effect transistors}.
\newblock \emph{\bibinfo{journal}{Applied Physics Letters}}
  \textbf{\bibinfo{volume}{47}}, \bibinfo{pages}{1195--1197}
  (\bibinfo{year}{1985}).

\bibitem{Stampfer2020phd}
\bibinfo{author}{Stampfer, B.}
\newblock \bibinfo{title}{Advanced electrical characterization of charge
  trapping in mos transistors: Advanced electrical characterization of charge
  trapping in mos transistors}.
\newblock \emph{\bibinfo{journal}{{PhD Thesis, TU Wien}}}
  (\bibinfo{year}{2020}).

\bibitem{hite2013}
\bibinfo{author}{Hite, D.} \emph{et~al.}
\newblock \bibinfo{title}{Surface science for improved ion traps}.
\newblock \emph{\bibinfo{journal}{MRS Bulletin}} \textbf{\bibinfo{volume}{38}},
  \bibinfo{pages}{826--833} (\bibinfo{year}{2013}).

\bibitem{faoro2006}
\bibinfo{author}{Faoro, L.} \& \bibinfo{author}{Ioffe, L.~B.}
\newblock \bibinfo{title}{Quantum {Two} {Level} {Systems} and {Kondo}-{Like}
  {Traps} as {Possible} {Sources} of {Decoherence} in {Superconducting}
  {Qubits}}.
\newblock \emph{\bibinfo{journal}{Physical Review Letters}}
  \textbf{\bibinfo{volume}{96}}, \bibinfo{pages}{047001}
  (\bibinfo{year}{2006}).

\bibitem{reshef2021}
\bibinfo{author}{Reshef, A.} \& \bibinfo{author}{Caspary~Toroker, M.}
\newblock \bibinfo{title}{Method for assessing atomic sources of flicker noise
  in superconducting qubits}.
\newblock \emph{\bibinfo{journal}{npj Computational Materials}}
  \textbf{\bibinfo{volume}{7}}, \bibinfo{pages}{1--6} (\bibinfo{year}{2021}).

\bibitem{Morioka2020NatComm}
\bibinfo{author}{Morioka, N.} \emph{et~al.}
\newblock \bibinfo{title}{Spin-controlled generation of indistinguishable and
  distinguishable photons from silicon vacancy centres in silicon carbide}.
\newblock \emph{\bibinfo{journal}{Nature Communications}}
  \textbf{\bibinfo{volume}{11}}, \bibinfo{pages}{2516} (\bibinfo{year}{2020}).

\bibitem{White2021Opt}
\bibinfo{author}{White, S.} \emph{et~al.}
\newblock \bibinfo{title}{Phonon dephasing and spectral diffusion of quantum
  emitters in hexagonal boron nitride}.
\newblock \emph{\bibinfo{journal}{Optica}} \textbf{\bibinfo{volume}{8}},
  \bibinfo{pages}{1153} (\bibinfo{year}{2021}).

\bibitem{Vajner2022AQT}
\bibinfo{author}{Vajner, D.~A.}, \bibinfo{author}{Rickert, L.},
  \bibinfo{author}{Gao, T.}, \bibinfo{author}{Kaymazlar, K.} \&
  \bibinfo{author}{Heindel, T.}
\newblock \bibinfo{title}{Quantum communication using semiconductor quantum
  dots}.
\newblock \emph{\bibinfo{journal}{Advanced Quantum Technologies}}
  \textbf{\bibinfo{volume}{5}}, \bibinfo{pages}{2100116}
  (\bibinfo{year}{2022}).

\bibitem{OrphalKobin2023PRX}
\bibinfo{author}{Orphal-Kobin, L.} \emph{et~al.}
\newblock \bibinfo{title}{Optically coherent nitrogen-vacancy defect centers in
  diamond nanostructures}.
\newblock \emph{\bibinfo{journal}{Physical Review X}}
  \textbf{\bibinfo{volume}{13}}, \bibinfo{pages}{011042}
  (\bibinfo{year}{2023}).

\bibitem{Pompili2021Sci}
\bibinfo{author}{Pompili, M.} \emph{et~al.}
\newblock \bibinfo{title}{Realization of a multinode quantum network of remote
  solid-state qubits}.
\newblock \emph{\bibinfo{journal}{Science}} \textbf{\bibinfo{volume}{372}},
  \bibinfo{pages}{259--264} (\bibinfo{year}{2021}).

\bibitem{Iwasaki2017PRL}
\bibinfo{author}{Iwasaki, T.} \emph{et~al.}
\newblock \bibinfo{title}{Tin-vacancy quantum emitters in diamond}.
\newblock \emph{\bibinfo{journal}{Physical Review Letters}}
  \textbf{\bibinfo{volume}{119}}, \bibinfo{pages}{253601}
  (\bibinfo{year}{2017}).

\bibitem{Goss1996PRL}
\bibinfo{author}{Goss, J.~P.}, \bibinfo{author}{Jones, R.},
  \bibinfo{author}{Breuer, S.~J.}, \bibinfo{author}{Briddon, P.~R.} \&
  \bibinfo{author}{{\"O}berg, S.}
\newblock \bibinfo{title}{The twelve-line 1.682 ev luminescence center in
  diamond and the vacancy-silicon complex}.
\newblock \emph{\bibinfo{journal}{Physical Review Letters}}
  \textbf{\bibinfo{volume}{77}}, \bibinfo{pages}{3041--3044}
  (\bibinfo{year}{1996}).

\bibitem{Iwasaki2015SciRep}
\bibinfo{author}{Iwasaki, T.} \emph{et~al.}
\newblock \bibinfo{title}{Germanium-vacancy single color centers in diamond}.
\newblock \emph{\bibinfo{journal}{Scientific Reports}}
  \textbf{\bibinfo{volume}{5}}, \bibinfo{pages}{12882} (\bibinfo{year}{2015}).

\bibitem{Niaz2016JPCC}
\bibinfo{author}{Niaz, S.} \& \bibinfo{author}{Zdetsis, A.~D.}
\newblock \bibinfo{title}{Comprehensive ab initio study of electronic, optical,
  and cohesive properties of silicon quantum dots of various morphologies and
  sizes up to infinity}.
\newblock \emph{\bibinfo{journal}{The Journal of Physical Chemistry C}}
  \textbf{\bibinfo{volume}{120}}, \bibinfo{pages}{11288--11298}
  (\bibinfo{year}{2016}).

\bibitem{Wang2021ACS}
\bibinfo{author}{Wang, P.}, \bibinfo{author}{Taniguchi, T.},
  \bibinfo{author}{Miyamoto, Y.}, \bibinfo{author}{Hatano, M.} \&
  \bibinfo{author}{Iwasaki, T.}
\newblock \bibinfo{title}{Low-temperature spectroscopic investigation of
  lead-vacancy centers in diamond fabricated by high-pressure and
  high-temperature treatment}.
\newblock \emph{\bibinfo{journal}{ACS Photonics}} \textbf{\bibinfo{volume}{8}},
  \bibinfo{pages}{2947--2954} (\bibinfo{year}{2021}).

\bibitem{ShkarinPRL2021}
\bibinfo{author}{Shkarin, A.} \emph{et~al.}
\newblock \bibinfo{title}{Nanoscopic {Charge} {Fluctuations} in a {Gallium}
  {Phosphide} {Waveguide} {Measured} by {Single} {Molecules}}.
\newblock \emph{\bibinfo{journal}{Physical Review Letters}}
  \textbf{\bibinfo{volume}{126}}, \bibinfo{pages}{133602}
  (\bibinfo{year}{2021}).

\bibitem{Stark1914}
\bibinfo{author}{Stark, J.}
\newblock \bibinfo{title}{Beobachtungen \"{u}ber den effekt des elektrischen
  feldes auf spektrallinien. i. quereffekt}.
\newblock \emph{\bibinfo{journal}{Annalen der Physik}}
  \textbf{\bibinfo{volume}{348}}, \bibinfo{pages}{965--982}
  (\bibinfo{year}{1914}).

\bibitem{Tamarat2006PRL}
\bibinfo{author}{Tamarat, P.} \emph{et~al.}
\newblock \bibinfo{title}{Stark shift control of single optical centers in
  diamond}.
\newblock \emph{\bibinfo{journal}{Physical Review Letters}}
  \textbf{\bibinfo{volume}{97}}, \bibinfo{pages}{083002}
  (\bibinfo{year}{2006}).

\bibitem{Ruhl2020ACS}
\bibinfo{author}{R{\"u}hl, M.}, \bibinfo{author}{Bergmann, L.},
  \bibinfo{author}{Krieger, M.} \& \bibinfo{author}{Weber, H.~B.}
\newblock \bibinfo{title}{Stark tuning of the silicon vacancy in silicon
  carbide}.
\newblock \emph{\bibinfo{journal}{Nano Letters}} \textbf{\bibinfo{volume}{20}},
  \bibinfo{pages}{658--663} (\bibinfo{year}{2020}).

\bibitem{Huxter2023NatPhys}
\bibinfo{author}{Huxter, W.~S.}, \bibinfo{author}{Sarott, M.~F.},
  \bibinfo{author}{Trassin, M.} \& \bibinfo{author}{Degen, C.~L.}
\newblock \bibinfo{title}{Imaging ferroelectric domains with a single-spin
  scanning quantum sensor}.
\newblock \emph{\bibinfo{journal}{Nature Physics}}
  \textbf{\bibinfo{volume}{19}}, \bibinfo{pages}{644--648}
  (\bibinfo{year}{2023}).

\bibitem{Maletinsky2012NatNT}
\bibinfo{author}{Maletinsky, P.} \emph{et~al.}
\newblock \bibinfo{title}{A robust scanning diamond sensor for nanoscale
  imaging with single nitrogen-vacancy centres}.
\newblock \emph{\bibinfo{journal}{Nature Nanotechnology}}
  \textbf{\bibinfo{volume}{7}}, \bibinfo{pages}{320--324}
  (\bibinfo{year}{2012}).

\bibitem{Budker2007NatPhys}
\bibinfo{author}{Budker, D.} \& \bibinfo{author}{Romalis, M.}
\newblock \bibinfo{title}{Optical magnetometry}.
\newblock \emph{\bibinfo{journal}{Nature Physics}}
  \textbf{\bibinfo{volume}{3}}, \bibinfo{pages}{227--234}
  (\bibinfo{year}{2007}).

\bibitem{Foy2020ACS}
\bibinfo{author}{Foy, C.} \emph{et~al.}
\newblock \bibinfo{title}{Wide-field magnetic field and temperature imaging
  using nanoscale quantum sensors}.
\newblock \emph{\bibinfo{journal}{ACS Applied Materials {\&} Interfaces}}
  \textbf{\bibinfo{volume}{12}}, \bibinfo{pages}{26525--26533}
  (\bibinfo{year}{2020}).

\bibitem{Kucsko2013Nat}
\bibinfo{author}{Kucsko, G.} \emph{et~al.}
\newblock \bibinfo{title}{Nanometre-scale thermometry in a living cell}.
\newblock \emph{\bibinfo{journal}{Nature}} \textbf{\bibinfo{volume}{500}},
  \bibinfo{pages}{54--58} (\bibinfo{year}{2013}).

\bibitem{li_atomic_2024}
\bibinfo{author}{Li, Z.} \emph{et~al.}
\newblock \bibinfo{title}{Atomic optical antennas in solids}.
\newblock \emph{\bibinfo{journal}{Nature Photonics}}
  \textbf{\bibinfo{volume}{18}}, \bibinfo{pages}{1113--1120}
  (\bibinfo{year}{2024}).

\bibitem{Hourahine2000}
\bibinfo{author}{Hourahine, B.} \emph{et~al.}
\newblock \bibinfo{title}{Identification of the hexavacancy in silicon with the
  ${B}^4_{80}$ optical center}.
\newblock \emph{\bibinfo{journal}{Physical Review B}}
  \textbf{\bibinfo{volume}{61}}, \bibinfo{pages}{12594--12597}
  (\bibinfo{year}{2000}).

\bibitem{gaubas_spectroscopy_2016}
\bibinfo{author}{Gaubas, E.}, \bibinfo{author}{Ceponis, T.},
  \bibinfo{author}{Meskauskaite, D.}, \bibinfo{author}{Grigonis, R.} \&
  \bibinfo{author}{Sirutkaitis, V.}
\newblock \bibinfo{title}{Spectroscopy of defects in hpht and cvd diamond by
  esr and pulsed photo-ionization measurements}.
\newblock \emph{\bibinfo{journal}{Journal of Instrumentation}}
  \textbf{\bibinfo{volume}{11}}, \bibinfo{pages}{C01017--C01017}
  (\bibinfo{year}{2016}).

\bibitem{girolami_transport_2019}
\bibinfo{author}{Girolami, M.} \emph{et~al.}
\newblock \bibinfo{title}{Transport properties of photogenerated charge
  carriers in black diamond films}.
\newblock \emph{\bibinfo{journal}{Ceramics International}}
  \textbf{\bibinfo{volume}{45}}, \bibinfo{pages}{9544--9547}
  (\bibinfo{year}{2019}).

\bibitem{Trusheim2020PRL}
\bibinfo{author}{Trusheim, M.~E.} \emph{et~al.}
\newblock \bibinfo{title}{Transform-limited photons from a coherent tin-vacancy
  spin in diamond}.
\newblock \emph{\bibinfo{journal}{Physical Review Letters}}
  \textbf{\bibinfo{volume}{124}}, \bibinfo{pages}{023602}
  (\bibinfo{year}{2020}).

\bibitem{Goerlitz2022NPJ}
\bibinfo{author}{G{\"o}rlitz, J.} \emph{et~al.}
\newblock \bibinfo{title}{Coherence of a charge stabilised tin-vacancy spin in
  diamond}.
\newblock \emph{\bibinfo{journal}{npj Quantum Information}}
  \textbf{\bibinfo{volume}{8}}, \bibinfo{pages}{1--9} (\bibinfo{year}{2022}).

\bibitem{Kambs2018NJP}
\bibinfo{author}{Kambs, B.} \& \bibinfo{author}{Becher, C.}
\newblock \bibinfo{title}{Limitations on the indistinguishability of photons
  from remote solid state sources}.
\newblock \emph{\bibinfo{journal}{New Journal of Physics}}
  \textbf{\bibinfo{volume}{20}}, \bibinfo{pages}{115003}
  (\bibinfo{year}{2018}).

\bibitem{Torun2021APL}
\bibinfo{author}{Torun, C.~G.} \emph{et~al.}
\newblock \bibinfo{title}{Optimized diamond inverted nanocones for enhanced
  color center to fiber coupling}.
\newblock \emph{\bibinfo{journal}{Applied Physics Letters}}
  \textbf{\bibinfo{volume}{118}}, \bibinfo{pages}{234002}
  (\bibinfo{year}{2021}).

\bibitem{Li2015NatComm}
\bibinfo{author}{Li, L.} \emph{et~al.}
\newblock \bibinfo{title}{Coherent spin control of a nanocavity-enhanced qubit
  in diamond}.
\newblock \emph{\bibinfo{journal}{Nature Communications}}
  \textbf{\bibinfo{volume}{6}}, \bibinfo{pages}{6173} (\bibinfo{year}{2015}).

\bibitem{Bopp2022Arxiv}
\bibinfo{author}{Bopp, J.~M.} \emph{et~al.}
\newblock \bibinfo{title}{{`Sawfish' Photonic Crystal Cavity for Near--Unity
  Emitter--to--Fiber Interfacing in Quantum Network Applications}}.
\newblock \emph{\bibinfo{journal}{Advanced Optical Materials}}
  \textbf{\bibinfo{volume}{12}}, \bibinfo{pages}{2301286}
  (\bibinfo{year}{2024}).

\bibitem{Bluvstein2019PRL}
\bibinfo{author}{Bluvstein, D.}, \bibinfo{author}{Zhang, Z.} \&
  \bibinfo{author}{Jayich, A. C.~B.}
\newblock \bibinfo{title}{Identifying and mitigating charge instabilities in
  shallow diamond nitrogen-vacancy centers}.
\newblock \emph{\bibinfo{journal}{Physical Review Letters}}
  \textbf{\bibinfo{volume}{122}}, \bibinfo{pages}{076101}
  (\bibinfo{year}{2019}).

\bibitem{ElementSix2022}
\bibinfo{author}{{Element Six}}.
\newblock \bibinfo{title}{{CVD Diamond Handbook}} (\bibinfo{year}{2022}).
\newblock \bibinfo{note}{{https://e6cvd.com/media/wysiwyg/pdf/
  Element\_Six\_CVD\_Diamond\_handbook\_2022.pdf}}.

\bibitem{yamamoto_extending_2013}
\bibinfo{author}{Yamamoto, T.} \emph{et~al.}
\newblock \bibinfo{title}{Extending spin coherence times of diamond qubits by
  high-temperature annealing}.
\newblock \emph{\bibinfo{journal}{Physical Review B}}
  \textbf{\bibinfo{volume}{88}}, \bibinfo{pages}{075206}
  (\bibinfo{year}{2013}).

\bibitem{Slepetz2014PhysChem}
\bibinfo{author}{Slepetz, B.} \& \bibinfo{author}{Kertesz, M.}
\newblock \bibinfo{title}{Divacancies in diamond: a stepwise formation
  mechanism}.
\newblock \emph{\bibinfo{journal}{Physical Chemistry Chemical Physics}}
  \textbf{\bibinfo{volume}{16}}, \bibinfo{pages}{1515--1521}
  (\bibinfo{year}{2014}).

\bibitem{onoda_diffusion_2017}
\bibinfo{author}{Onoda, S.} \emph{et~al.}
\newblock \bibinfo{title}{Diffusion of {Vacancies} {Created} by {High}-{Energy}
  {Heavy} {Ion} {Strike} {Into} {Diamond}}.
\newblock \emph{\bibinfo{journal}{physica status solidi (a)}}
  \textbf{\bibinfo{volume}{214}}, \bibinfo{pages}{1700160}
  (\bibinfo{year}{2017}).

\bibitem{favaro_de_oliveira_tailoring_2017}
\bibinfo{author}{Fávaro~de Oliveira, F.} \emph{et~al.}
\newblock \bibinfo{title}{Tailoring spin defects in diamond by lattice
  charging}.
\newblock \emph{\bibinfo{journal}{Nature Communications}}
  \textbf{\bibinfo{volume}{8}}, \bibinfo{pages}{15409} (\bibinfo{year}{2017}).

\bibitem{Lomer1973}
\bibinfo{author}{Lomer, J.~N.} \& \bibinfo{author}{Wild, A. M.~A.}
\newblock \bibinfo{title}{Electron spin resonance in electron irradiated
  diamond annealed to high temperatures}.
\newblock \emph{\bibinfo{journal}{Radiation Effects}}
  \textbf{\bibinfo{volume}{17}}, \bibinfo{pages}{37--44}
  (\bibinfo{year}{1973}).

\bibitem{Nguyen2019PRB}
\bibinfo{author}{Nguyen, C.~T.} \emph{et~al.}
\newblock \bibinfo{title}{An integrated nanophotonic quantum register based on
  silicon-vacancy spins in diamond}.
\newblock \emph{\bibinfo{journal}{Physical Review B}}
  \textbf{\bibinfo{volume}{100}}, \bibinfo{pages}{165428}
  (\bibinfo{year}{2019}).

\bibitem{Maity2018PRApp}
\bibinfo{author}{Maity, S.} \emph{et~al.}
\newblock \bibinfo{title}{Spectral alignment of single-photon emitters in
  diamond using strain gradient}.
\newblock \emph{\bibinfo{journal}{Physical Review Applied}}
  \textbf{\bibinfo{volume}{10}}, \bibinfo{pages}{024050}
  (\bibinfo{year}{2018}).

\bibitem{Sipahigil2014PRL}
\bibinfo{author}{Sipahigil, A.} \emph{et~al.}
\newblock \bibinfo{title}{Indistinguishable photons from separated
  silicon-vacancy centers in diamond}.
\newblock \emph{\bibinfo{journal}{Physical Review Letters}}
  \textbf{\bibinfo{volume}{113}}, \bibinfo{pages}{113602}
  (\bibinfo{year}{2014}).

\bibitem{WangNatMat2023}
\bibinfo{author}{Wang, C.}, \bibinfo{author}{You, L.}, \bibinfo{author}{Cobden,
  D.} \& \bibinfo{author}{Wang, J.}
\newblock \bibinfo{title}{Towards two-dimensional van der waals
  ferroelectrics}.
\newblock \emph{\bibinfo{journal}{Nature Materials}}
  \textbf{\bibinfo{volume}{22}}, \bibinfo{pages}{542--552}
  (\bibinfo{year}{2023}).

\bibitem{2020SciPy-NMeth}
\bibinfo{author}{Virtanen, P.} \emph{et~al.}
\newblock \bibinfo{title}{{{SciPy} 1.0: Fundamental Algorithms for Scientific
  Computing in Python}}.
\newblock \emph{\bibinfo{journal}{Nature Methods}}
  \textbf{\bibinfo{volume}{17}}, \bibinfo{pages}{261--272}
  (\bibinfo{year}{2020}).

\bibitem{Avni1976}
\bibinfo{author}{Avni, Y.}
\newblock \bibinfo{title}{Energy spectra of x-ray clusters of galaxies}.
\newblock \emph{\bibinfo{journal}{The Astrophysical Journal}}
  \textbf{\bibinfo{volume}{210}}, \bibinfo{pages}{642} (\bibinfo{year}{1976}).

\bibitem{Brown2019}
\bibinfo{author}{Brown, K.~J.}, \bibinfo{author}{Chartier, E.},
  \bibinfo{author}{Sweet, E.~M.}, \bibinfo{author}{Hopper, D.~A.} \&
  \bibinfo{author}{Bassett, L.~C.}
\newblock \bibinfo{title}{{Cleaning diamond surfaces using boiling acid
  treatment in a standard laboratory chemical hood}}.
\newblock \emph{\bibinfo{journal}{J. Chem. Heal. Saf.}}
  \textbf{\bibinfo{volume}{26}}, \bibinfo{pages}{40--44}
  (\bibinfo{year}{2019}).

\bibitem{Atikian2014}
\bibinfo{author}{Atikian, H.~A.} \emph{et~al.}
\newblock \bibinfo{title}{{Superconducting nanowire single photon detector on
  diamond}}.
\newblock \emph{\bibinfo{journal}{Applied Physics Letters}}
  \textbf{\bibinfo{volume}{104}}, \bibinfo{pages}{122602}
  (\bibinfo{year}{2014}).

\bibitem{Binder2017SoftwareX}
\bibinfo{author}{Binder, J.~M.} \emph{et~al.}
\newblock \bibinfo{title}{{Qudi: A modular python suite for experiment control
  and data processing}}.
\newblock \emph{\bibinfo{journal}{SoftwareX}} \textbf{\bibinfo{volume}{6}},
  \bibinfo{pages}{85--90} (\bibinfo{year}{2017}).

\bibitem{Zenodo2025}
\bibinfo{author}{Pieplow, G.} \emph{et~al.}
\newblock \bibinfo{title}{{Quantum Electrometer for Time-Resolved Material
  Science at the Atomic Lattice Scale}}.
\newblock \emph{\bibinfo{journal}{Zenodo}}  (\bibinfo{year}{2025}).
\newblock \bibinfo{note}{{https://zenodo.org/records/15704695}}.

\end{thebibliography}

\section*{Acknowledgments}
    The authors would like to thank Viviana Villafane for sample annealing, Alex Kühlberger for his support with implantation, Lilian Hughes for discussions regarding the formation of vacancy complexes in diamond during annealing, Özgün Ozan Nacitarhan and Kilian Unterguggenburger with their support on experimental control software, David Hunger, Boris Naydenov, Matthew Trusheim, Laura Orphal-Kobin and Julian Bopp for their feedback on the manuscript.
    
    Funding for this project was provided by the European Research Council (ERC, Starting Grant project QUREP, No. 851810, granted to T.S.), the German Federal Ministry of Education and Research (BMBF, project DiNOQuant, No. 13N14921, granted to T.S.; project QPIS, No. 16KISQ032K, granted to G.P. and T.S.; project QPIC-1, No. 13N15858, granted to T.S.), the Einstein Foundation Berlin (Einstein Research Unit on Quantum Devices, granted to T.S.) and the Alexander von Humboldt Foundation (granted to J.H.D.M.).

\section*{Author Contributions Statement}
    G.P. was responsible for the modeling and executed a range of Monte-Carlo and stochastic simulations. 
    C.G.T constructed the confocal microscope under the guidance of J.H.D.M and designed the overall experimental infrastructure and protocols. 
    C.G. developed the EOM-based rapid optical readout setup. 
    C.G.T. acquired the data from S1, and conducted the spectral diffusion and autocorrelation measurements.
    C.G. and C.G.T. collected the experimental data for S2 and C.G. analyzed the temporal data.
    F.M.H. performed the Rabi oscillation control measurements with a home-built SHG laser system, which she constructed. A.T. implanted the Sn atoms into the sample. T.P. lead the sample preparation process and fabricated the nanopillars. T.S. developed the idea and supervised the project. G.P., C.G.T, and T.S. wrote the original manuscript with the input of all authors.

\section*{Competing Interest Statement}

Patent applications with numbers DE 10 2024 003 454.4 and US 19/036,421 have been submitted by Humboldt-Universität zu Berlin, listing C.G.T., G.P., and T.S. as inventors. The patented aspects include the temporal analysis of the PLE data for identifying the charge trap configuration and the application of Monte Carlo simulations for localizing individual proximal charge traps and determining the density of remote traps. The applications are currently pending approval. The other authors declare that they have no competing interests.

\end{document}


\title{Supplementary Information to Quantum Electrometer for Time-Resolved Material Science at the Atomic Lattice Scale}

\author{Gregor Pieplow}
\altaffiliation{These authors contributed equally to this work}
\affiliation{Department of Physics, Humboldt-Universit\"{a}t zu Berlin, 12489 Berlin, Germany}

\author{Cem G\"{u}ney Torun}
\altaffiliation{These authors contributed equally to this work}
\affiliation{Department of Physics, Humboldt-Universit\"{a}t zu Berlin, 12489 Berlin, Germany}

\author{Charlotta Gurr}
\affiliation{Department of Physics, Humboldt-Universit\"{a}t zu Berlin, 12489 Berlin, Germany}

\author{Joseph H. D. Munns}
\affiliation{Department of Physics, Humboldt-Universit\"{a}t zu Berlin, 12489 Berlin, Germany}

\author{Franziska Marie Herrmann}
\affiliation{Department of Physics, Humboldt-Universit\"{a}t zu Berlin, 12489 Berlin, Germany}

\author{Andreas Thies}
\affiliation{Ferdinand-Braun-Institut (FBH), Gustav-Kirchhoff-Str. 4, 12489 Berlin, Germany}

\author{Tommaso Pregnolato}
\affiliation{Department of Physics, Humboldt-Universit\"{a}t zu Berlin, 12489 Berlin, Germany}
\affiliation{Ferdinand-Braun-Institut (FBH), Gustav-Kirchhoff-Str. 4, 12489 Berlin, Germany}

\author{Tim Schr\"{o}der}
\email[Corresponding author: ]{tim.schroeder@physik.hu-berlin.de}
\affiliation{Department of Physics, Humboldt-Universit\"{a}t zu Berlin, 12489 Berlin, Germany}
\affiliation{Ferdinand-Braun-Institut (FBH), Gustav-Kirchhoff-Str. 4, 12489 Berlin, Germany}

\maketitle

\section*{Simulation details}
\subsection*{Relative electric field sensitivity}
\label{sec:ap_sensitiviy}

\begin{figure}
    \centering
    \includegraphics[width=\linewidth]{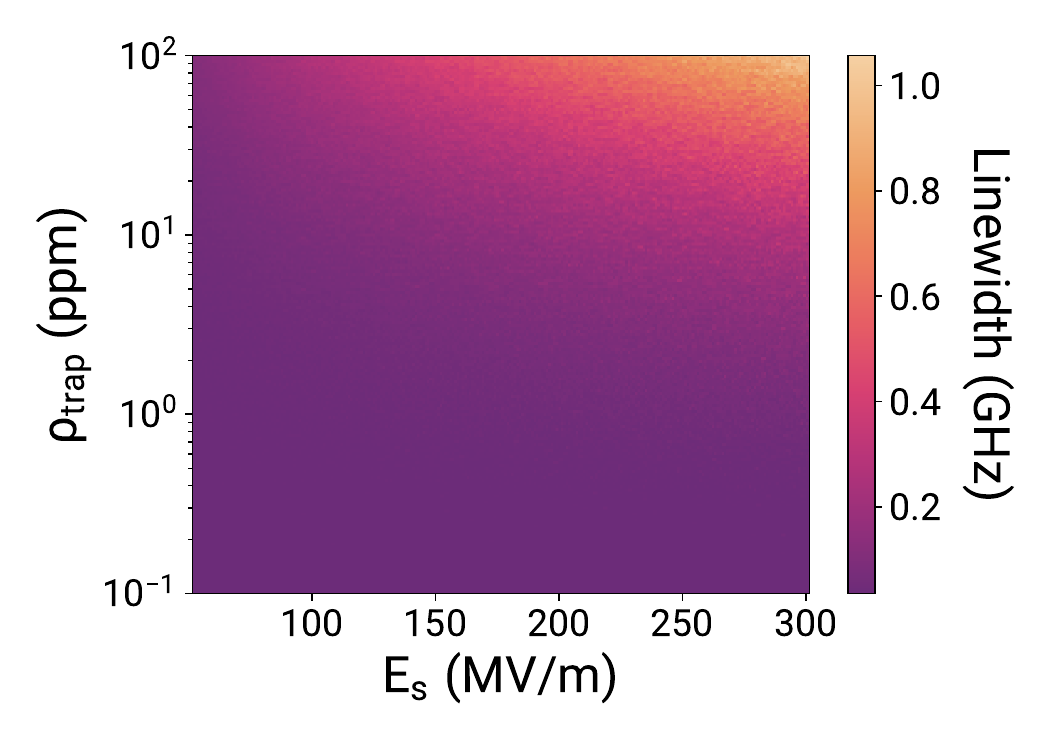}
    \caption{\textbf{Inhomogeneous broadening as a function of the local bias field $E_{\rm s}$}. The linewidth increases due to an increasing $E_{\rm s}$ or $\rho_{\rm trap}$.}
    \label{fig:ap_linewidth}
\end{figure}

The relative electric field sensitivity $\delta \epsilon = \Delta E / E_{\rm s}$ at a given $\rho_{\rm trap}$ can be calculated by identifying $\Delta E$ corresponding to the smallest spectral shift $\Delta_{\rm Stark}$ that can be resolved according to a modified Rayleigh criterion \cite{rayleigh_xxxi_1879} as described below. 

We calculate $\Delta E$ in a two step procedure:
First we simulate the expected inhomogeneously broadened linewidth in the presence of an electric field $\Vec{E}_{\rm s}$ (see Supplementary Fig.~\ref{fig:ap_linewidth}), which is generated by either a charged proximity trap or by a non-neutral charge state of the entire spatial trap configuration. The total field at the sensor position can be separated into two components $\vec{E} = \Vec{E}_{\rm s} + \delta \vec{E}_{\rm s}$, where $\delta \Vec{E}_{\rm s}$ is a fluctuating electric field produced by the varying charge states of the remote trap configuration. Similar modeling has been performed in \citen{ShkarinPRL2021}. 

The average value of the non-linear Stark shift is given by $\langle \Delta_{\rm Stark} \rangle = -\Delta\alpha (E_{\rm s}^2 + \sigma^2)$. Its variance is $\sigma_{\Delta_{\rm Stark}} = \Delta\alpha^2 (4 E_{\rm s}^2 \sigma^2 + 2 \sigma^4)$ (assuming $\delta \Vec{E}_{\rm s}$ is normally distributed with variance $\sigma$). The expressions demonstrate that a field induces both a discrete spectral shift and a quasi-permanent dipole moment, resulting in inhomogeneous broadening of lines dependent on the magnitudes of $E_{\rm s}$ and $\sigma$.

In the second step, we calculate $\Delta E$ at a given $\rho_{\rm trap}$ by using a modified Rayleigh criterion: Two spectral peaks originating from distinct fields $E_{\rm s}$ and $E_{\rm s}'$ are considered separable if the sum of the two individually normalized lineshapes resulting from $\vec{E} = \vec{E}_{\rm s} + \delta \vec{E}_{\rm s}$ and $\vec{E} = \vec{E}_{\rm s}' + \delta \vec{E}_{\rm s}$ exhibit a contrast of at least 26.3\% between their local maxima.

For Fig.~\ref{fig:sensor}\hyperlink{fig:principleL}{c} from the main text, we choose $\vec{E}_{\rm s} = (0, 0, E_{\rm s})$. To determine the $\langle \Delta_{\rm Stark} \rangle$ and the inhomogeneously broadened linewidth we employ a Monte Carlo simulation as outlined in the simulation overview. The traps generating $\delta \vec{E}_{\rm s}$ were placed at a fixed density $\rho_{\rm trap}$ in a conical volume $z >0$ with an opening angle of $45^\circ$, mimicking the anisotropic distribution of traps produced by implantation and annealing (e.g. Fig.~\ref{fig:annealing} from the main text). The conical volume was capped at $z = 30$~nm. A spherical volume with a radius of $2.5$ nm, was left empty of traps to reduce the occurrence of exaggerated multimodal spectral features.

The averaged linewidths required for Fig.~\ref{fig:principle}\hyperlink{fig:principleL}{c} from the main text are calculated using $\gamma_{\rm FWHM}  = a \sigma_{\rm hom} + (b \sigma_{\rm hom} ^ 2 + \sigma_{\rm inhom}^2)^{1/2}$, where $\sigma_{\rm hom}$ and $\sigma_{\rm inhom}$ are the full width half maximum of the Lorentzian and Gaussian Contribution to the Voigt profile and $a = 0.5346, b = 0.2166$ \cite{Olivero1977}.
In total we average the Gaussian and Lorentzian components of 100 different spatial trap configurations at a given $\rho_{\rm trap}$. For each $\rho_{\rm trap}$ we sample 2500 randomly generated charge states to simulate a single spectrum. The averaged spectral profiles corresponding to $\vec{E} = \vec{E}_{\rm s} + \delta \vec{E}_{\rm s}$ and $\vec{E}' = \vec{E}_{\rm s}' + \delta \vec{E}_{\rm s}$ are then used to determine $\Delta \vec{E}_{\rm s} = |E_{\rm s} - E_{\rm s}'|$ using the Rayleigh criterion.

Finally, we calculate the relative sensitivity shown in Fig.~\ref{fig:sensor}\hyperlink{fig:sensorL}{c} from the main text by dividing $\delta\epsilon = \Delta E / E_{\rm s}$ at a given $\rho_{\rm trap}$.

\subsection*{Impact of noise}
\label{sec:app_noise}

\begin{figure}
    \centering
    \includegraphics[width = \linewidth]{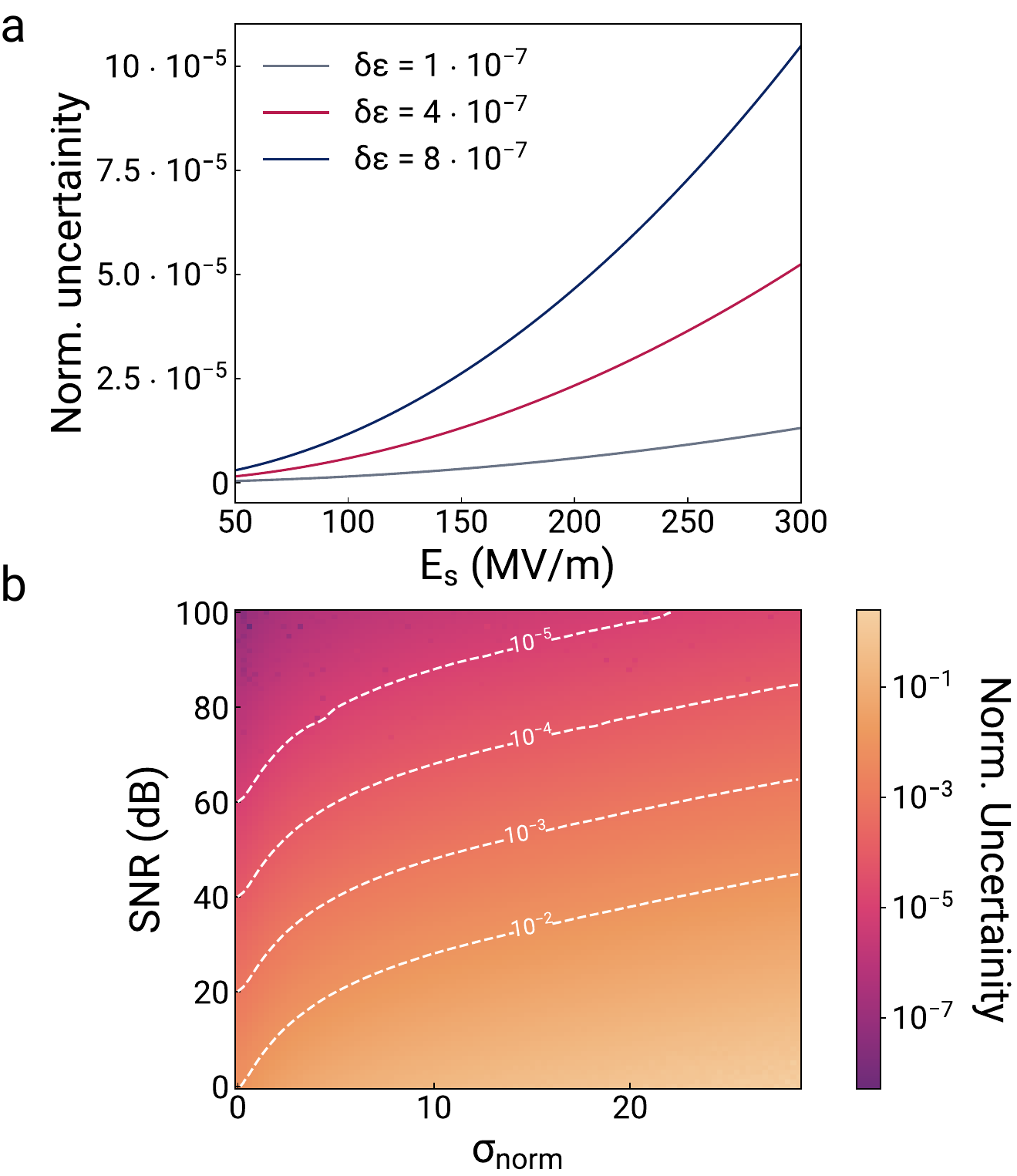}
    \caption{\textbf{Normalized uncertainty.} 
    \protect\hypertarget{fig:app_noiseL}{}
    \textbf{a} The uncertainty normalized to the homogeneous linewidth for three different values of the relative electric field sensitivity $\delta\epsilon = 1,4,7 \cdot 10^{-7}$. 
    \textbf{b} The uncertainty extracted from a fit as a function of the SNR and the Gaussian component of the Voigt profile, both normalized to the homogeneous linewidth.}
    \label{fig:app_noise}
\end{figure}

The relative electric field sensitivity shown in Fig.~\ref{fig:principle}\hyperlink{fig:principleL}{c} from the main text depends on how well the center frequency of a spectral peak can be determined. The determination of the peak position is affected by uncertainties induced by noise other than the stochastic shifts of the C-transition. Sources for such noise can be dark counts of the detector or undesired background fluorescence. In this section we estimate signal to noise ratios (SNR) that are required to enable the relative sensitivities shown in Fig.~\ref{fig:principle}\hyperlink{fig:principleL}{c} (main text). We first assume that the $\alpha$ dominates the response of the sensor's interaction with an electric field, so that the relative Stark shift (Eq.~\eqref{eq:nonlin_stark}, main text) produced by two distinct resolvable electric fields $E_1$ and $E_2$ becomes 
\begin{align}
    |\delta \omega| &\approx \frac{\alpha}{2}|E_1^2 - E_2^2| \\
     & = \alpha \delta\epsilon E_1^2~,
\end{align}
where we used the definition of the relative electric field sensitivity $\delta\epsilon = |E_1 - E_2|/E_1$ and assumed that $E_1 + E_2 \approx 2 E_1$. We define the normalized uncertainty as $\varLambda = |\delta \omega| / \gamma_{\rm hom}$, where we chose the homogeneous linewidth of the SnV $\gamma_{\rm hom} = 35$ MHz as a reference.
$\varLambda$ is the smallest Stark shift difference that has to be resolved, so that a relative electric field sensitivity of $\delta \epsilon$ can be reached. In Supplementary Fig.~\ref{fig:app_noise}\hyperlink{fig:app_noiseL}{a} we show the normalized uncertainty for three values of $\delta\epsilon$, which are representative values picked from Fig.~\ref{fig:principle}\hyperlink{fig:principleL}{c} from the main text. For the range of relevant field strengths, we find that $2.5 \cdot 10^{-5} < \varLambda \leq 10^{-4}$. To make sense of the normalized uncertainty we simulate the normalized uncertainty of the central peak position $\delta\omega_0/\gamma_{\rm hom}$ of a spectral fit with a centered Voigt profile $V(\omega - \omega_0, \gamma_{\rm hom}, \sigma)$ with $\omega_0 = 0$, a Lorentzian component $\gamma_{\rm hom}$ and Gaussian component $\sigma$ in the presence of noise. If $\delta \omega_0/ \gamma_{\rm hom}$ produced by the fit does not exceed the threshold demanded by $\varLambda$ we assume the corresponding relative electric field sensitivity to be achievable. In Supplementary Fig.~\ref{fig:app_noise}\hyperlink{fig:app_noiseL}{b} we show the result of the simulations. We normalize the Gaussian component of the Voigt profile according to $\sigma_{\rm norm} = \sigma/\gamma_{\rm hom}$. We calculate ${\rm SNR} = 10 \log_{10}(A^2 / \delta_{\rm noise}^2)$, where the amplitude of the Voigt profile $A = 1$ and $\delta_{\rm noise}^2$ is the amplitude of the white noise: $S(\omega) = V(\omega, \gamma_{\rm hom}, \sigma) + \delta_{\rm noise}$. Supplementary Fig.~\ref{fig:app_noise}\hyperlink{fig:app_noiseL}{b} shows the required SNR as a function of $\sigma_{\rm norm}$. Even though the requirements are challenging, they are not a fundamental limitation of our proposed sensor. For the multimodal spectrum in Fig.~\ref{fig:sensor} from the main text, the normalized uncertainties are between $10^{-2}\leq \varLambda < 7\cdot 10^{-1}$. The poor $\varLambda$ in our experiment is mostly due to experimental imperfections, and not a fundamental constraint of the sensor principle. 

Even though the $\varLambda$ in our implementation does not reach the simulated requirement to produce the simulated limit of the relative sensitivity of our proposed electrometer, they are sufficient for the claimed Angstrom resolution of the sensor. 
We can perform a similar estimation of the normalized sensitivity as a function of the relative resolution $\delta \epsilon_r = (r_1 - r_2) / r_1$ where we find making similar assumptions ($r_1 + r_2 \approx 2 r_1$) as in the paragraph above so that 
\begin{equation}
    \varLambda =  2 \delta \epsilon_r a^2  \frac{\alpha}{\gamma} \left( \frac{1}{r_{\rm bias}^2 r_1^2} + \frac{1}{r_1^4} \right)~,
\end{equation}
where $a = 1/ 4 \pi \epsilon_0 \epsilon_r$.
We find that $ 18 < \varLambda < 166$, for $\delta \epsilon_r = 1$, $r_{\rm bias} = 10~\AA$  and $r_1 \in (10, 30)$ $\AA$, which far exceeds the relative fit uncertainties provided in the paragraph above.

\subsection*{Resolution}
\label{sec:ap_res}

For determining the spatial resolution $\Delta_{\rm r} = |\vec{r} - \vec{r}'|$, where $\vec{r}$ and $\vec{r}'$ are two distinct positions of point-like charges, we perform the same calculation as for the relative electric field sensitivity but additionally assume that the charges generate electric fields
\begin{equation}
    \vec{E}(q, \vec{r}) =  \frac{q_i}{4 \pi \epsilon_0 \epsilon_r} \frac{\vec{r}}{r^{3}}~,
    \label{eq:app_bulk_electric_field}
\end{equation}
where $\epsilon_0$ is the vacuum permittivity and $\epsilon_r = 5.5$ is the relative permittivity of diamond. Using the bulk expression and neglecting surface contributions is justified because of the pillar dimensions $r>40$ nm (Fig.~\ref{fig:broadening}\hyperlink{fig:broadeningL}{b,c} from the main text), if the SnV is located on the pillar's symmetry axis. 

In Fig.~\ref{fig:principle}\hyperlink{fig:principleL}{d} from the main text, we present the sensor's resolution in the presence of a constant static field $\vec{E}_{r_0}$ produced by a negatively charged trap situated at the fixed location $\vec{r}_0 = (0,0, 0.8)$ nm. 

As described in the previous section, Fig.~\ref{fig:principle}\hyperlink{fig:principleL}{d} (main text) was generated in a two step procedure: First the expected spectral profiles were computed for a given $\rho_{\rm trap}$ and $\vec{E} = \vec{E}(-1, \vec{r}_0) + \vec{E}(-1. \vec{r}_1) + \delta \vec{E}_{\rm s}$. Then the resolution was calculated by employing the Rayleigh criterion.

We place $\vec{r}_1 = (0,0,d)$ in line with $\vec{r}_0$. 
The averaged profiles are then used to determine the smallest resolvable distance $\Delta_{\rm r} = |\vec{r} - \vec{r}'|$ from the spectral profiles corresponding to the fields $\vec{E}_{\rm bias} = \vec{E}(-1, \vec{r}_0) + \vec{E}(-1, \vec{r}_1) + \delta \vec{E}_{\rm s}$ and $\vec{E}_{\rm bias} = \vec{E}(-1, \vec{r}_0) + \vec{E}(-1, \vec{r}_1') + \delta \vec{E}_{\rm s}$.

\subsection*{Most likely spatial trap configuration}
\label{sec:ap_config}

\begin{figure}
    \centering
    \includegraphics[width=\linewidth]{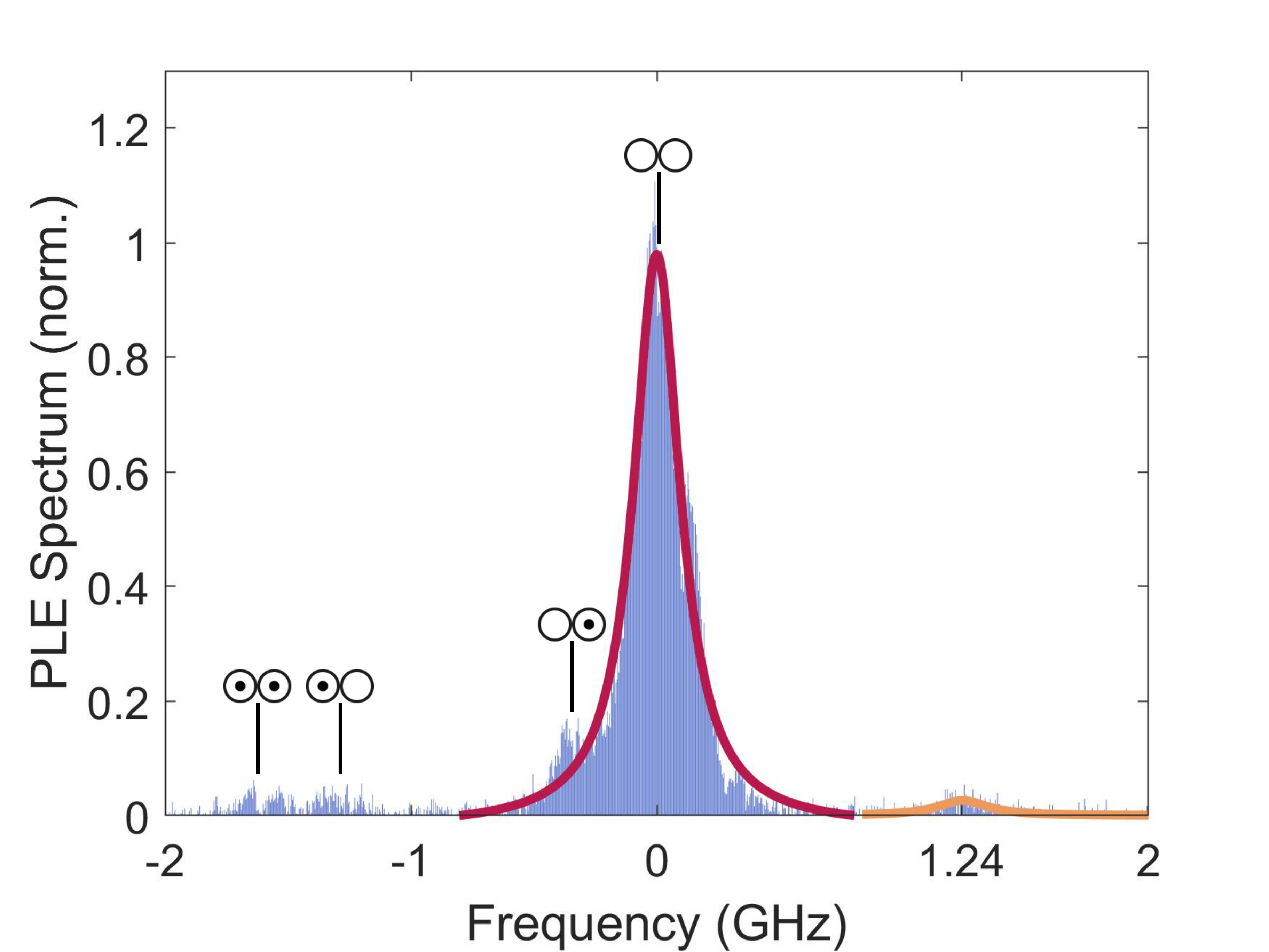}
    \caption{\label{fig:biasAnalysis}
        \textbf{Experimental estimation of the bias field on our sensor.} Background subtracted integrated spectrum of the linescans between (0-200) from the Fig. \ref{fig:sensor}\protect\hyperlink{fig:sensorL}{b} from the main text. The fit centered at 1.24 GHz indicates the existence of an extra charge trap which is ionized most of the time.}
\end{figure}

The integrated multimodal spectrum in Fig.~\ref{fig:sensor}\hyperlink{fig:sensorL}{b} from the main text can arise from distinct spatial charge configurations, leading to identical results. Nonetheless, it is possible to narrow down potential proximity charge configurations.

The integrated spectrum of Fig.~\ref{fig:sensor}\hyperlink{fig:sensorL}{b} (main text) shows four peaks. The two simplest configurations producing such a spectrum are: A) three traps, where one trap is permanently charged and the other two can be in the states $[\medcirc\medcirc, \medcirc\odot, \odot\medcirc, \odot\odot]$ or B) four proximity traps with one trap being permanently charged and the other three in the charge states $[{\medcirc\medcirc\medcirc},{\medcirc\medcirc\odot}, {\medcirc\odot\medcirc},{\odot\medcirc\medcirc}]$. 
In both cases, a bias field / permanently charged trap is required to explain the inhomogeneous broadening of the rightmost peak. Many more trap configurations could in principle produce the same features, but we deem them less likely, given that they require more and more traps, where only a subset of all possible charge state combinations then contributes to the observed spectrum. Of the two scenarios, scenario A) requires the fewest additional assumptions.          

The strongest argument in favor of A) is based on the transition probability $p({\medcirc\medcirc \rightarrow \odot \odot }) = 3(1)\%$ (see Supplementary Fig.~\ref{fig:dynamics_s1}). 
If one assumes that traps independently ionize with a probability $P$, then the corresponding rates for B) $p({\medcirc\medcirc\medcirc}{ \rightarrow} {\odot\medcirc\medcirc}) \approx P$. However, it is one of the least likely processes. Scenario A) would require two ionization events which would be of order $P^2$, which is much closer to the observation. The same argument can be constructed for $p({\medcirc\odot \rightarrow \odot \odot}) = 33(6)\%$. For B) the corresponding event would be $p({\medcirc\odot\medcirc \rightarrow \odot\medcirc\medcirc}) \approx P^2$, which should be unlikely. However, the single ionization event $p({\medcirc\odot \rightarrow \odot \odot})$ is more likely and therefore more consistent with the two trap scenario. 

We estimate the bias field experienced by the sensor by positioning a constantly ionized charge trap such that the inhomogeneous broadening of the simulations matches the observed linewidths. From the simulations, we find a bias field inducing a spectral shift of 1.3(4) GHz. We compare this result with the integrated spectrum for lines between 0 and 200 from the spectrum from Fig.~\ref{fig:sensor}\hyperlink{fig:sensorL}{b} (main text) and find evidence of a small blue shifted peak with a spectral shift of 1.24(2) GHz (Supplementary Fig.~\ref{fig:biasAnalysis}) compared to the $\medcirc\medcirc$ peak. This experimentally substantiates that there is indeed a third charge trap which is ionized most of the time. Analyzing the inhomogenous linewidth with Monte Carlo simulations and experimental data independently affirm each other on the estimated magnitude of the bias field. This consistency further demonstrates that our simulations can replicate the charge environment and is able to detect traps that do not dynamically change their charge on the time scales of the remote traps.  

\subsubsection*{Generalized algorithm for trap configuration determination}

Here, we provide a generalized algorithm for determining a charge trap \textit{configuration} using the jumping probabilities between different spectral peaks, i.e. trap \textit{states}.
 
\paragraph*{Definitions:}

Trap state: A possible distinguishable combination of ionized traps.

Trap configuration: The total set of trap states the sensor detects.

Correlation matrix: The probability matrix of trap states changing from one to another, i.e., a spectral jump occurring from one peak to another. It has a total of M*(M-1) entries where we discard the diagonal (no state change) elements. We order the peaks (trap states) from the least red-shifted (no ionization) to most red-shifted.

\paragraph*{Algorithm:}

i.	Data acquisition:
Automatically perform a series of photoluminescence excitation (PLE) scans of the probe.

ii.	Spectral analysis:
Utilize a peak-finding algorithm to identify peaks within the integrated spectra for which each peak corresponds to a trap state. It is possible to return to step i, and apply RORO sequences for enhancing the temporal resolution.

iii.	Correlation matrix extraction:
Extract jumping probabilities from the individual and time-resolved PLE scans. The jumping probabilities are only associated to the peaks that were identified in step ii. 
This is explained in detail for the temporal analysis of the S1 data.

iv.	Determining the trap configuration:

The number of peaks in the spectrum = M

The number of detectable traps that can host a charge in the vicinity of the sensor = N

We assume single-ionization events exist and are more likely than multi-ionization events.
We ensure the scan speed is high enough such that a maximum of two peaks are observed in between scans. Based on this, we assume that a single ionization process could have happened in this time period. 
We attribute anything below a threshold (e.g. 3\%/M) as a multi-ionization event occurring in a  single scan and neglect it (set it to 0).
This reduces the complexity of detectable trap configurations.
From the M, we determine the minimal number of N with M$>2^{\rm N-1}$ and the maximal number, N=M.
To determine N, we calculate the order (number) of correlated (nonzero probability) steps of the jumping events. We analyze the jumping probabilities from the non-ionized case '0' (least red-shifted peak) and step by step reach a particular peak '$k$': p($0\rightarrow 1\rightarrow…{\rm k-3\:  steps}… \rightarrow k$). 

Example application for M peaks:

•	We expect the simplest configuration of traps to consist of N=M-1 traps with N trap states, i.e., no multi-ionized traps at the same time visible in the spectrum, and only single ionization events are likely. This corresponds to 2*(M-1) nonzero values in the correlation matrix that are all correlated to the least red-shifted peak. 

•	Under similar assumptions, we can construct a spectrum from N=M-2 traps where M-2 peaks are the results of ionizing M-2 single traps. One peak corresponds to no ionization. The remaining peak has to emerge from one trap state with two ionized traps. This can already be distinguished from the N=M-1 scenario, by identifying 4 entries in the correlation matrix that are bigger than 0 that are not correlated with the least red-shifted peak (first column, first row).  These nonzero elements uniquely correlate each peak to a trap state.

•	With the N=M-3 traps case, we have M-3 peaks correlated with the single ionized traps states. One peak corresponds to non-ionized trap state. The remaining two peaks can be associated to two configurations with either two distinguishable simultaneously ionized traps states or two trap states which include two and three simultaneously ionized trap states.
These two possible trap configurations from each other are further distinguished by the higher order correlations revealing whether or not subsequent ionization processes have occurred. By considering these higher-order correlations, it is in principle possible to further identify which set of trap states is responsible for the observed peaks.  

•	This method can be generalized to further reduce the number of participating traps, by identifying the highest-order correlation. 
By comparing the observed correlations with the correlation matrices we can identify a trap configuration with the experimentally observed data.

v.	Determining remote charge trap distribution:
The remote trap distribution (density, spatial distribution) can be predefined or left variable and fine-tuned through a series of preliminary Monte Carlo scans to model the integrated spectrum, without requiring manual input. In this paper, the spatial distribution of the remote trap density was estimated with a physical model of the implantation damage distribution. Other physical models can be also supplemented to the simulations. Without a physical model,  a predefined charge distribution (such as cylindrically symmetric) can also be used.

vi.	Estimating the proximity charge trap positions:
Based on the selected trap configuration there are two possible ways of determining the proximity charge trap positions. First comes the initialization of the trap positions at random positions with the distances corresponding to the spectral shifts of the assigned peaks. These distances are determined by a set of equations that govern the DC Stark shifts. The remaining free parameters (relative angles and distances) are further refined by minimizing a Chi-squared test as explained in detail in the corresponding sections. Based on this test, the most probable trap configuration and positions can be returned in a general and automated fashion.

\subsection*{Annealing}
\label{sec:ap_annealing}

The creation of V$_2$ is understood to be a consequence of implantation damage \cite{Slepetz2014PhysChem, Fu2019} and the annealing procedure:
Implantation damage occurs during the collision cascade in the diamond lattice that decelerates the implanted ion. Collisions with an energy above the displacement-threshold ($\approx 37.5 - 47.6$ eV \cite{koike1992} much smaller than typical implantation energies) dislodge carbon atoms and produce Frenkel pairs: a pair of V$_1$ and a dislocated carbon atom located at an interstitial lattice site. 
After the implantation, an annealing procedure is performed to create the color center through vacancy diffusion and to heal the lattice damage. At temperatures above 600 K interstitial carbon becomes mobile \cite{laidlaw2020} and at 800 K the V$_1$ \cite{Slepetz2014PhysChem} have a high degree of mobility. Consequently, during annealing, interstitial carbon can either recombine with the V$_1$ \cite{kiflawi2007, newton2002} or diffuse away from the damage site and eventually leave the sample through the boundaries. The V$_1$ that have not recombined with an interstitial carbon can form immobile V$_2$ \cite{Slepetz2014PhysChem}, vacancy clusters, or create a color center together with the implanted ion.  

Our model assumes one mobile species (V$_1$) and considers the formation of V$_2$ without multi-vacancy complexes. Since we do not consider multiple species, we forego assigning different hopping frequencies as in \cite{favaro_de_oliveira_tailoring_2017}. 
The initial number $N$ and the 3D distribution of the V$_1$ after implantation is estimated using a SRIM simulation \cite{Ziegler2010NIM}. Assuming a percentage of V$_1$ not being consumed by interstitial carbon, which we call yield quantified in \% of Frenkel pairs that have not recombined (V$_1$ yield in Fig.~\ref{fig:annealing} from the main text) we find a range of concentrations of V$_2$ shown in Fig.~\ref{fig:annealing} (main text) for the three atomic G4V species Si, Ge and Sn and varying implantation energies. 
We estimate the distribution of V$_2$ in the sample by using a kinetic Monte Carlo simulation. 
In each time step of the kinetic simulation, the V$_1$ can take a random step along any of the neighboring lattice sites. If two V$_1$ are adjacent to each other, they form a static V$_2$ which no longer diffuses.  
The initial distribution of V$_1$ is estimated using SRIM \cite{Ziegler2010NIM}. For each implantation energy we use 50000 implantation events for a given atomic species and implantation energy to find the probability distribution $p(z)$ of V$_1$ as a function of the depth $z$ measured relative to the diamond surface $(001)$. We then use the $p(z)$ to generate a realization of a spatial distribution of V$_1$ after a single implantation event. The V$_1$ are distributed on the diamond lattice according to the $p(z)$ along a narrow damage channel with a rectangular cross section of $2a \times 2a$ as described in the main text. The loss of V$_1$, that do not contribute to the formation of V$_2$ due to recombination with interstitial carbon atoms is by reducing the initial amount of V$_1$ as determined by the SRIM simulation by a fixed percentage. 

\subsection*{Bulk charges \label{sec:ap_bulk}}

Based on our model, we provide an overview of the charge trap densities and the resulting inhomogeneous broadening with certain thresholds enabling 90\% interference visibility and  $> 87\%$ entanglement fidelity according to\citen{Kambs2018NJP}. First, we use our Monte Carlo simulation to determine the distributions of linewidths for a given trap density $\rho$. We assume a carbon density of $\rho_{\rm C} = 8 / a^3$ in bulk
and an isotropic distribution of traps in the environment of the SnV at a given density $\rho$. For each $\rho$ we consider 500 spatial trap configurations that produce single peaked spectra for $\rho \in (1, 100)$ ppm. We use Supplementary Eq.~\eqref{eq:app_bulk_electric_field} in the sum Eq.~\eqref{eq:mc_spectrum} (main text, methods) to calculate the spectra.

\subsection*{Surface charges \label{sec:ap_surface}}

The surface density for both the semi-infinite half space and the cylindrical geometry given in ppm is calculated with respect to a carbon density of $\rho_c = 2 / a ^ 2$ [(001) plane]. For the semi infinite half space the traps are randomly positioned on a square with a $100$ nm edge length. The cylindrical surface has a height of $100$ nm. 
The simulation of the inhomogeneous linewidth was performed in both cases with the previously explained Monte Carlo method using 5000 different charge configurations for a single spatial configuration of traps. 
We also use the electrostatic fields of a point charge on a surface taking the respective boundary conditions into account.
The electric field of a charge located on the surface of the semi infinite half space is 
\begin{equation}
    \vec{E}(q, \vec{r}_q) = \frac{q}{2 \pi \epsilon_0 (\epsilon_r + 1)} \frac{\vec{r}_q}{r_q^3} 
\end{equation}
For the cylindrical surface, we assume a diamond cylinder with radius $R$ extending to $z = \pm \infty$. 
We used the expression in\citen{Cui_2006} [suppl. in\citen{OrphalKobin2023PRX}] for the electric field of a charge on the cylindrical diamond surface.
Here, we do not consider band bending, which can be advantageous for eliminating surface noise through screening. 
We also neglect free carrier screening, because we do not see the stark reduction of sensitivity to charge-noise that would be expected even for moderate screening lengths of tens of nanometers.

\section*{Experimental details}

\begin{figure}[H]
    \centering
    \includegraphics[width=\linewidth]{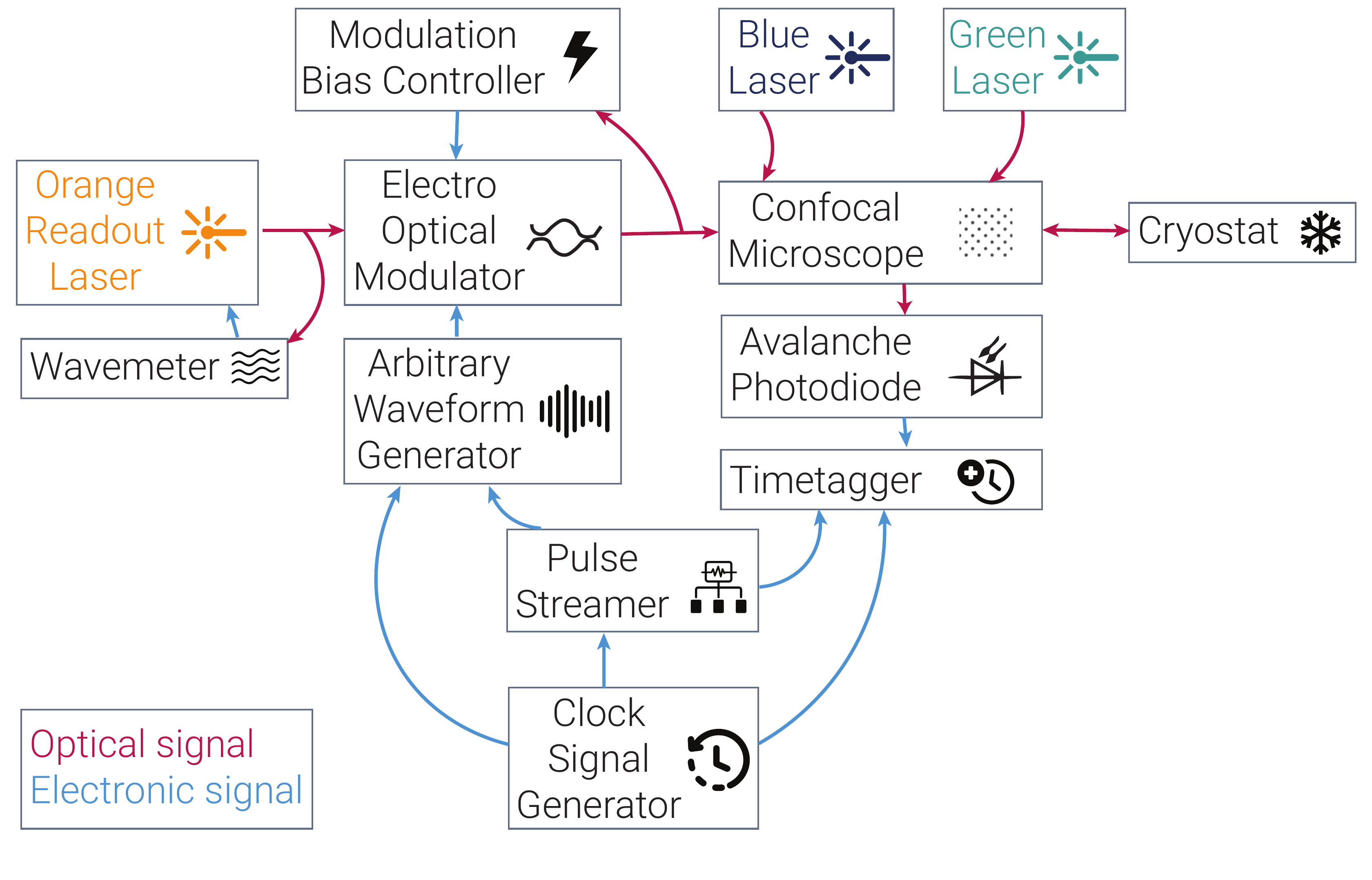}
    \caption{\label{fig:eomSETUP}
\textbf{A simplified illustration of the devices and connections that are used to collect signals from SnV electrometers.} See Methods section in the main text for detailed explanations.}
\end{figure}

\subsection*{RORO data analysis steps}
    
    \begin{enumerate}
        \item Using the periodic calibration signal and a previously measured constant delay between the calibration signal and arrival time of the photons, the starting timestamp with 0 value is determined. The rest of the timestamps are corrected accordingly. Timestamps (ps resolution) are rounded to ns.
        \item The 1D timestamp data is organized in a 2D matrix with: The X-axis records 1 to 300 ns (RORO sequence length); the Y-axis  indexes the individual RORO sequences; The matrix elements represent: 1 for registered counts, 0 otherwise. 
        \item We separate the X-axis of the 2D matrix, into three `frequency windows' according to the elapsed time within a  RORO sequence. The first 60 ns is off-resonant $\omega_0$, followed by 120 ns $\omega_1$ (including the $\omega_{\rm 1s}$ application period), and a final 120 ns $\omega_2$ (including the $\omega_{\rm 2s}$ application period), following our experimental configuration.
        \item Counts in the first 23 ns of each of the five windows are set to 0. This step is required due to the $\sim$10 ns spontaneous emission lifetime of the SnV leaking into the next frequency window.
        \item The matrix values are then binned into three data points on the X-axis corresponding to each frequency window.
        \item For each registered count, we implement a noise filtering step. The time between consecutive events is compared with previously characterized background count rates. If the inverse time between events is smaller than the background count rate, the earlier event is rejected and deleted from the matrix. Otherwise, the count is accepted. 
        \item Consecutive accepted events in the same frequency bin count towards the bright time where the emitter is considered resonant with the corresponding frequency. 
        \item If there is a period with counts comparable to the background at any frequency window, i.e. a completely dark period when the SnV itself is doubly ionized, is identified, the time period is excluded from the spectral jump analysis.
        \item Bright time durations after which the SnV switches to the other resonance for each frequency window are histogrammed. The fit results show negligible changes for bin widths smaller than a characteristic value (found by scanning a range of bin widths). We then choose a bin width smaller than the characteristic value producing the smallest fit uncertainty.
        \item The histogram is firstly fit to bi-exponential function. If it fails, the histogram is fit to an exponential function with a single rate.
    \end{enumerate}

\paragraph*{Selective analysis of bright and dark window rates:} In the main text, we compare the jumping rates $\varGamma_{\omega_1\rightarrow\omega_2}$ from a selected resonance during the periods when the registered counts are predominantly received from the investigated resonance $\omega_1$, or the other resonance $\omega_2$. To collect enough statistics for data analysis, we conduct the experiment shown in Fig 3(d) for an extended 90 seconds. Then, we manually select periods when the $\omega_1$ is mainly bright or mainly dark. Next, we combine the bright times in a histogram for both cases. From these histograms, $\varGamma_{\omega_1\rightarrow\omega_2}$ is extracted with a single exponential decay fit.

\paragraph*{Note on the extracted fast rates and photon count rate limitations:} 
We attribute the inability to determine the fast rates of several kHz on all of our measurements due to the overall photon count rates of the same order of magnitude. We also note the possibility of a systematic error on the extracted fast rates, as some jumps might not have been registered. This, however, is not a fundamental temporal resolution limitation of our method and is purely related to the photon count rates, as we are already modulating the readout laser at rates faster than a MHz and up to $\sim$ GHz is electronically possible. In Supplementary Table \ref{tab:countRates}, we provide  demonstrated Purcell enhancement factors and optimized collection rated from the literature, and estimate achievable temporal resolutions while applying RORO sequences.

\begin{table*} []
    \centering
    \caption{\label{tab:countRates} \textbf{Measured count rates in the literature, and estimation of achievable RORO acquisition rates.}
            \textbf{a} Time constants and rates from our experimental configuration.
            \textbf{b} Achievable count rates from literature using nanostructures optimized for enhanced emission and efficient collection. A factor of 2 of the lifetime leaking into the other probing window (see RORO data analysis details), and for\cite{Kuruma2021} \&\cite{Rugar2021PRX} a waveguide to fiber coupling efficiency of 90\% \cite{Burek2017}, detector efficiency of 84.7\% \cite{Wang2019}, quantum efficiency of 80\% \cite{Iwasaki2017PRL} are assumed for calculating the anticipated acquisition times.}   
    \begin{tabular}{{p{0.2\textwidth}p{0.18\textwidth}p{0.44\textwidth}}}
    \bf{(a)} & & \\
        \toprule 
            \bf{Specification} & \bf{Time constant} & \bf{Rate} \\ \hline
            SnV lifetime in bulk & 5 ns & 200 MHz  \\
            SnV lifetime in nanopillar & 10 ns & 100 MHz   \\ 
            RORO acquisition time & 60 ns & 16.7 MHz   \\ \bottomrule
        \end{tabular}
            \begin{tabular}{{p{0.1\textwidth}p{0.07\textwidth}p{0.12\textwidth}p{0.12\textwidth}p{0.08\textwidth}p{0.09\textwidth}p{0.08\textwidth}p{0.12\textwidth}}}
            \\
                \bf{(b)} & & & & & & \\
            \toprule
            \bf{Reference} & \bf{Purcell factor} & \bf{Fluorescence enhancement} & \bf{Excited \newline state lifetime} & \bf{Emission rate} & \bf{Cavity coupling} & \bf{Setup efficiency} & \bf{Anticipated acquisition time} 
            \\ \hline
            \citen{Wan2018} (NV) & - & - & 12.67 ns  & 79 MHz & - & 41\% & 56.34 ns  \\
            \citen{Kuruma2021} (SnV)  & 16 & 12 & 0.38 ns & 2.63 GHz & 95\% & - & 1.42 ns   \\ 
            \citen{Rugar2021PRX} (SnV) & 25 & 40 & 0.685 ns & 1.46 GHz & 90.1\% & - & 2.62 ns  \\ 
            \bottomrule
        \end{tabular}
        
\end{table*}

\section*{Control experiments}

\subsection*{Verification of emission from a single transition}
\label{sec:ap_control}

       \begin{figure*}  
        \centering
        \includegraphics[width=\linewidth]{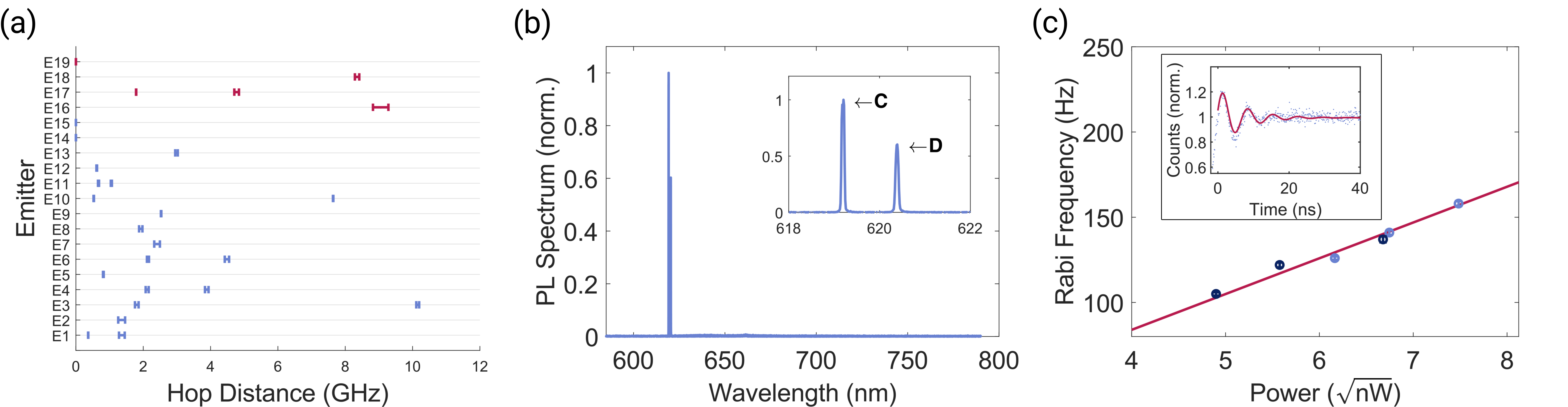}
        \caption{\label{fig:}
        \textbf{Control experiments to support that the analyzed emission from the SnV belongs to the same transition.}
        \protect\hypertarget{fig:controlL}{}
            \textbf{a} Spectral hop ranges of characterized emitters.
            Emitters without hopping were stable during linewidth scans that happened at different time scales between minutes to an hour.
            Red data points are from a different sample E002 with a higher implantation fluence (2.5 $\times 10^{11}$ cm$^{-2}$). The error bars show the 95\% confidence intervals of the central frequency distances extracted from the fits. 
            \textbf{b} Photolimunescence spectrum taken from E1 (emitter from the main text) showing $\sim$850 GHz splitting between C and D transitions. Inset: Zoom-in
            \textbf{c} Rabi frequencies of the emitter E2. Values are extracted from a damped oscillation function at different resonant (between levels $\ket{1}$ and $\ket{3}$) excitation powers.
            Dark blue and light blue data points are taken before and after a spectral jump, and therefore at different frequencies.
            Slope of the combined data (red) is extracted as 21.0(5) Hz/$\sqrt{\rm nW}$ where the lower frequency (light blue) resonance had a slope of 20.9(9) Hz/$\sqrt{\rm nW}$ and higher frequency (dark blue) data is a 21.2(1.8) Hz/$\sqrt{\rm nW}$. The uncertainties and error bars represent 95\% confidence intervals extracted from the fits. 
            INSET: Example Rabi oscillations observed with 45.5 nW power at the higher frequency resonance. 
            }
            \label{fig:control}
    \end{figure*}

An important test for our sensor is the verification that the multimodal spectral fingerprint is originating from the same transition. Here, we provide four characterization measurements, under zero magnetic field to exclude Zeeman splitting, that indicate that the signal originates from a single source and transition.
    
\subsubsection*{Distribution of jump distances} \label{ap:distribution}
Among the 19 characterized emitters, hopping distances varying from a few hundred MHz to a few GHz were found. 
On the investigated samples, either one or two distinct jump processes or their combinations are found, fully consistent with the number of estimated lattice defects.
The distribution of these distances is presented in Supplementary Fig.~\ref{fig:control}\hyperlink{fig:controlL}{a}.
Therefore, the existence of unknown levels with quasi-forbidden transition rules seems unlikely as the hop distances appear to be random for each emitter.
    
\subsubsection*{PL Spectrum}
The photoluminescence emission spectrum (Supplementary Fig.~\ref{fig:control}\hyperlink{fig:controlL}{b} measured under 520 nm excitation light at 4 K shows a typical SnV spectrum with discernible spectrometer-limited peaks attributed to C (between levels $\ket{1}$-$\ket{3}$, Fig.~\ref{fig:sensor}\hyperlink{fig:sensorL}{b} from main text) and D ($\ket{2}$-$\ket{3}$) transitions. Since they are $\sim850$ GHz apart, we can safely claim that multiple peaks from the PLE scan do not correspond to these transitions.
    
\subsubsection*{Autocorrelation measurement}
Autocorrelation measurements presented in control experiments for the single-photon ionization charge dynamics model are taken from the emitter investigated in the main text. The likelihood of multiple emitters contributing to the spectrum is made highly improbable by an autocorrelation measurement with $g^{(2)}(0) = 0.12(9) < 0.5$ close to the theoretical expected value of $g^{(2)}(0) = 0$.
    
\subsubsection*{Rabi frequencies of different resonances:} \label{ap:Rabi}

Here, we demonstrate Rabi oscillations between levels $\ket{1}$ and $\ket{3}$ (C transition) of an SnV, on emitter E2 at two different resonance frequencies before and after a spectral jump event.
In Supplementary Fig.~\ref{fig:control}\hyperlink{fig:controlL}{c}, Rabi frequencies at different powers from both resonances and an example measurement are provided.
Oscillations are obtained via resonant excitation after a green stabilization pulse.
Data after the resonant laser rise-time is fit to a damped oscillation function.
After repeating the measurement at different powers, the lower frequency resonance had a slope of 20.9(9) Hz/$\sqrt{\rm nW}$, higher frequency one had a 21.2(1.8) Hz/$\sqrt{\rm nW}$ and combined data set had 21.0(5) Hz/$\sqrt{\rm nW}$ on a linear frequency-$\sqrt{\rm Power}$ line.
The fact that slopes for three data sets remained within the fitting error range strongly suggests the dipole moment did not change between the spectral jumps and the same transition is being addressed between the two measurements.

\subsection*{Demonstration of ionization processes via single-photon processes using autocorrelation measurements} \label{sec:ap_bunching}

    \begin{figure*}
        \centering       \includegraphics[width=0.90\linewidth]{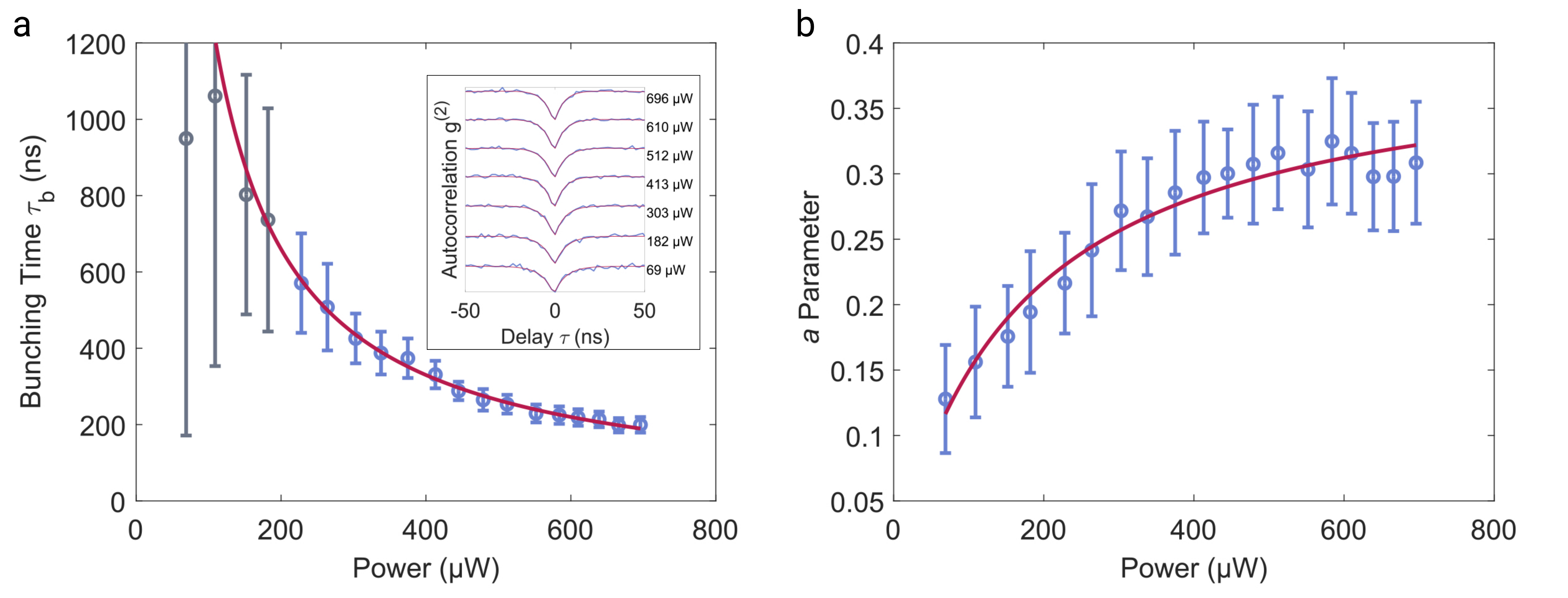}
      \caption{\label{fig:bunching}
        \textbf{Extracted parameters from the autocorrelation measurements for the emitter E1.} The error bars represent 95\% confidence intervals derived from the fits (functions provided in the text). These measurements indicate that shelving and deshelving processes can be modeled as single-photon events. 
      \protect\hypertarget{fig:bunchingL}{}
            \textbf{a} Bunching time at different powers. Grey points are excluded from the fit as they are not expected to behave according to the approximated model at low powers and they have large errors. Solid line: Fit to a $1/(c P)$ function. Inset: Selected example measurements. 
            \textbf{b} $a$ parameter at different powers. Solid line: Fit to a saturation curve.}
    \end{figure*}

The photophysical picture behind the charge dynamics investigated in the main text has been previously explained in\citen{Goerlitz2022NPJ}. One of the events that can occur during laser irradiation is the group-IV vacancy (G4V) emitters transitioning to a dark state.
This manifests as shoulder-like bunching features around the antibunching dip in autocorrelation measurements.
Here, applying the single-photon process assumption from the proposed model, we find a linear power dependence for both hole creation/capture and electron promotion processes.
These experiments show that the charge transfer picture presented in the main text is consistent with photon statistics measurements.
We base our analysis on the derivation of the autocorrelation function and the rate equations provided in the reference\citen{Kitson1998PRA}.

We assume a three level system where level 1 is the ground state, 2 is the excited state, and 3 is a nonradiative shelving state, which is identified as G4V$^{-2}$.
Such a system's $g^{(2)}$ with nonzero background obeys the following equation:

 \begin{equation}
     g^{(2)}=1+p^2[1-(1+a)\exp(-\frac{\tau}{\tau_{\rm a}})+a\exp(-\frac{\tau}{\tau_{\rm b}})]
 \end{equation}
 
\noindent where $p$ determines the background contribution, $\tau_{\rm a}$ is the antibunching time related to the dip at 0 delay, $\tau_{\rm b}$ is the bunching time which determines the shoulders surrounding the antibunching dip, and the $a$ parameter is related to the transition rates.
To test the model, $g^{(2)}$ measurements of an SnV center at different powers ($P$) are fit to this equation and the parameters are extracted.
The following power relations are then assumed to predict the transition rates ($k_{\rm Initial \,Final}$):

\begin{itemize}
    \item $k_{12}$ (incoherent excitation) is assumed to have a linear dependence on power  `$\delta P$', as it is a single-photon process promoting an electron from the ground state to quasi-continuous phononic bands of the excited state. 
    \item $k_{21}$ (spontaneous emission) is modelled with a constant rate `$\varGamma$'.
    \item $k_{23}$ (shelving) is assumed to have a linear power dependence `$\alpha P$', because this process is known to be a single-photon process promoting electrons from the valance band to an excited G4V \cite{Goerlitz2022NPJ}.
    \item $k_{31}$ (deshelving) is also modelled to be linearly proportional to power `$\beta P$': Here, hole donation is assumed to be a single-photon process induced by promoting an electron from valance band to a V$_{\rm n}$.
    Previously, this rate has been modeled with a saturation curve \cite{Neu2011NJP}, which may be attributed to the limited amount of contributing V$_{\rm n}$.
    However for this case, our Monte Carlo simulations predict too high V$_n$ density for saturation to occur. Therefore, a linear model can capture the data well.
    We also want to note that a saturation curve imitates a linear relationship at low powers, and both models can work at different regimes consistently.
\end{itemize}

The bunching time $\tau_{\rm b}$ relates to the transition rates through the equation:
 \begin{equation}
   \tau_{\rm b}=\frac{1}{k_{31}+k_{23}\frac{k_{12}}{k_{12}+k_{21}}}
 \end{equation}
 
 If $k_{12}$ is assumed to be much larger than $k_{21}$ -- as expected at higher powers --, then $k_{12}/(k_{12}+k_{21})$ approaches 1. Then:
\begin{equation}  \label{eq:bunching}
   \tau_{\rm b}=\frac{1}{k_{31}+k_{23}}=\frac{1}{(\alpha+\beta)P}
 \end{equation}
 
 This shows that $\tau_{\rm b}$ is effectively determined by the total rate of $k_{31}$ and $k_{23}$ at higher powers. When a 1/x model is fit to the extracted $\tau_{\rm b}$s at the Supplementary Fig.~\ref{fig:bunching}\hyperlink{fig:bunchingL}{a}, it can be seen that the model captures the data well and $\alpha+\beta$ is extracted as 7.5(1) kHz/µW.
 The total charge cycle rate of 1 MHz at $\sim150$ µW also seems reasonable as we expect the charge transfer process to be slower than spontaneous emission or excitation. 
 
 For estimating the rate coefficients separately, we can determine the $a$ parameter which is governed by:
 \begin{equation} \label{eq:ratio}
   a=\frac{k_{23}}{k_{31}}\frac{k_{12}}{k_{12}+k_{21}}=\frac{\alpha}{\beta}\frac{\delta P}{\delta P+\varGamma}~.
 \end{equation}
 
At high powers, $a$ parameter will asymptotically reach $\alpha / \beta$.
In Supplementary Fig.~\ref{fig:bunching}\hyperlink{fig:bunchingL}{b}, the extracted $a$ values from the measurements follow a saturation curve where the fit asymptotically approaches 0.40(3). Furthermore, using this relation in Eq.~(\ref{eq:bunching}), we can deduce the shelving and deshelving rates at each power with $\alpha=$ 2.2(2) kHz/µW and $\beta=$ 5.4(2) kHz/µW. Because the linear power dependence assumption is consistent with the observed data, we argue that single-photon processes are the main driver of the charge dynamics in the sample.

\subsection*{Temporal analysis of the S1 Data}

\begin{figure}[]
    \centering
    \includegraphics[width=\linewidth]{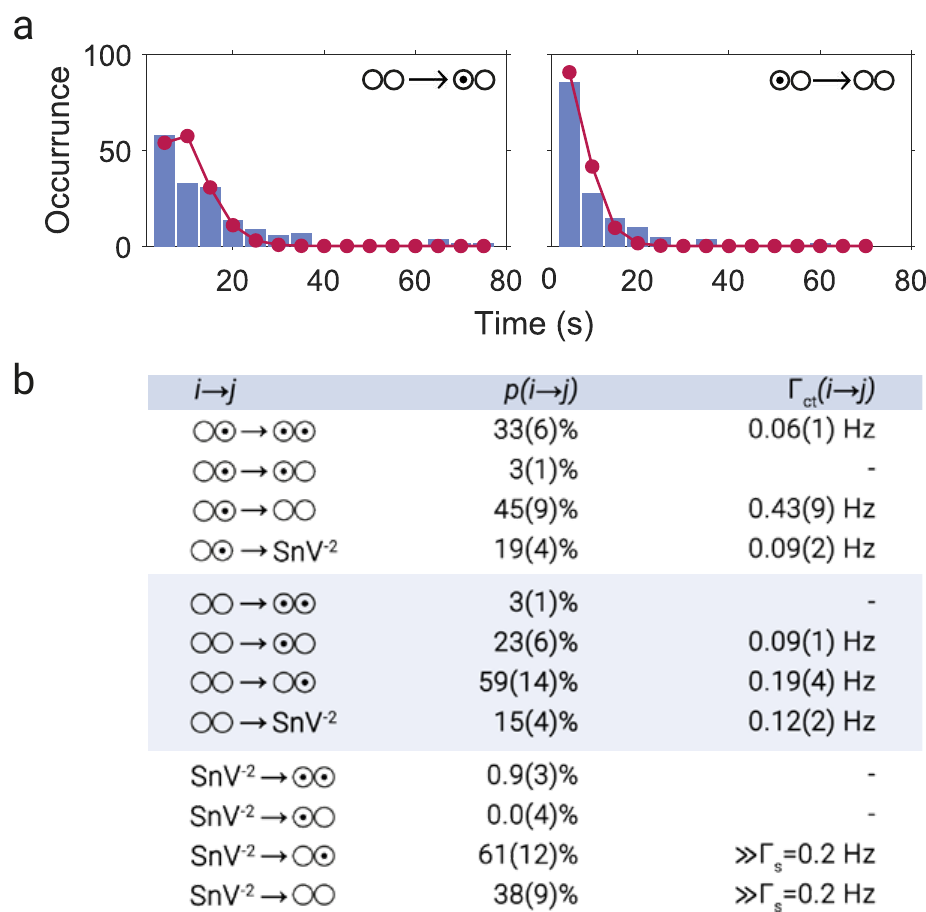}
    \caption{
    \textbf{Charge dynamics of the sensor S1.} 
   \protect\hypertarget{fig:dynamicsL}{}
    \textbf{a} Example histograms showing the duration resonance remains in a charge state until it switches to another. The data are fit to a Poisson distribution to estimate a mean time.
    \textbf{b} Extracted temporal values from the sensor data
    presented in Fig.~\ref{fig:sensor}\protect\hyperlink{fig:sensorL}{b} from the main text.
    $p(i{\rightarrow}j)$ and $\varGamma_{\rm ct}{(i{\rightarrow} j)}$ represent conditional spectral jump probabilities and rates respectively, where $i,j$ are the initial and final charge state configurations.
    Missing rates are due to insufficient data points. $\varGamma_{\rm s}$ is the scanning rate. The uncertainties are estimated from the overlap of individual peaks for $p(i{\rightarrow}j)$ and  95\% confidence intervals extracted from the fits for $\varGamma_{\rm ct}{(i{\rightarrow} j)}$.
    }
    \label{fig:dynamics_s1}
\end{figure}

For identifying the position of charge traps, we have used \textit{accumulated} spectral fingerprints that reflect the integrated spectrum for the entire set of charge states $\mathcal{U}_C =  \{\medcirc\medcirc, \medcirc\odot, \odot\medcirc, \odot\odot, {\rm SnV^{-2}}\}$, including the dark state $\rm SnV^{-2}$. 
Comparing \textit{individual} read-out events of our electrometer, i.e., single PLE linescans between different charge configurations within $\mathcal{U}_C$, gives access to time-resolved charge transfer dynamics.

To characterize the local charge environment and dynamics, we introduce the charge state transition probabilities $p(i{\rightarrow}j)$ and the conditional transfer rates $\varGamma_{\rm ct}{(i{\rightarrow} j)}$ between charge states $i$ and $j$ of the proximity traps, where $i, j \in \mathcal{U}_C$. We extract $p(i{\rightarrow}j)$ and $\varGamma_{\rm ct}{(i{\rightarrow} j)}$ by histogramming the charge transfer events and intervals between them (Supplementary Fig.~\ref{fig:dynamics_s1}\hyperlink{fig:dynamicsL}{a,b}). In addition, we define the lifetimes of each configuration as $\tau{(i)}$. 

We begin the analysis by quantifying the smallest $p(i{\rightarrow}j)$. The occurrence of a charge exchange event, considering our present linescan time of 5 seconds, given by $p({\medcirc\odot \rightarrow\odot\medcirc})$=0.03(1) indicates an improbable direct transfer between the two proximity traps.
Moreover, the occurrence of a two-trap charging processes $p({\medcirc\medcirc\rightarrow\odot\odot})$=0.03(1) is also unlikely, demonstrating that these events are not correlated.

We further explore the relationship between the reinitialization of the bright SnV charge state ${\rm SnV}^{-2}{\rightarrow}{\rm SnV}^{-1}$ \cite{Goerlitz2022NPJ,Gardill2021NanoLet} and the trap's charge states. The probabilities $p({{\rm SnV}^{-2}\rightarrow\medcirc\odot})$=0.61(12) and $p({{\rm SnV}^{-2}\rightarrow\medcirc\medcirc})$=0.38(9) are close to the corresponding peak intensities observed in the spectrum (0.63(5) and 0.31(3), respectively), indicating that the trap states are not correlated with the SnV's charge state.

Next, we compare the ionization and neutralization rates for a single trap,\\
$\sum_{X=\medcirc,\odot}\varGamma_{\rm ct}{({\medcirc X\rightarrow\odot X})}/2$=0.075(1)~Hz and $\sum_{X=\medcirc,\odot}\varGamma_{\rm ct}{({\odot X\rightarrow\medcirc X})}/2{\gg}\varGamma_{\rm s}{=}0.2$~Hz, respectively, with $\varGamma_{\rm s}$ the scanning rate. The more than 3-fold higher ionization rate possibly reflects the distinct underlying physical mechanism compared to neutralization.
It is notable that the ionization rates of the other trap abruptly change in time: for linescans 0-250 $\varGamma_{\rm ct}{({\medcirc\medcirc\rightarrow \medcirc\odot})}$=0.07(2) Hz and 250 - 500 $\varGamma_{\rm ct}{({\medcirc\medcirc\rightarrow \medcirc\odot})}{\gg}\varGamma_{\rm s}{=}0.2$ Hz. For the neutralization rate, the trend is inverted. We attribute this intriguing change of rates to discrete changes in the trap environment, however, a more detailed analysis is beyond the scope of this work \cite{Bluvstein2019PRL}. Furthermore, the distinct ionization rates, ${\varGamma_{\rm ct}{(\medcirc\medcirc\rightarrow\medcirc\odot})}$=0.09(1) Hz and $\varGamma_{\rm ct}{({\medcirc\medcirc\rightarrow \odot\medcirc})}$=0.19(4) Hz, observed under the same illumination laser field, indicate either large variations of the local electrostatic potentials in a $\sim$1~nm range modifying charge dynamics or the presence of multiple charge trap species. A future study could therefore help to differentiate among the various V$_{\rm n}$. 

Lastly, we determine and interpret overall charge state lifetimes $\tau{(i)}$ which provide a figure of merit for experiments that require spectral stability. We find $\tau{(\medcirc\odot)} = 2.3 (1)$~s and $\tau{(\medcirc\medcirc)} = 4 (1)$~s, approximately corresponding to the duration of a linescan.
Our measurement procedure involves a blue 445 nm charge initialization pulse between each linescan, accompanied by continuous orange 619 nm laser illumination. These timescales imply that the primary driver for changes in charge trap states is the blue laser (see Supplementary Fig.~\ref{fig:stabilizationMethod}), suggesting the potential for maintaining trap state stability during optical operations resonant with SnV transitions.
Since the trap states are stable much longer than the measured SnV ionization time of 50~ms \cite{Goerlitz2022NPJ} and spin coherence time of about 1~ms \cite{Debroux2021PRX}, although not deterministically, the emitter may still act as an optically coherent spin-photon interface.

\subsubsection*{Data analysis steps}

\begin{enumerate}
    \item Wavemeter correction: During scans, the frequency of the laser is controlled by applying an external voltage signal. We monitor the laser frequency through a pick-off path directed at a wavemeter. PLE spectra are initially recorded as a set of voltages and fluorescence signals. The voltage can then be converted to frequencies by matching the time stamps. Any nonlinear frequency changes occurring during the scan are therefore accounted for.
    \item Binning: Individual scans are mapped to a frequency axis by selecting an individual linescan and subsequent frequency binning. If multiple data points fall into the same bin, they are averaged. If a bin remains empty, the average of the previous and next bin is used.
    \item Histogramming scans: Binned scans are summed and normalized to create the histogrammed PLE spectra.
    \item Configuration identification: A peak finder algorithm (MATLAB: findpeaks) is used to identify the frequencies of the four peaks. These peaks are then labeled and used for averaging the spectral location corresponding to a particular charge configuration of proximity traps.
    \item Configuration ranges: We separate state configurations by assigning a spectral range to each central peak position. This range is half the spectral distance between two adjacent peaks. 
    \item Scanwise peak identification: The same peak finder algorithm is used on each individual scan to find peaks.
    \item Scanwise configuration identification: The identified peaks are then matched with a charge state configuration depending on their central frequencies.
    \item Determining the brighttime durations: The 'brighttime' is determined by the amount of time a peak is associated to the same charge configuration until a change occurs. Each brighttime is recorded together with the changes in the charge state configuration.
    \item Histogramming brighttimes: The brighttime durations are histogrammed according to the occurrence they were observed, with the goal of extracting averaged lifetimes and switching rates. 
    \item Charge state change probability $p(i{\rightarrow}j)$:  The amount of times that a spectral jump from a charge state $i$ to another $j$ has occurred is recorded. They are then normalized to the total amount of jumps from configuration $i$ to  obtain a probability.
    
    There are two factors that limit the quantification of uncertainties. 
    First, jumping events depend on the individual identification of peak locations per line. The implemented peak finder algorithm locates the maximum of a line for each scan. Due to spectral diffusion, it is not possible to fit every individual line and extract a central frequency uncertainty.
    Secondly, there are overlaps of individual spectral peaks in the integrated spectrum in Fig.~\ref{fig:sensor}\hyperlink{fig:sensorL}{b} from the main text. Even though we select a cut-off position in the middle of the peaks, some of the identified peaks could actually belong to the neighboring spectral peak's tail, instead of where we identified its position. Therefore, we assign an overall uncertainty factor by computing the overlap of the individual integrated fits of the individual peaks. We then multiply these factors with the extracted probabilities.
    \item Poisson fitting: The histograms are converted to probability densities and then fit with a Poisson distribution. After the fit, the histogram and the fit are scaled back to the original occurrences. The brighttimes are then converted to real time units by the duration of a single scan.
    \item Lifetime $\tau(i)$ and conditional spectral jump rate $\varGamma_{\rm ct}(i{\rightarrow}j)$ extraction: Mean values of the Poisson distribution fits and their uncertainties are provided as proximity charge configuration lifetimes, or their inverse as state switching rates between configurations.
\end{enumerate}

\subsection*{Charge trap-illumination field interactions}
\label{sec:ap_diff}

In the main text, we show the influence of charge trap densities on spectral diffusion. Here, we provide experimental results on how the properties of the laser can affect spectral diffusion. Since the illumination induces the ionization events in the sample, we show that the interactions and observed phenomena are consistent with the existence of charge traps.

\subsubsection*{Position dependency of the stabilization laser and subdiffraction drift sensing}

    \begin{figure*}  
    \centering
    \includegraphics[width=\linewidth]{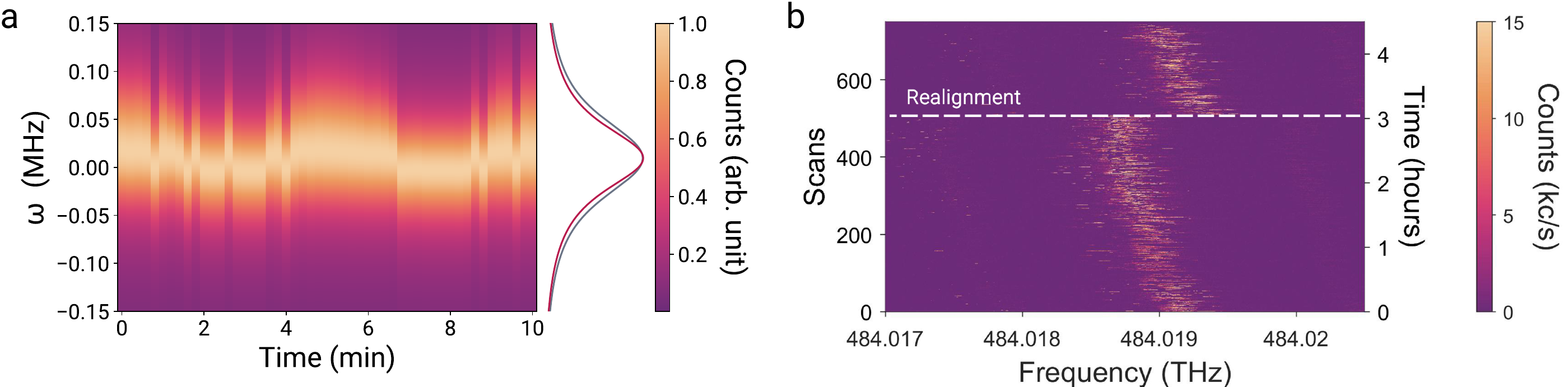}
    \caption{\label{fig:misalignmentDrift} 
    \textbf{Influence of laser misalignment on the PLE spectra.}
    \protect\hypertarget{fig:misalignmentDriftL}{}
    \textbf{a} Simulation of the temporal change of the central position and linewidth of the C-transition caused by a periodic change in the alignment of the charge state stabilization laser, which can result from temperature changes in the experiment \cite{Goerlitz2022NPJ}. We assume a power broadened homogeneous linewidth of ${\rm FWHM}_{\rm hom} = 88$ MHz (red line) and find an inhomogeneous linewidth of ${\rm FWHM}_{\rm hom} = 103.3$ MHz (extracted from fit with a Voigt profile, dark gray line) for the charge state polarization laser parameters provided in the main Suppl. text.
    \textbf{b} An exemplary PLE measurement, showing a drift of the resonance frequency. After optimizing the xyz position of the sample and laser spot, the central frequency returns to the original position.
    }
    \end{figure*}

A peculiar property of the ZPL of an SnV was reported in\citen{Goerlitz2022NPJ}, Fig. 3B where the spectral line drifted in correlation with the laboratory air conditioning cycle. In order to investigate this further, we conduct simulations to reproduce the periodic changes and inhomogeneous broadening of the reported PLE measurement. Our simulations involve introducing a periodic misalignment of the laser by varying the participating remote charge density. 

We assume that the blue stabilization laser has a Gaussian intensity distribution in the $z$-direction that oscillates in time
\begin{equation}
    I(z,t) = I_0 e^{ - [z - z_0(t)]^2 / 2 \sigma} ~, 
\end{equation}
 where $I_0$ is the laser's peak intensity at the focal point, $\sigma$ the focal width and 
 \begin{equation}
     z_0(t) = a \sin(\omega t)~.
 \end{equation}
 The amplitude $a$, describing the magnitude of misalignment due to the changes in temperature of the setup is not known. The frequency $\omega = 2 \pi / T$ corresponds to the $T = 10$ min cycle of the air conditioning described in\citen{Goerlitz2022NPJ}. 
 To perform the Monte Carlo simulation we follow the previously outlined steps with the field produced by an ionized trap given by 
  \begin{equation}
    \vec{E}(q, \vec{r}) =  \frac{q_i}{4 \pi \epsilon_0 \epsilon_r} \frac{\vec{r}}{r^{3}}~.
\end{equation}
 We randomly distribute traps with a density of $\rho = 22.7$ ppm in a cubic volume with an edge length of 100 nm. 
 The trap density well reproduces the inhomogeneous broadened linewidth of $\approx 103$ MHz shown in Fig. 3b of\citen{Goerlitz2022NPJ} for a power broadened homogeneous linewidth of $\approx 88$ MHz. 
 We assume that the probability of a trap participating in being ionized is given by $P(t) = P(z,t) + P_0$, where $P(z,t) \propto I(z,t)$ and $P_0 = 0.1$ is a constant background ionization probability. For the Monte Carlo simulation of the inhomogeneous linewidth at each time step $t$ we use 2500 different charge configurations. We found very good agreement with the results reported in\citen{Goerlitz2022NPJ} for a laser with a focal spot width of ${\rm FWHM} = 240$ nm ($\sigma = {\rm FWHM} / 2 \sqrt{2 \log(2)}$) and an oscillation amplitude of $a = 200$ nm. The results can be seen in Supplementary Fig.~\ref{fig:misalignmentDrift}\hyperlink{fig:misalignmentDriftL}{a}.

To further confirm our model, we performed a long-term PLE scan (Supplementary Fig.~\ref{fig:misalignmentDrift}\hyperlink{fig:misalignmentDriftL}{b}) and utilized the xyz control of the confocal microscopy setup to optimize the fluorescence signal. By monitoring the changes in the spectral line, we were able to measure a $\sim$200 MHz drift over a 3-hour period, which corresponds to a $\sim$50 nm shift according to our position optimizer. By realigning the setup, we were able to retrieve the initial position of the resonance, providing further evidence in support of our hypothesis.

We extracted spectral drift relations $\sim$0.2 MHz/nm for Supplementary Fig.~\ref{fig:misalignmentDrift}\hyperlink{fig:misalignmentDriftL}{a} and $\sim$4 MHz/nm for Supplementary Fig.~\ref{fig:misalignmentDrift}\hyperlink{fig:misalignmentDriftL}{b}. This means, depending on the surrounding charge density, it would be reasonable to estimate MHz/nm correspondence of the laser position drift to the emission central frequency. We propose a spectral test such as this could prove useful for estimating the positional drifts under the diffraction limit. It has been demonstrated that chirped pulses from an EOM can scan 200 MHz range under a second \cite{Martinez2022PRL}. Therefore, using a spectral approach would also allow a higher bandwidth exceeding read-out rates from the fluorescence intensity-based schemes \cite{Mortensen2010Nat}.

Overall, an SnV, or emitters with inversion symmetry in general, can be used to temporally resolve the participating remote charge trap density at each moment. By correlating the central frequency, spatial drifts in experimental systems can be tracked. 

\subsubsection*{Stabilization method of the emitter:} \label{sec:ap_stabilization}

\begin{figure*}[]  
    \centering
    \includegraphics[width=0.90\linewidth]{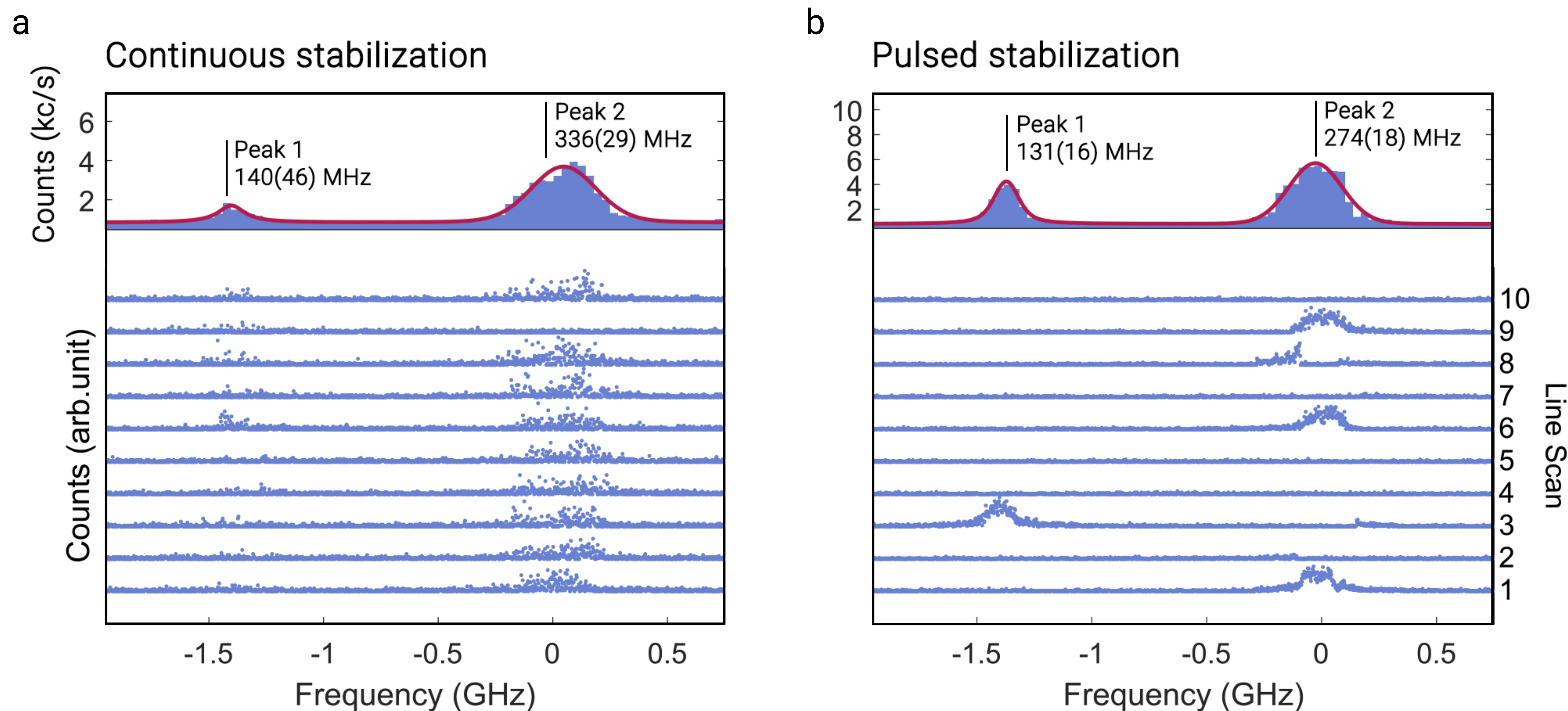}
    \caption{\label{fig:stabilizationMethod}
        \textbf{Photoluminescence excitation (PLE) scans of the C transition under different charge stabilization schemes on the emitter E1.}
        During the scans, a hopping between two different resonances is observed. The
        resonant laser has a power of 0.7 nW, and the blue laser has a power of 300 nW.
        The spectra are fit using Voigt profiles. The uncertainties represent 95\% confidence intervals extracted from the fits.       \protect\hypertarget{fig:stabilizationMethodL}{}
        \textbf{a} The blue laser at 450 nm is illuminating the sample continuously during the  resonant laser scans.
        The continuous operation of the blue laser induces hopping that is faster than a single line scan, which results in  both peaks being observable in every individual scan.
        \textbf{b} A 4 ms blue laser pulse is sent at the beginning of each scan.
        Without the help of the blue laser, the cycling between the two resonances is slower but still present due to the laser that is resonant with the C transition.
    }
\end{figure*}

We also investigate the spectral properties of our emitters using different charge stabilization procedures involving blue laser light to study its interaction with the V$_{\rm n}$. Supplementary Fig.~\ref{fig:stabilizationMethod} shows PLE scans and spectra using two different stabilization schemes with a charge stabilization laser at 450 nm and 300 nW average power. The first scheme uses continuous-wave (CW) laser light during each PL scan (continuous stabilization). The second is a pulsed scheme: before each PLE scan a blue laser pulse of 4 ms duration irradiates the sample (pulsed stabilization). PLE scans were performed on emitter E1 with a resonant power of 0.7 nW, which lies below the saturation power ($> 20$ nW), and is also low enough to avoid ionization during the scan.

Supplementary Fig.~\ref{fig:stabilizationMethod}\hyperlink{fig:stabilizationMethodL}{a} clearly shows both two resonance peaks ($\sim1.4$ GHz apart), that are also present in each individual linescan. The individual PLE scans of the pulsed scheme in Supplementary Fig.~\ref{fig:stabilizationMethod}\hyperlink{fig:stabilizationMethodL}{b} reveal that both resonances correspond to two distinct spectral positions of the C transition, which we attribute to Stark shift (Eq.~\eqref{eq:nonlin_stark}, main text) induced by two distinct charge configurations of ionized V$_{\rm n}$ in the vicinity of the SnV. A quasi-continuous fluorescence signal with the resonant laser is observable showing the full inhomogeneous linewidth. Continuous stabilization will cycle the charge state of the environment with a spectral jump rate $\varGamma_{\rm SH} \gg \varGamma_{\rm scan}$ much higher than the PLE scan rate, leading to two recognizable peaks during individual PLE scans. 

Another clear signal of the increased ionization of V$_{\rm n}$ is the more pronounced inhomogeneous broadening of the resonance lines under continuous stabilization. 450 nm CW light will cause more traps in the environment to participate in creating the fluctuating electric field at the emitter's position during each individual scan. Just as predicted by our Monte Carlo simulations, an increased activity of charge traps leads to increased inhomogeneous broadening.

\subsubsection*{Wavelength dependency of the stabilization laser:}
\label{sec:ap_wavelength}

    \begin{figure*}
        \centering
        \includegraphics[width=0.90\linewidth]{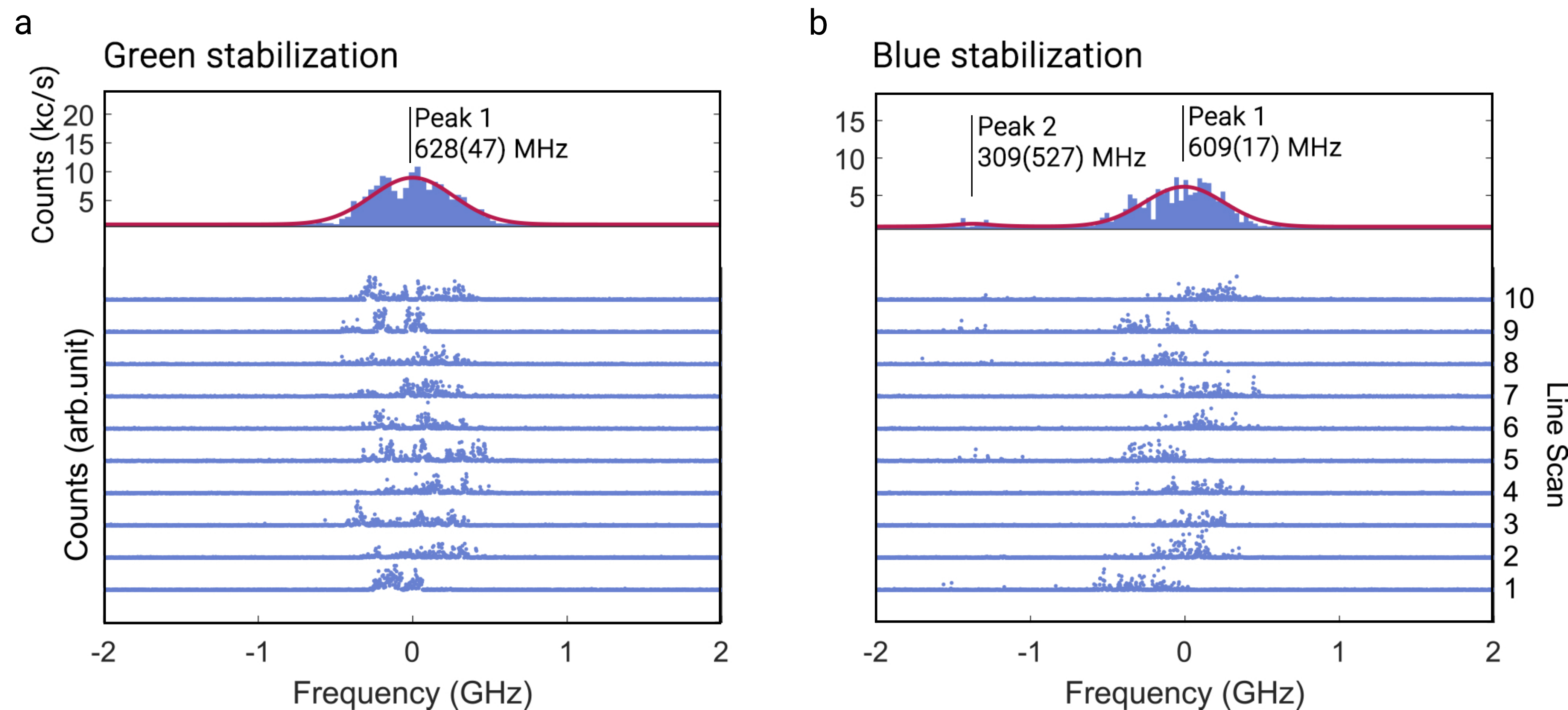}
        \caption{ \label{fig:greenVSblue}
            \textbf{Photolimunescence excitation (PLE) scans of the C transition under different colored stabilization schemes of the emitter E2.}
            C laser had a power of 1 nW. The spectra were fit using bimodal Voigt profiles. Uncertainties represent 95\% confidence intervals extracted from the fits.           \protect\hypertarget{fig:greenVSblueL}{}
            \textbf{a} 500 nW green laser at 520 nm is kept continuously on as the resonant laser scans. Continuous fluorescence from smaller peaks were sometimes observed. The
            secondary peak was not observable in this configuration.
            The specta were fit using a Voigt profile.
            \textbf{b} 500 nW blue laser at 450 nm is kept continuously on as the resonant laser scans. The blue laser produces a spectral jump resulting in a secondary peak.
            The spectrum was fit using a bimodal Voigt profile.
    }
    \end{figure*}

    An indication that charge dynamics and occupation of nearby traps playing a role in the spectral jumping phenomena comes from comparing charge stabilization with blue (450 nm) and green lasers (520 nm).
    Here in Supplementary Fig.~\ref{fig:greenVSblue}, PLE spectra of the emitter E2 taken under the same resonant and stabilization laser powers show a single peaked behavior with the green laser whereas a smaller second peak (although weakly) can be observed when the blue laser is used.
    It was shown that blue laser irradiation is more efficient for charge trap ionization \cite{Goerlitz2022NPJ}.
    Using that information one can deduce that a previously inaccessible charge trap is activated with the blue laser, resulting in the new discrete spectral jump.
    Spectroscopy of the charge trap transition rates could help in identifying ionization energies for individual charge trap species. For example, it is possible to observe, although only qualitatively, a more quickly switching fluorescence signal within the PLE acquisition resolution with the blue laser at each single line resulting from rapid spectral jumps.

\subsubsection*{Power dependency of the stabilization laser:} \label{ap:powerDiffusion}

    \begin{figure*}
        \centering
     \includegraphics[width=0.95\linewidth]{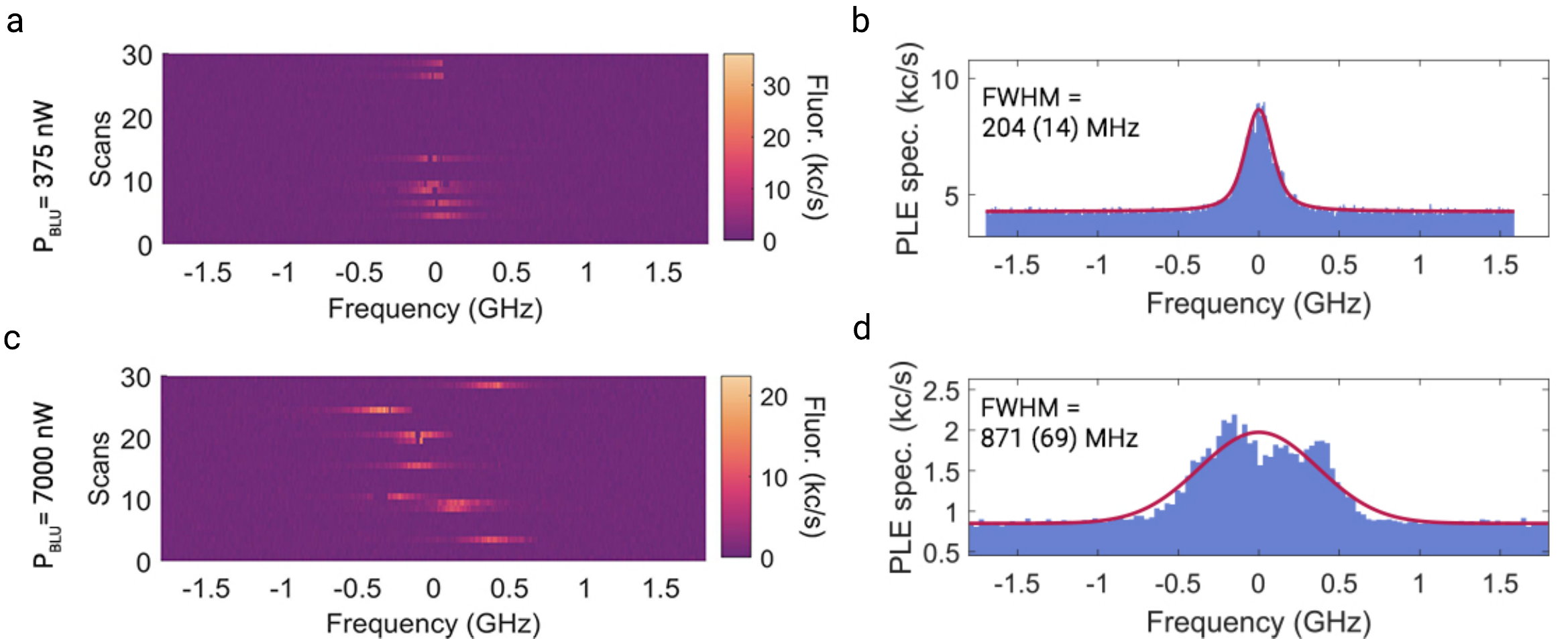}
        \caption{\label{fig:initPower}
            \textbf{Comparison of different stabilization pulse powers illuminating the emitter E14.}
            Both measurements are done under 0.5~nW resonant laser excitation.
            Two different blue laser powers (P$_{\rm BLU}$) of 375~nW and 7000~nW are used.
            Higher power resulted in a broadened inhomogeneous linewidth and more pronounced spectral drifts.
            The spectra were fit using a Voigt profile. The uncertainties represent 95\% confidence intervals extracted from the fits.             \protect\hypertarget{fig:initPowerL}{}
            \textbf{a},\textbf{c} Measured fluorescence at each cycle as the laser frequency was scanned.
            \textbf{b},\textbf{d} Histogrammed counts for linewidth determination using a Voigt fit.}
    \end{figure*}

    \begin{figure}
        \centering
        \includegraphics[width=\linewidth]{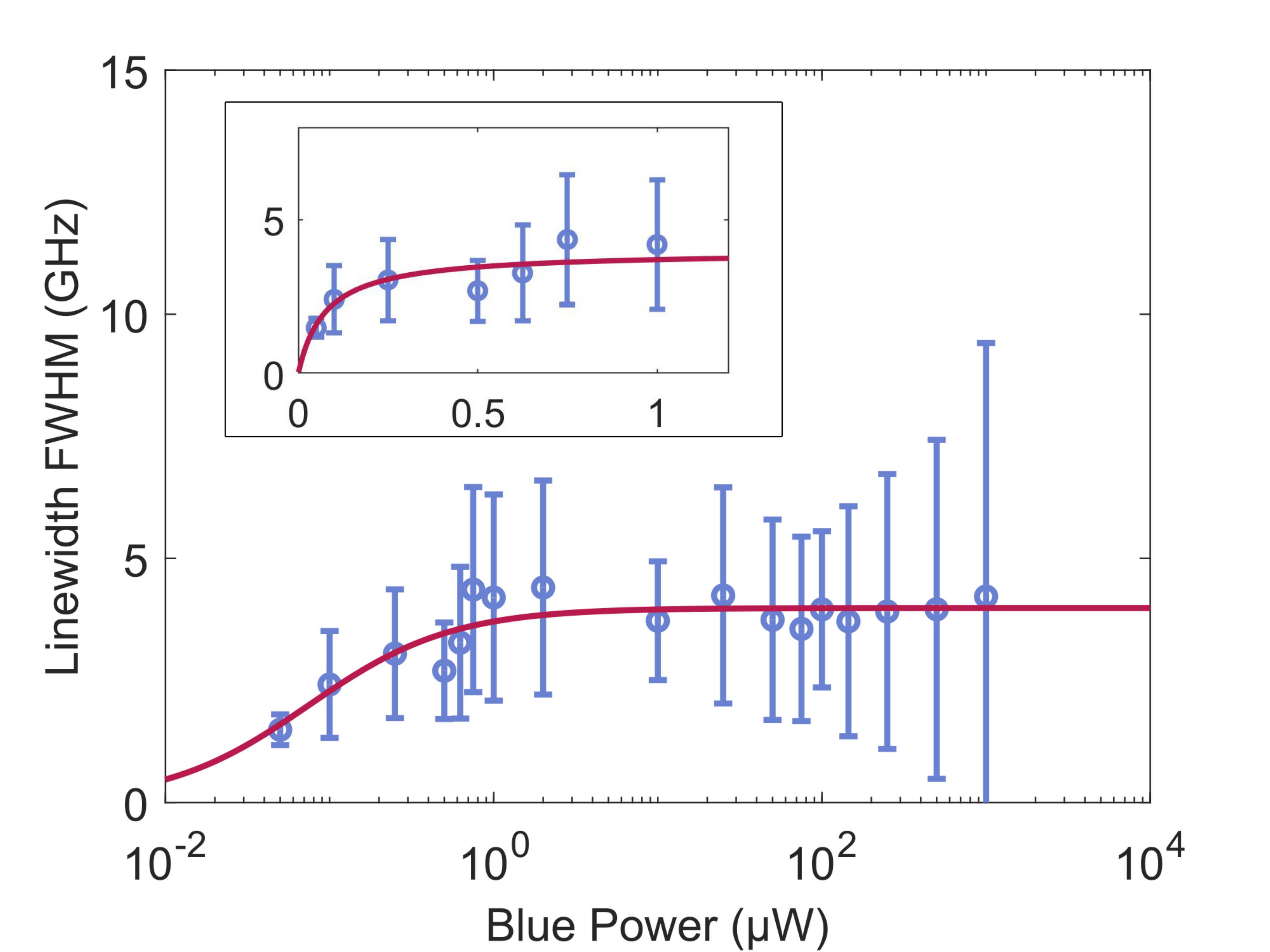}
        \caption{\label{fig:snorlaxDiff}
            \textbf{Comparison of different stabilization powers on the emitter E20.}
            Measurements are taken on another sample which had a five times higher Sn implantation dose compared to the sample used in the main text and was co-implanted with sulfur.
            The same resonant excitation power of 5 nW is used in all measurements while the power of the blue continuous stabilization laser is varied.
            Increasing blue power broadens the linewidth reaching an asymptotic limit.
            The inset shows a zoom-in of the smaller powers to better demonstrate the saturation trend. 
            Error bars are the 95\% fit confidence intervals which are heavily influenced by the background fluorescence induced by the comparatively large blue laser powers.}
    \end{figure}

    Extended resonant excitation of SnVs induces a dark state transition.
    This has been connected to a change of its charge state through the promotion of an electron from the valance band.
    A hole capture process induced by blue or green lasers can reinitialize the SnV back to its bright state.
    Using higher powers or longer illumination periods increase the probability of charge state stabilization and reemission \cite{Goerlitz2022NPJ}.
    This is enabled by ionizing or charging the defects around the quantum emitter which act as charge/hole donors.
    As a result, illumination changes the charge distribution around the color center and induces spectral diffusion.
    Therefore, charge stabilization and inhomogeneous broadening become competing effects where one has to optimize both for high quality emission.
    
    In Supplementary Fig.~\ref{fig:initPower}, a demonstration of this trade-off in a measurement on emitter E14 is provided.
    Using a power of 7000 (375) nW results with 30\%, 9/30 (23\%, 7/30) of the time bright lines and a histogrammed linewidth of 871 (204) MHz.
    Because charge traps play such a crucial role in the stabilization of the bright state, a competition between spectral diffusion and the charge state stabilization efficiency $\eta_{\rm bright}$ is expected. The ideal V$_{\rm n}$ (or hole donors, in general) density can then be determined from a compromise between minimizing spectral diffusion and maximizing $\eta_{\rm bright}$.

    Additionally, we tried to identify the contribution of blue laser to the spectral diffusion with respect to its power.
    In Supplementary Fig.~\ref{fig:snorlaxDiff}, we present measurements from the emitter E20, that is on the sample E014 which has a similar fabrication recipe with E001 with five times higher Sn implantation dosage and an extra Sulphur co-implantation step.
    It can be clearly seen that higher blue laser light powers introduce more pronounced broadening, displaying a degree of saturation, which again is in accordance with both our Monte Carlo simulations and our previous work on spectral diffusion \cite{OrphalKobin2023PRX}.

\section*{Additional discussion}
 
 \subsection*{Current limitations of the electrometer and how to overcome them}   \label{sec:ap_limitations}
 
Besides the many exciting opportunities our electrometer offers, we must also critically reflect on the present limitations. In addition to the limited time resolution discussed in the main text, the calibration process can also be improved.  We have used an experimentally determined scalar value for the induced polarizability, and not a full tensor, requiring us to infer the direction of the background electric field from consideration of the implantation created defects. The same limitation also prohibits perfectly localizing charge traps in three dimensions. 

Furthermore, we assume that the position of the observed spectral features is purely caused by induced Stark shifts.  We cannot finally exclude any other mechanisms such as strain or phononic contributions, however, the experimental results and our model agree for a variety of local defect configurations, and we therefore conclude that our method is self-affirmative. 
Any dynamical strain environment cannot also be fully excluded. However, measurements of lifetime-limited lines without any dynamical broadening in implanted samples have been reported \cite{Goerlitz2022NPJ, Narita.2023}; therefore, it can be claimed that such an effect did not play a role in previously measured spectra. For the GeV \cite{li_atomic_2024}, dynamic strain stemming from changes in the Jahn-Teller configuration of surrounding defects has been ruled out as the source of the observed spectra due to a significant mismatch between the predicted and observed hopping rates. 

In addition, we have not calibrated the power broadening effects on the natural linewidth of the emission since a saturation measurement is not feasible when a constant signal cannot be retrieved from a spectrally unstable emitter. We can, however, estimate that the effect is negligible since the broadening follows the following equation\cite{Cohen1998}:

\begin{equation}
    \gamma_{\rm broadened} = \gamma_{\rm natural} \times \sqrt{1 + \frac{P}{P+P_{\rm saturation}}} 
\end{equation}
where $\gamma$ represents the linewidths and $P$ represents excitation powers. Assuming an under-estimated saturation power $P_{\rm saturation}$ of 10 nW (it was measured to be 4.2 µW in a similarly prepared sample in\citen{Trusheim2020PRL}), our measurement power of 0.5 nW corresponds to a broadening less than 2.5\%.

Moreover, we are not able to distinguish between all the trap configurations that would yield the same integrated spectral fingerprint.  Using the time resolved charge dynamics in conjunction with a detailed physical model of the trap ionization, including the implied charge dynamics and thermalization as well as other possible charge localization mechanisms such as Anderson localization \cite{anderson_absence_1958} can improve on this ambiguity. 

Although the SRIM simulations yield valuable insights into the distribution of V$_1$ lattice defects during the implantation process, they do not provide an accurate estimation of the post-annealing distribution of higher-order V$_{\rm n}$ defects. In this study, we take a preliminary step by simulating a V$_2$ distribution based on the initial V$_1$ distribution using stochastic methods. However, to obtain more realistic distributions and extend the estimation to multi-vacancy complexes, more advanced techniques such as molecular dynamics simulations, as demonstrated in\citen{lehtinen_molecular_2016}, or incorporating formation energies and diffusion paths on a 3D crystal lattice, as explored in\citen{Slepetz2014PhysChem}, can be employed. These sophisticated approaches have the potential to improve the modeling of realistic distributions and enhance the generalization of estimations for multi-vacancy complexes.

\bibliography{refz}